\documentclass[journal=jacsat,manuscript=article]{achemso}
\setkeys{acs}{maxauthors=10,etalmode=truncate,chaptertitle =true,articletitle=true}

\usepackage{amssymb}
\usepackage{amsmath}
\usepackage[usenames,dvipsnames]{xcolor}
\usepackage{graphicx}
\usepackage{float}
\usepackage[normalem]{ulem}
\usepackage{caption}
\usepackage{subcaption}
\usepackage{natbib}
\usepackage{multirow}
\usepackage{bm}
\usepackage[version=4]{mhchem} 
\SectionNumbersOn

\title{Fast and Scalable GPU-Accelerated Quantum Chemistry for Periodic Systems with Gaussian Orbitals: Implementation and Hybrid Density Functional Theory Calculations}
\author{Yuanheng Wang}
\affiliation
{Department of Chemistry and The PULSE Institute, Stanford University, Stanford, California 94305, United States}
\alsoaffiliation{SLAC National Accelerator Laboratory, Menlo Park, California 94024, United States}
\altaffiliation{These authors contributed equally to this work.}
\author{Diptarka Hait}
\affiliation
{Department of Chemistry and The PULSE Institute, Stanford University, Stanford, California 94305, United States}
\alsoaffiliation{SLAC National Accelerator Laboratory, Menlo Park, California 94024, United States}
\altaffiliation{These authors contributed equally to this work.}
\author{Pablo A. Unzueta}
\affiliation
{Department of Chemistry and The PULSE Institute, Stanford University, Stanford, California 94305, United States}
\alsoaffiliation{SLAC National Accelerator Laboratory, Menlo Park, California 94024, United States}
\author{Juncheng Harry Zhang}
\affiliation
{Department of Chemistry and The PULSE Institute, Stanford University, Stanford, California 94305, United States}
\alsoaffiliation{SLAC National Accelerator Laboratory, Menlo Park, California 94024, United States}
\author{Todd J. Mart{\'i}nez}
\email{todd.martinez@stanford.edu; toddjmartinez@gmail.com}
\affiliation
{Department of Chemistry and The PULSE Institute, Stanford University, Stanford, California 94305, United States}
\alsoaffiliation{SLAC National Accelerator Laboratory, Menlo Park, California 94024, United States}

\begin{document}
\maketitle
\begin{abstract}
    Efficient hybrid DFT simulations of solid state materials would be extremely beneficial for computational chemistry and materials science, but is presently bottlenecked by difficulties in computing Hartree-Fock (HF) exchange with plane wave orbital bases. We present a GPU-accelerated, Gaussian orbital based integral algorithm for systems with periodic boundary conditions, which takes advantage of Ewald summation to efficiently compute electrostatic interactions. We have implemented this approach into the TeraChem software package within the $\Gamma$ point approximation, enabling simulation of unit cells with hundreds or thousands of atoms at the HF or hybrid DFT level on a single GPU card. Our implementation readily parallelizes over multiple GPUs and paves the road to accurate simulation of the properties and dynamics of extended materials in both the ground and excited states.  
\end{abstract}

\section{Introduction}
Accurate \textit{ab initio} simulation of extended systems would enable efficient \textit{in silico} design of better materials for chemical applications like heterogenous catalysis, electrochemistry, or optoelectronics. The significant computational cost for modeling the condensed phase (as opposed to small gas phase molecules) has led to extensive use of density functional theory\cite{ksdft,review_dft_solid,dft_energy_review} (DFT) with local functionals\cite{pbe_functional,SCAN_functional}, although simulations with hybrid DFT functionals employing some proportion of Hartree-Fock (HF) exchange, \cite{becke_hybrid,pbe0_functional,hse06_functional} and even correlated wavefunction techniques like coupled cluster theory \cite{cc_mp2_1d, ccsd_pbc_1, ccsd_pbc_2, eom_ccsd_pbc,cc_metal,local_periodic_cc} are becoming more frequent in recent years. 

The practical feasibility of applying various \textit{ab initio} modeling methods to extended materials is strongly tied to the choice of the basis for representing orbitals. Plane wave basis sets have long been utilized for solid state simulations as they can naturally take advantage of the translational symmetry of crystals\cite{bloch_original} and are well suited for representing highly delocalized electron densities. Plane waves are also intrinsically orthonormal and have a particularly simple form for the Coulomb interactions \cite{plane_wave_dft}, leading to continuing popularity of this approach in software packages \cite{quantum_espresso_original, quantum_espresso_recent, quantum_espresso_plane_wave_exchange, vasp_1, vasp_2, vasp_3, vasp_recent, abinit_original, abinit_recent, gpaw_2010, gpaw_recent, fhi_aims_2009}. However, the delocalized nature of individual plane waves necessitate the use of a very large number of plane wave basis functions for spatially compact orbitals, like those for inner-shell electrons. This is generally addressed through the use of pseudopotentials \cite{pseudopotential_GTH_1, pseudopotential_GTH_2, pseudopotential_GTH_3, pseudopotential_norm_conserving_1, pseudopotential_norm_conserving_2, pseudopotential_norm_conserving_3,vanderbilt_soft_pseudopotential}, augmented plane wave \cite{augmented_plane_wave_original, linear_augmented_plane_wave_original, projector_augmented_plane_wave_original, projector_augmented_plane_wave_2, projector_augmented_plane_wave_3, projector_augmented_plane_wave_nice_equation} or projection methods \cite{plane_wave_projection_to_gaussian_1, plane_wave_exchange_projection, plane_wave_exchange_exact_2}, but these approximations may also lead to significant errors \cite{pseudopotential_error}. The evaluation of HF exchange is also computationally demanding with plane wave basis functions \cite{quantum_espresso_plane_wave_exchange, plane_wave_exchange_projection, plane_wave_exchange_exact, plane_wave_exchange_exact_2}, resulting in continued widespread use of local DFT methods that have been largely superseded by more accurate hybrid functionals in studies of isolated molecular systems\cite{molecular_dft_review}. 

The use of atom-centered Gaussian basis functions addresses many of these challenges as their local form permits \textit{ab initio} description of inner shell electrons without recourse to additional approximations or large numbers of basis functions. This has consequently led to greater interest in using such bases for simulation of solid state materials in recent years \cite{periodic_hf_1, crystal_17, crystal_23, pyscf_mixed_density_fitting, pyscf_periodic_new, cp2k_recent, qchem_periodic_1, qchem_periodic_2, qchem_periodic_3, mpqc_real_density_fitting, periodic_integral_gamma}. The Hamiltonian integrals with Gaussian type basis functions are however much more complicated and the evaluation of electrostatics for extended systems is challenging due to the long-ranged nature of the Coulomb operator. 
There are several possible approaches that can be utilized to efficiently compute long-range electrostatic interactions involving Gaussian basis functions, such as Ewald summation, \cite{ewald_original, periodic_integral_gamma} plane wave density fitting, \cite{pyscf_mixed_density_fitting, cp2k_mixed_basis, qchem_periodic_1, qchem_periodic_2, pyscf_range_separated_gaussian_density_fitting}  and multipole expansion \cite{crystal_bipolar_expansion, fast_multipole_method, very_fast_multipole_method, fast_multipole_method_linear}.
In particular, the Ewald summation method\cite{ewald_original} separates the electrostatic interactions into short- and long-range components, which are summed over in the real and reciprocal space respectively, to reduce total computational cost.

General-purpose graphics processing units (GPUs) have emerged as powerful computational tools with massively parallel processing capabilities and high memory bandwidth, making them well-suited for the demanding computational requirements of Hamiltonian integral evaluations. The TeraChem electronic structure package pioneered the use of GPUs for accelerating \textit{ab initio} calculations on discrete molecular systems with Gaussian type basis functions \cite{terachem_gpu_1, terachem_gpu_2, terachem_gpu_3, terachem_2013, terachem_2021,terachem_f}, and was followed by other packages providing GPU support \cite{brianqc, gaussian_program, gamess_spd, fermions_original, quick, libintx_up_to_iiii, pyscf_gpu, quantum_supercharger_library, turbomole, wuhan_electronic_structure_package}. 
TeraChem accelerates HF, DFT, coupled cluster \cite{terachem_ccsd_gpu, terachem_rank_reduce_cc_1, terachem_rank_reduce_cc_2, terachem_rank_reduce_cc_3, terachem_rank_reduce_ccsd} and multireference quantum chemistry methods\cite{terachem_casscf, terachem_fomo_casci, terachem_direct_ci_1, terachem_direct_ci_2, terachem_direct_ci_3, terachem_rank_reduced_ci}; permitting efficient simulation of \textit{ab initio} molecular dynamics. Expanding the TeraChem codebase to support simulation of periodic systems therefore represents a natural route towards GPU-accelerated, accurate \textit{ab initio} simulation of the properties and dynamics of extended materials. In particular, GPU-acceleration will facilitate the modeling of systems with large unit cells, such as solid state point defects\cite{point_defect_review} and liquids. Large unit cells would also assist in simulating excited state dynamics by permitting greater spatial separation between localized excitons. In passing, we note that software packages providing GPU acceleration for pure plane wave or plane-wave density fitting calculations are available at present \cite{vasp_gpu_1, vasp_gpu_2, vasp_gpu_3, pwmat_gpu_1, pwmat_gpu_2, cp2k_gpu, quantum_espresso_gpu_1, quantum_espresso_gpu_2, abinit_2016_gpu, gpaw_gpu,west_program}, but the use of GPUs for accelerating \textit{ab initio} integral evaluations for periodic systems with Gaussian type orbitals appears to be underexplored. 

In this work, we present our GPU-accelerated integral algorithm for evaluation of Hamiltonian integrals with periodic symmetry along all three spatial dimensions, as well as their utilization for HF and DFT simulations in extended materials. We treat long-range electrostatic interactions with Ewald summation, and apply the $\Gamma$-point approximation to avoid having to explicitly sample reciprocal space. We note that our interest in simulating molecular dynamics furthermore leads us to treat all lattices with the $P_1$ space group symmetry and not consider any rotational or inversion symmetries. We present the definitions and GPU implementable formulae for the basis functions, most of the integrals, and other relevant terms involved in periodic HF or DFT calculations in Section \ref{sec:theory}. In Section \ref{sec:implementation} we describe the detailed aspects of implementing periodic integrals that differ from integrals for discrete molecular systems. Finally, in Section \ref{sec:performance} we show the performance of our implementation for all-electron HF/DFT simulations of model systems, examine parallelization over multiple GPUs, and compare to other packages.

\section{Theory\label{sec:theory}}
Electronic structure calculations with periodic boundary conditions require implementation of various integral routines, including overlap, electron kinetic energy, nuclear-electron attraction, DFT local exchange-correlation, and electron repulsion integrals. The definition and implementation of these integrals have been previously presented in a number of publications, \cite{periodic_hf_1, periodic_integral_gamma, crystal_23} but not to a great level of detail, in our opinion. Therefore, in this section, we present a comprehensive and implementable list of formulae for all aforementioned integrals.

\subsection{Periodic Gaussian basis function}

We express the molecular crystalline orbitals $\psi_{i, \vec{k}}(\vec{r})$, corresponding to some reciprocal space vector $\vec{k}$ inside the first Brillouin zone, as linear combinations of atomic crystalline orbitals $\phi_{\mu, \vec{k}}(\vec{r})$:
\begin{align}
\psi_{i, \vec{k}}(\vec{r}) &= \sum_{\mu}^{n_{AO}} C_{\mu,i}^{\vec{k}} \phi_{\mu, \vec{k}}(\vec{r})
\label{eq:crystalline_molecular_orbital_definition}
\end{align}

The atomic crystalline orbitals are defined as:
\begin{align}
\phi_{\mu, \vec{k}}(\vec{r}) &= \sum_{\vec{R} \in \mathbb{Z}^3} e^{i \vec{k}\cdot\vec{R}} \phi_{\mu}^{local}(\vec{r} - \vec{R})
\label{eq:crystalline_atomic_orbital_definition}
\end{align}
In this equation, $\phi_{\mu}^{local}(\vec{r})$ is a local function defined in one unit cell and $\displaystyle\sum_{\vec{R} \in \mathbb{Z}^3}$ denotes a summation over all possible crystal lattice vectors $\{\vec{R}\}$ (i.e. $\vec{R} \in \{n_a \vec{R}_a + n_b \vec{R}_b + n_c \vec{R}_c , \forall n_a, n_b, n_c \in \mathbb{Z}\}$, where $\vec{R}_a$, $\vec{R}_b$ and $\vec{R}_c$ are primitive cell vectors). 
This form of $\phi_{\mu, \vec{k}}(\vec{r})$ satisfies Bloch's theorem \cite{bloch_original} as:
\begin{align}
\phi_{\mu, \vec{k}}(\vec{r} + \vec{R}) = \phi_{\mu, \vec{k}}(\vec{r})e^{i\vec{k} \cdot \vec{R}}
\label{eq:bloch_theorem}
\end{align}
As a result, $\psi_{i, \vec{k}}(\vec{r})$ will also satisfy Bloch's theorem.

We choose contracted atom-centered Cartesian Gaussian type orbitals (GTO) as $\phi^{local}_{\mu}$, which in turn are linear combinations of primitive Gaussian type orbitals (pGTOs):
\begin{align}
\phi^{local}_{\mu}(\vec{r}) = \sum_m^{N_{contraction}} C_m^{contraction} \mu_m^{local}(\vec{r})
\label{eq:local_contracted_basis}
\end{align}
where each primitive Gaussian type orbital (pGTO) has the form
\begin{align}
\mu^{local}(\vec{r};\vec{i},a,\vec{A}) = C^{normalization} (x - A_x)^{i_x} (y - A_y)^{i_y} (z - A_z)^{i_z} e^{-a \left| \vec{r} - \vec{A} \right|^2}
\label{eq:local_primtive_basis}
\end{align}
All parameters in the pGTO definition have the same meaning as in non-periodic calculations (as mentioned in Ref \citenum{terachem_f}): $a$ represents the exponent of the primitive Gaussian function, the angular momentum index vector $\vec{i}$ (with components $i_x$, $i_y$ and $i_z$) determines the shape of the atomic orbital, and $\vec{A}$ is the center location of the atomic orbital in any lattice image. The normalization constant
\begin{align}
C^{normalization} = \left(\frac{2a}{\pi}\right)^{3/4} \left( \frac{(8a)^{i_x+i_y+i_z} i_x! i_y! i_z!}{(2i_x)! (2i_y)! (2i_z)!} \right)^{1/2}
\label{eq:local_normalization}
\end{align}
only normalizes the local pGTO (i.e. $\iiint_\infty d\vec{r} |\mu^{local}(\vec{r};\vec{i},a,\vec{A})|^2 = 1$). The atomic crystalline orbital is not normalized by $C^{normalization}$ due to the presence of periodic images that need to be considered as well. In other words, the diagonal of overlap matrix in the atomic crystalline orbital basis is not 1. We note that the contraction coefficients $C_m^{contraction}$ are also chosen to be real, and $\phi^{local}(\vec{r})$ is thus always real-valued. 

Since the contraction and local normalization of pGTOs is independent of the $\vec{k}$ vector or summing over the lattice vectors $\vec{R}$ in equation \ref{eq:crystalline_atomic_orbital_definition}, it still follows that the integrals in the contracted crystalline GTO basis are just summations over integrals in terms of crystalline pGTOs. We will therefore only consider crystalline pGTOs and condense all constant prefactors (contraction, normalization etc.) into a single real number $C_\mu$ per pGTO $\mu$, similar to the case of non-periodic calculations\cite{terachem_f}.

\subsection{Overlap integral and pair distribution}

For a pair of crystalline pGTOs: 
\begin{align}
\mu_{\vec{k}_\mu}(\vec{r}; \vec{i},a,\vec{A}) &= \sum_{\vec{R}_\mu \in \mathbb{Z}^3} e^{i \vec{k}_\mu \cdot \vec{R}_\mu} \mu^{local}(\vec{r} - \vec{R}_\mu; \vec{i},a,\vec{A}) \notag \\
    &= \sum_{\vec{R}_\mu \in \mathbb{Z}^3} e^{i \vec{k}_\mu \cdot \vec{R}_\mu} \left( \prod_{\tau \in \{x,y,z\}} (\tau - R_{\mu, \tau} - A_\tau)^{i_\tau} \right) e^{-a \left| \vec{r} - \vec{R}_\mu - \vec{A} \right|^2} \label{eq:pbc_mu_definition} \\
\nu_{\vec{k}_\nu}(\vec{r}; \vec{j},b,\vec{B}) &= \sum_{\vec{R}_\nu \in \mathbb{Z}^3} e^{i \vec{k}_\nu \cdot \vec{R}_\nu} \nu^{local}(\vec{r} - \vec{R}_\nu; \vec{j},b,\vec{B}) \notag \\
    &= \sum_{\vec{R}_\nu \in \mathbb{Z}^3} e^{i \vec{k}_\nu \cdot \vec{R}_\nu} \left( \prod_{\tau \in \{x,y,z\}} (\tau - R_{\nu, \tau} - B_\tau)^{j_\tau} \right) e^{-a \left| \vec{r} - \vec{R}_\nu - \vec{B} \right|^2} \label{eq:pbc_nu_definition}
\end{align}

The overlap integral between them is defined as
\begin{align}
S_{\mu\nu} &= \iiint_{\infty} d\vec{r} \ \mu_{\vec{k}_\mu}^*(\vec{r}; \vec{i},a,\vec{A}) \nu_{\vec{k}_\nu}(\vec{r}; \vec{j},b,\vec{B}) \notag \\
    &= \sum_{\vec{R}_\mu \in \mathbb{Z}^3} e^{-i \vec{k}_\mu \cdot \vec{R}_\mu} \sum_{\vec{R}_\nu \in \mathbb{Z}^3} e^{i \vec{k}_\nu \cdot \vec{R}_\nu} \iiint_{\infty} d\vec{r} \ \mu^{local}(\vec{r} - \vec{R}_\mu; \vec{i},a,\vec{A}) \nu^{local}(\vec{r} - \vec{R}_\nu; \vec{j},b,\vec{B})
\label{eq:pbc_overlap_definition}
\end{align}

The above expression can be simplified through the use of certain mathematical relationships. The following invariances hold for arbitrary functions $f(\vec{r})$ and arbitrary lattice vectors $\vec{R}^\prime$:
\begin{align}
\sum_{\vec{R} \in \mathbb{Z}^3} f(\vec{R} + \vec{R}^\prime) &= \sum_{\vec{R} \in \mathbb{Z}^3} f(\vec{R})
\label{eq:summation_translation_invariance} \\
\sum_{\vec{R} \in \mathbb{Z}^3} f(-\vec{R}) &= \sum_{\vec{R} \in \mathbb{Z}^3} f(\vec{R})
\label{eq:summation_inversion_invariance} \\
\iiint_{\infty} d\vec{r} \ f(\vec{r} + \vec{R}^\prime) &= \iiint_{\infty} dr \ f(\vec{r}) \label{eq:integral_translation_invariance}
\end{align}
as long as the corresponding sum and integral operations lead to convergent results. We also possess the
Fourier transform relationship \cite{yellow_book_eik_summation}:
\begin{align}
\sum_{\vec{R} \in \mathbb{Z}^3} e^{i(\vec{k}_\mu - \vec{k}_\nu) \cdot \vec{R}} = \delta_{\vec{k}_\mu, \vec{k}_\nu}
\end{align}
where both $\vec{k}_\mu$ and $\vec{k}_\nu$ are inside the first Brillouin zone, and $\delta_{\vec{k}_\mu, \vec{k}_\nu}$ is a Kronecker delta. 

The overlap integral then reduces to:
\begin{align}
S_{\mu\nu} &= \delta_{\vec{k}_\mu, \vec{k}_\nu} \sum_{\vec{R} \in \mathbb{Z}^3} e^{i \vec{k}_\mu \cdot \vec{R}} \iiint_{\infty} d\vec{r} \ \mu^{local}(\vec{r}; \vec{i},a,\vec{A}) \nu^{local}(\vec{r} - \vec{R}; \vec{j},b,\vec{B})
\label{eq:pbc_overlap_to_local}
\end{align}
In this equation the integral part is local and real-valued, and can be evaluated in the same manner as in discrete molecular systems.

In order to reduce equation \ref{eq:pbc_overlap_to_local} further into the molecular integral form, we define the local pGTO pair distribution $\mu\nu^{local}(\vec{r}, \vec{R})$ as:
\begin{align}
\mu\nu^{local}(\vec{r}, \vec{R}) &= \mu^{local}(\vec{r}; \vec{i},a,\vec{A}) \nu^{local}(\vec{r} - \vec{R}; \vec{j},b,\vec{B}) \notag \\
    &= C_\mu C_\nu \left( \prod_{\tau \in \{x,y,z\}} (\tau - A_\tau)^{i_\tau} (\tau - R_\tau - B_\tau)^{j_\tau} \right) e^{-a \left| \vec{r} - \vec{A} \right|^2} e^{-b \left| \vec{r} - \vec{R} - \vec{B} \right|^2}
\label{eq:local_pair_definition}
\end{align}

We combine the lattice vector $\vec{R}$ with the atom center location $\vec{B}$ into $\vec{B}(\vec{R}) = \vec{B} + \vec{R}$. The pair distribution center is then defined as $\vec{P}(\vec{R}) = \frac{a\vec{A} + b(\vec{B} + \vec{R})}{a+b}$. With these definitions, we can express the local pair distribution $\mu\nu^{local}(\vec{R})$ in the exact same way as in the molecular case, which is to transform it into the Hermite Gaussian basis following the McMurchie-Davidson scheme \cite{McMurchie_Davidson_original}:
\begin{align}
\mu\nu^{local}(\vec{r}, \vec{R}) = C_\mu C_\nu \left( \prod_{\tau \in \{x,y,z\}} \sum^{i_\tau + j_\tau}_{t_\tau = 0} E^{i_\tau, j_\tau}_{t_\tau, \tau}(A_\tau, B_\tau(R_\tau), p) \left( \frac{\partial}{\partial P_\tau} \right)^{t_\tau} \right) e^{-p \left| \vec{r} - \vec{P}(\vec{R}) \right|^2}
\label{eq:local_pair_cartesian_to_hermite}
\end{align}
Here $p = a+b$ is the exponent of pair distribution, and $E^{i_\tau, j_\tau}_{t_\tau, \tau}(A_\tau, B_\tau, p)$ terms are the Cartesian Gaussian to Hermite Gaussian transformation coefficients, which can be obtained from the McMurchie-Davidson recurrence
relationship:
\begin{align}
E^{i_\tau + 1, j_\tau}_{t_\tau, \tau} &= \frac{1}{2p} E^{i_\tau, j_\tau}_{t_\tau - 1, \tau} + (P_\tau(R_\tau) - A_\tau) E^{i_\tau, j_\tau}_{t_\tau, \tau} + (t_\tau + 1) E^{i_\tau, j_\tau}_{t_\tau + 1, \tau} \label{eq:E_recursion_i} \\
E^{i_\tau, j_\tau + 1}_{t_\tau, \tau} &= \frac{1}{2p} E^{i_\tau, j_\tau}_{t_\tau - 1, \tau} + (P_\tau(R_\tau) - B_\tau(R_\tau)) E^{i_\tau, j_\tau}_{t_\tau, \tau} + (t_\tau + 1) E^{i_\tau, j_\tau}_{t_\tau + 1, \tau} \label{eq:E_recursion_j} \\
E^{0,0}_{0,\tau} &= e^{-\frac{ab}{a+b} (A_\tau - B_\tau(R_\tau))^2} \label{eq:E_recursion_0} \\
E^{i_\tau, j_\tau}_{t_\tau, \tau} &= 0 \qquad \text{ if } t_\tau < 0 \text{ or } t_\tau > i_\tau + j_\tau \label{eq:E_recursion_out}
\end{align}

Expressing the local pair distribution in Hermite Gaussian form, and the Gaussian integral formula $\iiint_{\infty} d\vec{r} \ e^{-p \left| \vec{r} \right|^2} = \left( \frac{\pi}{p} \right)^{3/2}$, the overlap integral of crystalline pGTOs reduces to:
\begin{align}
S_{\mu\nu} &= \delta_{\vec{k}_\mu, \vec{k}_\nu} \sum_{\vec{R} \in \mathbb{Z}^3} e^{i \vec{k}_\mu \cdot \vec{R}} C_\mu C_\nu E^{i_x, j_x}_{0, x}(A_x, B_x+R_x, p) E^{i_y, j_y}_{0, y}(A_y, B_y+R_y, p) E^{i_z, j_z}_{0, z}(A_z, B_z+R_z, p) \left( \frac{\pi}{p} \right)^{3/2}
\label{eq:pbc_overlap_formula}
\end{align}

Since each $E^{i_\tau, j_\tau}_{t_\tau, \tau}$ term contains a Gaussian factor $e^{-\frac{ab}{a+b} (A_\tau - (B_\tau + R_\tau))^2}$ (equation \ref{eq:E_recursion_0}), the term within lattice summation decays as fast as a Gaussian function in $\vec{R}$. This rapid decay allows us to perform a cutoff for the lattice summation. For example, we can define a lattice vector cutoff threshold for pair distribution $\delta_{pair-R}$, such that only the lattice vectors $\vec{R}$ satisfying the following equation are included in the lattice summation:
\begin{align}
\left| C_\mu C_\nu e^{-\frac{ab}{a+b} \left| (\vec{A} - \vec{B}) - \vec{R} \right|^2} \right| > \delta_{pair-R}
\label{eq:pbc_overlap_cutoff_definition}
\end{align}
Or equivalently:
\begin{align}
\left| \vec{R} - (\vec{A} - \vec{B}) \right| < \sqrt{ \frac{a+b}{ab} \ln\left(\frac{|C_\mu C_\nu|}{\delta_{pair-R}}\right) }
\label{eq:pbc_overlap_cutoff_implementation}
\end{align}

Equation \ref{eq:pbc_overlap_cutoff_implementation} is a straightforward scheme for defining a cutoff for lattice vector summation, but we do not actually use it in our implementation. Instead, we utilize a less aggressive cutoff scheme that includes a sufficient number of lattice vectors for computing the long-ranged electron repulsion interaction. This is shown later in this work (equation \ref{eq:pbc_eri_pair_cutoff_implementation}).

In summary, an overlap integral of two crystalline pGTOs is computed as follows: we first compute the offset $\vec{A} - \vec{B}$ and the cutoff radius according to equation \ref{eq:pbc_overlap_cutoff_implementation}, then look for all lattice vectors inside the bounding sphere. For each lattice vector $\vec{R}$, we construct a local pair $\mu\nu^{local}(\vec{r}, \vec{R})$, and evaluate the local (non-periodic) overlap integral. At the end we sum over the local overlap integrals with the phase factor for each lattice vector to obtain the crystalline overlap integral.

The electron kinetic energy integral is treated in the exact same manner as the overlap integral, and is thus not discussed in detail here. We only provide the final expression in the Supporting Information.

\subsection{Nuclear attraction integral and Ewald summation}

In a periodic system, the potential generated by a nucleus (modeled as a point charge $q_C$ at location $\vec{C}$) together with its periodic images is:
\begin{align}
V_{nucleus}^{diverge}(\vec{r}) = \sum_{\vec{R}_C \in \mathbb{Z}^3} \frac{q_C}{|\vec{r} + \vec{R}_C - \vec{C}|}
\label{eq:nuclear_potential}
\end{align}
The resulting potential is divergent on account of the lattice vector summation and the $r^{-1}$ nature of the Coulomb potential. We thus label this term, as well as other terms where the lattice vector summation possibly leads to a divergence, as ``diverge". Techniques for removing such divergences will be discussed later. 

The interaction between electrons and point charges is accounted for with the nuclear-electron attraction operator, and its matrix element is the nuclear attraction integral, involving a pair of crystalline pGTOs and a periodic point charge potential:
\begin{align}
V_{\mu\nu C}^{diverge} &= \iiint_{\infty} d\vec{r} \ \mu_{\vec{k}_\mu}^*(\vec{r}; \vec{i},a,\vec{A}) \nu_{\vec{k}_\nu}(\vec{r}; \vec{j},b,\vec{B}) \sum_{\vec{R}_C \in \mathbb{Z}^3} \frac{q_C}{|\vec{r} + \vec{R}_C - \vec{C}|} \notag \\
    &= \sum_{\vec{R}_\mu \in \mathbb{Z}^3} e^{-i \vec{k}_\mu \cdot \vec{R}_\mu} \sum_{\vec{R}_\nu \in \mathbb{Z}^3} e^{i \vec{k}_\nu \cdot \vec{R}_\nu} \sum_{\vec{R}_C \in \mathbb{Z}^3} \notag \\
    &\quad \iiint_{\infty} d\vec{r} \ \mu^{local}(\vec{r} - \vec{R}_\mu; \vec{i},a,\vec{A}) \nu^{local}(\vec{r} - \vec{R}_\nu; \vec{j},b,\vec{B}) \frac{q_C}{|\vec{r} + \vec{R}_C - \vec{C}|}
\label{eq:pbc_v1e_definition}
\end{align}

Similar to the overlap integral case, $V_{\mu\nu C}$ is non-zero only if $\vec{k}_\mu = \vec{k}_\nu$:
\begin{align}
V_{\mu\nu C}^{diverge} &= \delta_{\vec{k}_\mu, \vec{k}_\nu} \sum_{\vec{R} \in \mathbb{Z}^3} e^{i \vec{k}_\mu \cdot \vec{R}} \sum_{\vec{R}_C \in \mathbb{Z}^3} \notag \\
    &\quad \iiint_{\infty} d\vec{r} \ \mu^{local}(\vec{r}; \vec{i},a,\vec{A}) \nu^{local}(\vec{r} - \vec{R}; \vec{j},b,\vec{B}) \frac{q_C}{|\vec{r} + \vec{R}_C - \vec{C}|}
\label{eq:pbc_v1e_simplify_1}
\end{align}

The base case, where both $\mu^{local}(\vec{r}; \vec{i},a,\vec{A})$ and $\nu^{local}(\vec{r}; \vec{j},b,\vec{B})$ are $s$-orbitals ($\vec{i} = \vec{j} = \vec{0}$), is:
\begin{align}
V_{ss C}^{diverge} &= \delta_{\vec{k}_\mu, \vec{k}_\nu} \sum_{\vec{R} \in \mathbb{Z}^3} e^{i \vec{k}_\mu \cdot \vec{R}} \sum_{\vec{R}_C \in \mathbb{Z}^3} q_C C_\mu C_\nu e^{-\frac{ab}{a+b} \left| \vec{A} - \vec{B}(\vec{R}) \right|^2} \iiint_{\infty} d\vec{r} \ e^{-p \left| \vec{r} - \vec{P}(\vec{R}) \right|^2} \frac{1}{|\vec{r} + \vec{R}_C - \vec{C}|} \notag \\
    &= \delta_{\vec{k}_\mu, \vec{k}_\nu} \sum_{\vec{R} \in \mathbb{Z}^3} e^{i \vec{k}_\mu \cdot \vec{R}} \sum_{\vec{R}_C \in \mathbb{Z}^3} q_C C_\mu C_\nu e^{-\frac{ab}{a+b} \left| \vec{A} - \vec{B}(\vec{R}) \right|^2} \frac{2\pi}{p} F_0\left( p\left| \vec{P}(\vec{R}) - \vec{C} + \vec{R}_C \right|^2 \right)
\label{eq:pbc_v1e_base_case}
\end{align}

The integral over $\vec{r}$ is reduced to the zeroth order Boys function $F_0(x)$ \cite{boys_original}:
\begin{align}
F_0(x) = \frac{\sqrt{\pi}}{2} \frac{\mathrm{erf}(\sqrt{x})}{\sqrt{x}}
\label{eq:boys_function_erf_form}
\end{align}

$F_0\left( p\left| \vec{P}(\vec{R}) - \vec{C} + \vec{R}_C \right|^2 \right)$ decays only as fast as $\dfrac{1}{|\vec{R}_C|}$ in the asymptotic limit of large $|\vec{R}_C|$. As a result, the summation over $\vec{R}_C$ is divergent. For later convenience, we define the summation over $\vec{R}_C$ in equation \ref{eq:pbc_v1e_base_case} as:
\begin{align}
S^{diverge}_{V_{ssC}, \vec{R}_C} = \displaystyle\sum_{\vec{R}_C \in \mathbb{Z}^3} F_0\left( p\left| \vec{P}(\vec{R}) - \vec{C} + \vec{R}_C \right|^2 \right)
\label{eq:diverge_summation_v1e}
\end{align}

This issue can be solved by introducing the Ewald summation scheme \cite{ewald_original}. We split the Boys function of zeroth order into a long-range and a short-range part \cite{periodic_integral_gamma}:
\begin{align}
F_0\left( p\left| \overrightarrow{PC}(\vec{R}) + \vec{R}_C \right|^2 \right) &= \frac{\omega}{\sqrt{p + \omega^2}} F_0\left( \frac{p\omega^2}{p + \omega^2} \left| \overrightarrow{PC}(\vec{R}) + \vec{R}_C \right|^2 \right) \notag \\
    &\quad + \left( F_0\left( p\left| \overrightarrow{PC}(\vec{R}) + \vec{R}_C \right|^2 \right) - \frac{\omega}{\sqrt{p + \omega^2}} F_0\left( \frac{p\omega^2}{p + \omega^2} \left| \overrightarrow{PC}(\vec{R}) + \vec{R}_C \right|^2 \right) \right)
\label{eq:boys_function_split}
\end{align}
where the range separation parameter $\omega$ is a positive value. The choice of $\omega$ will be discussed in the implementation section. We also use the shorthand notation $\overrightarrow{PC}(\vec{R}) = \vec{P}(\vec{R}) - \vec{C}$. The first term in equation \ref{eq:boys_function_split} represents the long-range part of the Boys function, and can be Fourier transformed to:
\begin{align}
&\quad \iiint_\infty d\vec{R}_C \ e^{-i 2\pi \vec{K}_C \cdot \vec{R}_C} \frac{\omega}{\sqrt{p + \omega^2}} F_0\left( \frac{p\omega^2}{p + \omega^2} \left| \overrightarrow{PC}(\vec{R}) + \vec{R}_C \right|^2 \right) \notag \\
&= \frac{1}{2\sqrt{\pi}} \frac{1}{\sqrt{p}} \frac{1}{|\vec{K}_C|^2} e^{i2\pi \vec{K}_C \cdot \overrightarrow{PC}(\vec{R})} e^{-\pi^2 (\frac{1}{p} + \frac{1}{\omega^2}) |\vec{K}_C|^2}
\label{eq:short_range_fourier}
\end{align}

The summation of long-range terms can be therefore performed in reciprocal space, according to the Poisson summation formula \cite{poisson_summation_formula_1}:
\begin{align}
\sum_{\vec{R} \in \mathbb{Z}^3} f(\vec{R}) = \frac{1}{V_{cell}} \sum_{\vec{K} \in \mathbb{Z}^3} \hat{f}\left(\frac{1}{2\pi} \vec{K}\right)
\label{eq:poisson_summation_formula}
\end{align}
where $f(\vec{r})$ is an arbitrary function that has a Fourier transform $\hat{f}(\vec{k}) = \iiint_\infty d\vec{r} \ e^{-i 2\pi \vec{k} \cdot \vec{r}} f(\vec{r})$, and both summations converge. The $\vec{K}$ summation iterates through a complete set of reciprocal space lattice vectors ($\vec{K} = n_1 \vec{K}_1 + n_2 \vec{K}_2 + n_3 \vec{K}_3$, where $\vec{K}_1$, $\vec{K}_2$ and $\vec{K}_3$ are primitive reciprocal cell vectors, and $n_1, n_2, n_3 \in \mathbb{Z}$). $V_{cell}$ is the volume of the real-space primitive cell.

The $\vec{R}_C$ summation (equation \ref{eq:diverge_summation_v1e}) can be reformulated as:
\begin{align}
S^{diverge}_{V_{ssC}, \vec{R}_C} &= \sum_{\vec{K}_C \in \mathbb{Z}^3} \frac{2\pi^{3/2}}{V_{cell}} \frac{1}{\sqrt{p}} \frac{1}{|\vec{K}_C|^2} \mathrm{Re}\left( e^{i\vec{K}_C \cdot \overrightarrow{PC}(\vec{R})} \right) e^{-\frac{1}{4}(\frac{1}{p} + \frac{1}{\omega^2})|\vec{K}_C|^2} \notag \\
    &\quad + \sum_{\vec{R}_C \in \mathbb{Z}^3} \left( F_0\left( p\left| \overrightarrow{PC}(\vec{R}) + \vec{R}_C \right|^2 \right) - \frac{\omega}{\sqrt{p + \omega^2}} F_0\left( \frac{p\omega^2}{p + \omega^2} \left| \overrightarrow{PC}(\vec{R}) + \vec{R}_C \right|^2 \right) \right)
\label{eq:pbc_V1e_RC_summation_2}
\end{align}

Although the imaginary number $i$ appears in the reciprocal space lattice summation, only the real part of the summation needs to be considered, because all other terms on both sides of the equation are real-valued.

The divergence arises from the reciprocal space term when $\vec{K}_C = 0$. In order to remove the divergence, we need to take the Fourier transform of the full range Boys function (left-hand side of equation \ref{eq:boys_function_split}):
\begin{align}
&\quad \iiint_\infty d\vec{R}_C \ e^{-i 2\pi \vec{K}_C \cdot \vec{R}_C} F_0\left( p \left| \overrightarrow{PC}(\vec{R}) + \vec{R}_C \right|^2 \right) \notag \\
&= \frac{1}{2\sqrt{\pi}} \frac{1}{\sqrt{p}} \frac{1}{|\vec{K}_C|^2} e^{i2\pi \vec{K}_C \cdot \overrightarrow{PC}(\vec{R})} e^{-\pi^2 \frac{1}{p} |\vec{K}_C|^2}
\label{eq:full_range_fourier}
\end{align}
and substract the diverging $\vec{K}_C = 0$ term of the full range Boys function (evaluated at $\frac{1}{2\pi}\vec{K}_C$) from the $\vec{K}_C = 0$ term of equation \ref{eq:pbc_V1e_RC_summation_2}, which will provide a difference term:
\begin{align}
&\quad \frac{2\pi^{3/2}}{V_{cell}} \frac{1}{\sqrt{p}} \lim_{\vec{K}_C \to \vec{0}} \left( \frac{1}{|\vec{K}_C|^2} e^{i\vec{K}_C \cdot \overrightarrow{PC}(\vec{R})} e^{-\frac{1}{4}(\frac{1}{p} + \frac{1}{\omega^2})|\vec{K}_C|^2} - \frac{1}{|\vec{K}_C|^2} e^{i \vec{K}_C \cdot \overrightarrow{PC}(\vec{R})} e^{- \frac{1}{4p} |\vec{K}_C|^2} \right) \notag \\
&= - \frac{2\pi^{3/2}}{V_{cell}} \frac{1}{\sqrt{p}} \frac{1}{4\omega^2}
\label{eq:v1e_self_term}
\end{align}
This term is usually referred to as the ``self-interaction" term in Ewald summation language. 

The removal of the divergent term can be justified, as similar divergences arise from nuclear-nuclear repulsion and electron-electron repulsion but with opposite signs. For a charge neutral periodic system, the divergent terms from these three sources will cancel each other, providing a finite total energy. Otherwise, classical electrostatics would not permit the corresponding ensemble of nuclei and electrons to be stable.

As a result, the non-divergent $\vec{R}_C$ Ewald summation replacement of equation \ref{eq:pbc_V1e_RC_summation_2} is:
\begin{align}
S_{V_{ssC}, \vec{R}_C} &= \frac{2\pi^{3/2}}{V_{cell}} \frac{1}{\sqrt{p}} \sum_{\vec{K}_C \in \mathbb{Z}^3, \vec{K}_C \neq \vec{0}} \frac{1}{|\vec{K}_C|^2} \mathrm{Re}\left( e^{i\vec{K}_C \cdot \overrightarrow{PC}(\vec{R})} \right) e^{-\frac{1}{4}(\frac{1}{p} + \frac{1}{\omega^2})|\vec{K}_C|^2} \notag \\
    &\quad + \sum_{\vec{R}_C \in \mathbb{Z}^3} \left( F_0\left( p\left| \overrightarrow{PC}(\vec{R}) + \vec{R}_C \right|^2 \right) - \frac{\omega}{\sqrt{p + \omega^2}} F_0\left( \frac{p\omega^2}{p + \omega^2} \left| \overrightarrow{PC}(\vec{R}) + \vec{R}_C \right|^2 \right) \right) \notag \\
    &\quad - \frac{2\pi^{3/2}}{V_{cell}} \frac{1}{\sqrt{p}} \frac{1}{4\omega^2}
\label{eq:pbc_V1e_ewald_summation}
\end{align}

And the non-divergent base case of the nuclear attraction integral is:
\begin{align}
V_{ss C} &= \delta_{\vec{k}_\mu, \vec{k}_\nu} q_C C_\mu C_\nu \frac{2\pi}{p} \sum_{\vec{R} \in \mathbb{Z}^3} e^{i \vec{k}_\mu \cdot \vec{R}} e^{-\frac{ab}{a+b} \left| \vec{A} - \vec{B}(\vec{R}) \right|^2} \notag \\
    &\quad \left( \frac{2\pi^{3/2}}{V_{cell}} \frac{1}{\sqrt{p}} \sum_{\vec{K}_C \in \mathbb{Z}^3, \vec{K}_C \neq \vec{0}} \frac{1}{|\vec{K}_C|^2} \mathrm{Re}\left( e^{i\vec{K}_C \cdot (\vec{P}(\vec{R}) - \vec{C})} \right) e^{-\frac{1}{4}(\frac{1}{p} + \frac{1}{\omega^2})|\vec{K}_C|^2} \right. \notag \\
    &\qquad + \sum_{\vec{R}_C \in \mathbb{Z}^3} \left( F_0\left( p\left| \vec{P}(\vec{R}) - \vec{C} + \vec{R}_C \right|^2 \right) - \frac{\omega}{\sqrt{p + \omega^2}} F_0\left( \frac{p\omega^2}{p + \omega^2} \left| \vec{P}(\vec{R}) - \vec{C} + \vec{R}_C \right|^2 \right) \right) \notag \\
    &\qquad \left. - \frac{2\pi^{3/2}}{V_{cell}} \frac{1}{\sqrt{p}} \frac{1}{4\omega^2} \right)
\label{eq:pbc_v1e_base_case_final}
\end{align}

The truncation scheme for the first summation in equation \ref{eq:pbc_v1e_base_case_final} (over $\vec{R}$) is similar to the pair distribution lattice vector cutoff scheme for the overlap integral (equation \ref{eq:pbc_overlap_cutoff_implementation}), since the summand decays as fast as a Gaussian function. The only difference is the additional $\dfrac{2\pi}{p}$ term in the prefactor.

The truncation scheme for the long-range summation over $\vec{K}_C$ in equation \ref{eq:pbc_v1e_base_case_final} can be performed by utilizing the following inequality:
\begin{align}
\left|\frac{1}{|\vec{K}_C|^2} \mathrm{Re}\left( e^{i\vec{K}_C \cdot (\vec{P}(\vec{R}) - \vec{C})} \right) e^{-\frac{1}{4}(\frac{1}{p} + \frac{1}{\omega^2})|\vec{K}_C|^2} \right| \leq \frac{1}{|\vec{K}_{min}|^2} e^{-\frac{1}{4}(\frac{1}{p} + \frac{1}{\omega^2})|\vec{K}_C|^2}
\label{eq:pbc_v1e_reciprocal_cutoff_1}
\end{align}
where $\vec{K}_{min}$ is the non-zero reciprocal space lattice vector with minimal norm, which is usually just the primitive reciprocal space lattice vector with minimal norm. We define an overall prefactor $C_{1e} = C_\mu C_\nu \frac{2\pi}{p} e^{-\frac{ab}{a+b} \left| \vec{A} - \vec{B}(\vec{R}) \right|^2}$ and a cutoff threshold $\delta_{Ewald-K}$, such that only the reciprocal lattice vectors $\vec{K}_C$ satisfying the following equation are included in the lattice summation:
\begin{align}
\left| C_{1e} \frac{2\pi^{3/2}}{V_{cell}} \frac{1}{\sqrt{p}} \frac{1}{|\vec{K}_{min}|^2} e^{-\frac{1}{4}(\frac{1}{p} + \frac{1}{\omega^2})|\vec{K}_C|^2} \right| > \delta_{Ewald-K}
\label{eq:pbc_v1e_reciprocal_cutoff_definition}
\end{align}
Or equivalently:
\begin{align}
|\vec{K}_C| < \sqrt{ \frac{4p\omega^2}{p + \omega^2} \ln\left(\frac{2\pi^{3/2} |C_{1e}|}{V_{cell} \sqrt{p} |\vec{K}_{min}|^2 \delta_{Ewald-K}}\right) }
\label{eq:pbc_v1e_reciprocal_cutoff_implementation}
\end{align}

The truncation scheme for the short-range summation over $\vec{R}_C$ in equation \ref{eq:pbc_v1e_base_case_final} can be performed by expressing the Boys function in terms of the error function, and utilizing the following inequality:
\begin{align}
&\quad F_0\left( p\left| \overrightarrow{PC}(\vec{R}) + \vec{R}_C \right|^2 \right) - \frac{\omega}{\sqrt{p + \omega^2}} F_0\left( \frac{p\omega^2}{p + \omega^2} \left| \overrightarrow{PC}(\vec{R}) + \vec{R}_C \right|^2 \right) \notag \\
    &= \frac{1}{\sqrt{p}} \frac{1}{\left|\overrightarrow{PC}(\vec{R}) + \vec{R}_C\right|} \int_{ \sqrt{\frac{p\omega^2}{p+\omega^2}} \left|\overrightarrow{PC}(\vec{R}) + \vec{R}_C\right| }^{ \sqrt{p} \left|\overrightarrow{PC}(\vec{R}) + \vec{R}_C\right| } e^{-t^2}dt \notag \\
&<  \frac{1}{\sqrt{p}} \frac{1}{\left|\overrightarrow{PC}(\vec{R}) + \vec{R}_C\right|} \int_{ \sqrt{\frac{p\omega^2}{p+\omega^2}} \left|\overrightarrow{PC}(\vec{R}) + \vec{R}_C\right| }^{ \sqrt{p} \left|\overrightarrow{PC}(\vec{R}) + \vec{R}_C\right| } e^{-\frac{p\omega^2}{p+\omega^2} \left|\overrightarrow{PC}(\vec{R}) + \vec{R}_C\right|^2} dt \notag \\
&= \frac{1}{\sqrt{p}} \frac{1}{\left|\overrightarrow{PC}(\vec{R}) + \vec{R}_C\right|} e^{-\frac{p\omega^2}{p+\omega^2} \left|\overrightarrow{PC}(\vec{R}) + \vec{R}_C\right|^2} \left( \sqrt{p} \left|\overrightarrow{PC}(\vec{R}) + \vec{R}_C\right| - \sqrt{\frac{p\omega^2}{p+\omega^2}} \left|\overrightarrow{PC}(\vec{R}) + \vec{R}_C\right| \right) \notag \\
    &= \left(1 - \frac{\omega}{\sqrt{p+\omega^2}} \right) e^{-\frac{p\omega^2}{p+\omega^2} \left|\overrightarrow{PC}(\vec{R}) + \vec{R}_C\right|^2}
\label{eq:pbc_v1e_real_cutoff_1}
\end{align}
It is clear from this inequality that the short-range term decays at the rate of a Gaussian function in $\vec{R}_C$ as well. We define another cutoff threshold $\delta_{Ewald-R}$, such that only the lattice vectors $\vec{R}_C$ satisfying the following equation are included in the lattice summation:
\begin{align}
\left| C_{1e} \left(1 - \frac{\omega}{\sqrt{p+\omega^2}} \right) e^{-\frac{p\omega^2}{p+\omega^2} \left|\vec{R}_C - (\vec{C} - \vec{P}(\vec{R}))\right|^2} \right| > \delta_{Ewald-R}
\label{eq:pbc_v1e_real_cutoff_definition}
\end{align}
Or equivalently:
\begin{align}
\left|\vec{R}_C - (\vec{C} - \vec{P}(\vec{R}))\right| < \sqrt{ \left(\frac{1}{p} + \frac{1}{\omega^2}\right) \ln\left(\frac{|C_{1e}|}{\delta_{Ewald-R}} \left(1 - \frac{\omega}{\sqrt{p+\omega^2}} \right) \right) }
\label{eq:pbc_v1e_real_cutoff_implementation}
\end{align}

Putting all the pieces together, a base case nuclear attraction integral involving two crystalline pGTOs and one point charge is computed as follows:
we first search for all lattice vectors $\vec{R}$ for the pair lattice summation, following the same procedure as in the overlap integral case. For every $\vec{R}$, we construct a local pair with center location $\vec{P}(\vec{R})$. We then loop through every local pair, and obtain a bounding sphere for reciprocal space long-range summation according to equation \ref{eq:pbc_v1e_reciprocal_cutoff_implementation}, as well as a bounding sphere for real space short-range summation according to equation \ref{eq:pbc_v1e_real_cutoff_implementation}. For every local pair, we compute the real and reciprocal space lattice summation within the corresponding bounding sphere, add them together and append the self-interaction term, according to equation \ref{eq:pbc_v1e_base_case_final}. Finally we sum over the list of local pairs ($\vec{R}$) with the phase factor $e^{-i \vec{k}_\mu\cdot\vec{R}}$, and we obtain the base case crystalline nuclear attraction integral. Notice that the only source of complex number is the phase factor; the Ewald summation part is completely real-valued.

For the general case where $\mu^{local}(\vec{r}; \vec{i},a,\vec{A})$ and $\nu^{local}(\vec{r}; \vec{j},b,\vec{B})$ have arbitrary angular momentum indices, we express the local pGTO pairs in terms of Hermite Gaussians (equation \ref{eq:local_pair_cartesian_to_hermite}) in the nuclear repulsion integral (equation \ref{eq:pbc_v1e_simplify_1}):
\begin{align}
V_{\mu\nu C}^{diverge} &= \delta_{\vec{k}_\mu, \vec{k}_\nu} q_C C_\mu C_\nu \frac{2\pi}{p} \sum_{\vec{R} \in \mathbb{Z}^3} e^{i \vec{k}_\mu \cdot \vec{R}} \left( \prod_{\tau \in \{x,y,z\}} \sum^{i_\tau + j_\tau}_{t_\tau = 0} E^{i_\tau, j_\tau}_{t_\tau, \tau}(A_\tau, B_\tau(R_\tau), p) \left( \frac{\partial}{\partial P_\tau} \right)^{t_\tau} \right) \notag \\
    &\quad \sum_{\vec{R}_C \in \mathbb{Z}^3} F_0\left( p\left| \vec{P}(\vec{R}) - \vec{C} + \vec{R}_C \right|^2 \right)
\label{eq:pbc_v1e_general_to_hermite}
\end{align}

The second summation over $\vec{R}_C$ is the same as the divergent summation in base case nuclear attraction integral, and we can replace it with the Ewald summation form (equation \ref{eq:pbc_V1e_ewald_summation}), and the resulting equation is:
\begin{align}
V_{\mu\nu C} &= \delta_{\vec{k}_\mu, \vec{k}_\nu} q_C C_\mu C_\nu \frac{2\pi}{p} \sum_{\vec{R} \in \mathbb{Z}^3} e^{i \vec{k}_\mu \cdot \vec{R}} \left( \prod_{\tau \in \{x,y,z\}} \sum^{i_\tau + j_\tau}_{t_\tau = 0} E^{i_\tau, j_\tau}_{t_\tau, \tau}(A_\tau, B_\tau(R_\tau), p) \left( \frac{\partial}{\partial P_\tau} \right)^{t_\tau} \right) \notag \\
    &\quad \left( \frac{2\pi^{3/2}}{V_{cell}} \frac{1}{\sqrt{p}} \sum_{\vec{K}_C \in \mathbb{Z}^3, \vec{K}_C \neq \vec{0}} \frac{1}{|\vec{K}_C|^2} \mathrm{Re}\left( e^{i\vec{K}_C \cdot (\vec{P}(\vec{R}) - \vec{C})} \right) e^{-\frac{1}{4}(\frac{1}{p} + \frac{1}{\omega^2})|\vec{K}_C|^2} \right. \notag \\
    &\qquad + \sum_{\vec{R}_C \in \mathbb{Z}^3} \left( F_0\left( p\left| \vec{P}(\vec{R}) - \vec{C} + \vec{R}_C \right|^2 \right) - \frac{\omega}{\sqrt{p + \omega^2}} F_0\left( \frac{p\omega^2}{p + \omega^2} \left| \vec{P}(\vec{R}) - \vec{C} + \vec{R}_C \right|^2 \right) \right) \notag \\
    &\qquad \left. - \frac{2\pi^{3/2}}{V_{cell}} \frac{1}{\sqrt{p}} \frac{1}{4\omega^2} \right)
\label{eq:pbc_v1e_general_summation_replace}
\end{align}

The derivatives with respect to $P_\tau$ for the long-range part ($\vec{K}_C$ summation) and self-interaction term are trivial. In order to carry out the derivatives with respect to $P_\tau$ for the short-range part ($\vec{R}_C$ summation), we follow the McMurchie-Davidson formalism and define auxiliary integrals $R_{t_x,t_y,t_z}^m$ as derivatives of Boys functions:
\begin{align}
R_{t_x,t_y,t_z}^m\left( p, \vec{L} \right) = \left( \frac{\partial}{\partial L_x} \right)^{t_x} \left( \frac{\partial}{\partial L_y} \right)^{t_y} \left( \frac{\partial}{\partial L_z} \right)^{t_z} \left( (-2p)^m F_m\left( p\left| \vec{L} \right|^2 \right) \right)
\label{eq:R_definition}
\end{align}
and the function form of $R_{t_x,t_y,t_z}^m$ can be obtained using the recurrence relationship \cite{McMurchie_Davidson_original}:
\begin{align}
R_{t_x + 1,t_y,t_z}^m\left( p, \vec{L} \right) &= t_x R_{t_x - 1,t_y,t_z}^{m + 1}\left( p, \vec{L} \right) + L_x R_{t_x,t_y,t_z}^{m + 1}\left( p, \vec{L} \right) \\
R_{t_x,t_y + 1,t_z}^m\left( p, \vec{L} \right) &= t_y R_{t_x,t_y - 1,t_z}^{m + 1}\left( p, \vec{L} \right) + L_y R_{t_x,t_y,t_z}^{m + 1}\left( p, \vec{L} \right) \\
R_{t_x,t_y,t_z + 1}^m\left( p, \vec{L} \right) &= t_z R_{t_x,t_y,t_z - 1}^{m + 1}\left( p, \vec{L} \right) + L_z R_{t_x,t_y,t_z}^{m + 1}\left( p, \vec{L} \right) \\
R_{0,0,0}^m\left( p, \vec{L} \right) &= (-2p)^m F_m\left( p\left| \vec{L} \right|^2 \right) \\
R_{t_x,t_y,t_z}^m\left( p, \vec{L} \right) &= 0 \qquad \text{ if } t_x < 0 \text{ or } t_y < 0 \text{ or } t_z < 0
\label{eq:R_recursion}
\end{align}

With the help of these auxiliary integrals, the final expression for the crystalline nuclear attraction integral is:
\begin{align}
V_{\mu\nu C} &= \delta_{\vec{k}_\mu, \vec{k}_\nu} q_C C_\mu C_\nu \frac{2\pi}{p} \sum_{\vec{R} \in \mathbb{Z}^3} e^{i \vec{k}_\mu \cdot \vec{R}} \notag \\
    &\quad \left( \frac{2\pi^{3/2}}{V_{cell}} \frac{1}{\sqrt{p}} \sum_{\vec{K}_C \in \mathbb{Z}^3, \vec{K}_C \neq \vec{0}} \frac{1}{|\vec{K}_C|^2} e^{-\frac{1}{4}(\frac{1}{p} + \frac{1}{\omega^2})|\vec{K}_C|^2} \right. \notag \\
    &\qquad \sum^{i_x + j_x}_{t_x = 0} E^{i_x, j_x}_{t_x, x}(A_x, B_x(R_x), p) \sum^{i_y + j_y}_{t_y = 0} E^{i_y, j_y}_{t_y, y}(A_y, B_y(R_y), p) \sum^{i_z + j_z}_{t_z = 0} E^{i_z, j_z}_{t_z, z}(A_z, B_z(R_z), p) \notag \\
    &\qquad \mathrm{Re}\left( i^{t_x+t_y+t_z} e^{i\vec{K}_C \cdot (\vec{P}(\vec{R}) - \vec{C})} \right) \left.K_{C,x}\right. ^{t_x} \left.K_{C,y}\right.^{t_y} \left.K_{C,z}\right.^{t_z} \notag \\
    &\qquad + \sum_{\vec{R}_C \in \mathbb{Z}^3} \sum^{i_x + j_x}_{t_x = 0} E^{i_x, j_x}_{t_x, x}(A_x, B_x(R_x), p) \sum^{i_y + j_y}_{t_y = 0} E^{i_y, j_y}_{t_y, y}(A_y, B_y(R_y), p) \sum^{i_z + j_z}_{t_z = 0} E^{i_z, j_z}_{t_z, z}(A_z, B_z(R_z), p) \notag \\
    &\qquad \left( R_{t_x,t_y,t_z}^0\left( p, \vec{P}(\vec{R}) - \vec{C} + \vec{R}_C \right) - \frac{\omega}{\sqrt{p + \omega^2}} R_{t_x,t_y,t_z}^0\left( \frac{\omega \sqrt{p}}{\sqrt{p + \omega^2}}, \vec{P}(\vec{R}) - \vec{C} + \vec{R}_C \right) \right) \notag \\
    &\qquad \left. - \frac{2\pi^{3/2}}{V_{cell}} \frac{1}{\sqrt{p}} \frac{1}{4\omega^2} E^{i_x, j_x}_{0, x}(R_x) E^{i_y, j_y}_{0, y}(R_y) E^{i_z, j_z}_{0, z}(R_z) \right)
\label{eq:pbc_v1e_general_final}
\end{align}
The order of equation reflects the order of implementation, that is, for example, for every summand of the real space lattice summation over $\vec{R}_C$, we compute and contract the $E^{i_\tau, j_\tau}_{t_\tau, \tau}$ terms with the $R_{t_x,t_y,t_z}^0$ terms, even though the $E^{i_\tau, j_\tau}_{t_\tau, \tau}$ terms have no dependency on $\vec{R}_C$. The purpose is to reduce the number of intermediate variables in order to reduce the register / memory usage.

\subsection{Electron repulsion integral}

In order to model the electron-electron repulsion interaction within a periodic system, we need to evaluate the crystalline electron repulsion integrals:
\begin{align}
(\mu\nu|\lambda\sigma)^{diverge} &= \iiint_{\infty} d\vec{r}_1 \iiint_{\infty} d\vec{r}_2 \ \mu_{\vec{k}_\mu}^*(\vec{r}_1; \vec{i},a,\vec{A}) \nu_{\vec{k}_\nu}(\vec{r}_1; \vec{j},b,\vec{B}) \notag \\
    &\hspace{120pt} \frac{1}{|\vec{r}_1 - \vec{r}_2|} \lambda_{\vec{k}_\lambda}^*(\vec{r}_2; \vec{\kappa},c,\vec{C}) \sigma_{\vec{k}_\sigma}(\vec{r}_2; \vec{l},d,\vec{D}) \notag \\
&= \sum_{\vec{R}_\mu \in \mathbb{Z}^3} e^{-i \vec{k}_\mu \cdot \vec{R}_\mu} \sum_{\vec{R}_\nu \in \mathbb{Z}^3} e^{i \vec{k}_\nu \cdot \vec{R}_\nu} \sum_{\vec{R}_\lambda \in \mathbb{Z}^3} e^{-i \vec{k}_\lambda \cdot \vec{R}_\lambda} \sum_{\vec{R}_\sigma \in \mathbb{Z}^3} e^{i \vec{k}_\sigma \cdot \vec{R}_\sigma} \notag \\
    &\quad \iiint_{\infty} d\vec{r}_1 \iiint_{\infty} d\vec{r}_2 \ \mu^{local}(\vec{r}_1 - \vec{R}_\mu; \vec{i},a,\vec{A}) \nu^{local}(\vec{r}_1 - \vec{R}_\nu; \vec{j},b,\vec{B}) \notag \\
    &\hspace{120pt} \frac{1}{|\vec{r}_1 - \vec{r}_2|} \lambda^{local}(\vec{r}_2 - \vec{R}_\lambda; \vec{\kappa},c,\vec{C}) \sigma^{local}(\vec{r}_2 - \vec{R}_\sigma; \vec{l},d,\vec{D})
\label{eq:pbc_eri_definition}
\end{align}
We denote the angular momentum indices for local pGTO $\lambda^{local}(\vec{r}; \vec{\kappa},c,\vec{C})$ as $\vec{\kappa}$ instead of $\vec{k}$ to avoid confusion.

After simplification using equations \ref{eq:summation_translation_invariance},\ref{eq:summation_inversion_invariance} and \ref{eq:integral_translation_invariance}, the expression reduces to:
\begin{align}
(\mu\nu|\lambda\sigma)^{diverge} &= \delta_{\vec{k}_\mu + \vec{k}_\lambda, \vec{k}_\nu + \vec{k}_\sigma} \sum_{\vec{R}_2 \in \mathbb{Z}^3} e^{i \vec{k}_\mu \cdot \vec{R}_2} \sum_{\vec{R}_3 \in \mathbb{Z}^3} e^{i \vec{k}_\lambda \cdot \vec{R}_3} \sum_{\vec{R}_1 \in \mathbb{Z}^3} e^{i (\vec{k}_\mu - \vec{k}_\nu) \cdot \vec{R}_1} \notag \\
    &\quad \iiint_{\infty} d\vec{r}_1 \iiint_{\infty} d\vec{r}_2 \ \mu^{local}(\vec{r}_1; \vec{i},a,\vec{A}) \nu^{local}(\vec{r}_1 - \vec{R}_2; \vec{j},b,\vec{B}) \notag \\
    &\hspace{120pt} \frac{1}{|\vec{r}_1 - \vec{r}_2|} \lambda^{local}(\vec{r}_2 - \vec{R}_1; \vec{\kappa},c,\vec{C}) \sigma^{local}(\vec{r}_2 - \vec{R}_1 - \vec{R}_3; \vec{l},d,\vec{D})
\label{eq:pbc_eri_simplify_1}
\end{align}
This is nonzero only when crystal momentum is conserved when an electron is scattered off another (i.e. $\vec{k}_\mu + \vec{k}_\lambda= \vec{k}_\nu + \vec{k}_\sigma$).

The base case, where all four pGTOs are $s$-orbitals, is:
\begin{align}
(ss|ss)^{diverge} &= \delta_{\vec{k}_\mu + \vec{k}_\lambda, \vec{k}_\nu + \vec{k}_\sigma} \sum_{\vec{R}_2 \in \mathbb{Z}^3} e^{i \vec{k}_\mu \cdot \vec{R}_2} \sum_{\vec{R}_3 \in \mathbb{Z}^3} e^{i \vec{k}_\lambda \cdot \vec{R}_3} \sum_{\vec{R}_1 \in \mathbb{Z}^3} e^{i (\vec{k}_\mu - \vec{k}_\nu) \cdot \vec{R}_1} \notag \\
    &\quad C_\mu C_\nu C_\lambda C_\sigma e^{-\frac{ab}{a+b} \left| \vec{A} - \vec{B}(\vec{R}_2) \right|^2} e^{-\frac{cd}{c+d} \left| \vec{C} - \vec{D}(\vec{R}_3) \right|^2} \notag \\
    &\quad \iiint_{\infty} d\vec{r}_1 \iiint_{\infty} d\vec{r}_2 \ e^{-p \left| \vec{r}_1 - \vec{P}(\vec{R}_2) \right|^2} e^{-q \left| \vec{r}_2 - \vec{R}_1 - \vec{Q}(\vec{R}_3) \right|^2} \frac{1}{\left| \vec{r}_1 - \vec{r}_2 \right|} \notag \\
&= \delta_{\vec{k}_\mu + \vec{k}_\lambda, \vec{k}_\nu + \vec{k}_\sigma} \sum_{\vec{R}_2 \in \mathbb{Z}^3} e^{i \vec{k}_\mu \cdot \vec{R}_2} \sum_{\vec{R}_3 \in \mathbb{Z}^3} e^{i \vec{k}_\lambda \cdot \vec{R}_3} \sum_{\vec{R}_1 \in \mathbb{Z}^3} e^{i (\vec{k}_\mu - \vec{k}_\nu) \cdot \vec{R}_1} \notag \\
    &\quad C_\mu C_\nu C_\lambda C_\sigma e^{-\frac{ab}{a+b} \left| \vec{A} - \vec{B}(\vec{R}_2) \right|^2} e^{-\frac{cd}{c+d} \left| \vec{C} - \vec{D}(\vec{R}_3) \right|^2}  \notag \\
    &\quad \frac{2\pi^{5/2}}{pq\sqrt{p+q}} F_0\left( \frac{pq}{p+q} \left| \vec{P}(\vec{R}_2) - \vec{Q}(\vec{R}_3) - \vec{R}_1 \right|^2 \right)
\label{eq:pbc_eri_base_case}
\end{align}

We have applied the shorthand notations $\vec{B}(\vec{R}_2) = \vec{B} + \vec{R}_2$ and $\vec{D}(\vec{R}_3) = \vec{D} + \vec{R}_3$. We also label the location and exponent of the local pair distribution $\mu\nu^{local}(\vec{r}_1, \vec{R}_2)$ (referred to as ``bra") as $\vec{P}(\vec{R}_2) = \frac{a\vec{A} + b\vec{B}(\vec{R}_2)}{a+b}$ and $p = a+b$, and the location and exponent of the local pair distribution $\lambda\sigma^{local}(\vec{r}_2, \vec{R}_3)$ (referred to as ``ket") as $\vec{Q}(\vec{R}_3) = \frac{c\vec{C} + d\vec{D}(\vec{R}_3)}{c+d}$ and $q = c+d$.

The summation over $\vec{R}_1$ has a summand decaying as slow as $\frac{1}{|\vec{R}_1|}$, which causes divergence. Following the same Ewald summation procedure as described in nuclear repulsion integral section, we obtain the following form of the base case electron repulsion integral:
\begin{align}
(ss|ss) &= \delta_{\vec{k}_\mu + \vec{k}_\lambda, \vec{k}_\nu + \vec{k}_\sigma} C_\mu C_\nu C_\lambda C_\sigma \frac{2\pi^{5/2}}{pq\sqrt{p+q}} \sum_{\vec{R}_2 \in \mathbb{Z}^3} e^{i \vec{k}_\mu \cdot \vec{R}_2} e^{-\frac{ab}{a+b} \left| \vec{A} - \vec{B}(\vec{R}_2) \right|^2} \sum_{\vec{R}_3 \in \mathbb{Z}^3} e^{i \vec{k}_\lambda \cdot \vec{R}_3} e^{-\frac{cd}{c+d} \left| \vec{C} - \vec{D}(\vec{R}_3) \right|^2} \notag \\
    &\quad \left( \frac{2\pi^{3/2}}{V_{cell}} \sqrt{\frac{1}{p} + \frac{1}{q}} \sum_{\vec{K}_1 \in \mathbb{Z}^3, \vec{K}_1 \neq \vec{0} \text{ if } \vec{k}_\mu = \vec{k}_\nu} \frac{1}{|\vec{K}_1  - (\vec{k}_\mu - \vec{k}_\nu)|^2} e^{-\frac{1}{4}(\frac{1}{p} + \frac{1}{q} + \frac{1}{\omega^2})|\vec{K}_1 - (\vec{k}_\mu - \vec{k}_\nu)|^2} \right. \notag \\
    &\hspace{200pt}  e^{i \left(\vec{K}_1 - (\vec{k}_\mu - \vec{k}_\nu)\right) \cdot \left(\vec{Q}(\vec{P}_3) - \vec{P}(\vec{P}_2)\right)} \notag \\
    &\qquad + \sum_{\vec{R}_1 \in \mathbb{Z}^3} e^{i (\vec{k}_\mu - \vec{k}_\nu) \cdot \vec{R}_1} \left( F_0\left( \frac{1}{\frac{1}{p} + \frac{1}{q}} \left|\vec{Q}(\vec{R}_3) - \vec{P}(\vec{R}_2) + \vec{R}_1\right|^2 \right) \right. \notag \\
    &\hspace{130pt} \left. - \frac{\omega\sqrt{p+q}}{\sqrt{ pq + \omega^2(p+q) }} F_0\left( \frac{1}{\frac{1}{p} + \frac{1}{q} + \frac{1}{\omega^2}} \left|\vec{Q}(\vec{R}_3) - \vec{P}(\vec{R}_2) + \vec{R}_1\right|^2 \right) \right) \notag \\
    &\qquad \left. - \delta_{\vec{k}_\mu, \vec{k}_\nu} \frac{2\pi^{3/2}}{V_{cell}} \sqrt{\frac{1}{p} + \frac{1}{q}} \frac{1}{4\omega^2} \right)
\label{eq:pbc_eri_base_case_final}
\end{align}

The divergence in reciprocal space lattice summation only occurs if $\vec{k}_\mu = \vec{k}_\nu$ and $\vec{k}_\lambda = \vec{k}_\sigma$. Since all the $\vec{k}$ vectors are within the first Brillouin zone, we do not need to consider the $\vec{k}_\mu - \vec{k}_\nu = 2\pi \vec{K}$ case, where $\vec{K}$ is a non-zero lattice vector.

For the summation over $\vec{R}_2$ and $\vec{R}_3$, the summand decays as fast as a Gaussian function, and thus we can obtain a cutoff scheme similar to equation \ref{eq:pbc_overlap_cutoff_implementation}. However we have an additional prefactor $\dfrac{2\pi^{5/2}}{pq\sqrt{p+q}}$, which relates to the exponent of all four pGTOs ($p = a+b$ and $q = c+d$). In practice we do not want to perform the bra local pair summation over $\vec{R}_2$ and the ket local pair summation over $\vec{R}_3$ inside a kernel function for the Ewald summation. We want to prepare a list of local pairs ahead of time, where the bra local pairs and ket local pairs will be the same list of local pairs. As a result, we want to perform the cutoff for bra ($\vec{R}_2$) and ket ($\vec{R}_3$) separately. The prefactor contribution $\dfrac{1}{\sqrt{p+q}}$ is not a simple multiplication of the pair exponents, and we resolve this problem by defining $\alpha_{min}$ as the minimal exponent among all primitive Gaussian functions of all atomic orbitals, and since $q = c + d \geq 2\alpha_{min}$, the upper bound of $\dfrac{1}{\sqrt{p+q}}$ is $\dfrac{1}{\sqrt{p + 2\alpha_{min}}}$. This way we separate the prefactor upper bound into the bra contribution $C_\mu C_\nu \dfrac{\sqrt{2}\pi^{5/4}}{p\sqrt{p + 2\alpha_{min}}}$ and ket contribution $C_\lambda C_\sigma \dfrac{\sqrt{2}\pi^{5/4}}{q\sqrt{2\alpha_{min} + q}}$, and obtain a bra pair cutoff scheme as follows:
\begin{align}
\left| C_\mu C_\nu \frac{\sqrt{2}\pi^{5/4}}{p\sqrt{p + 2\alpha_{min}}} e^{-\frac{ab}{a+b} \left| (\vec{A} - \vec{B}) - \vec{R}_2 \right|^2} \right| > \delta_{pair-R}
\label{eq:pbc_eri_pair_cutoff_definition}
\end{align}
Or equivalently:
\begin{align}
\left| \vec{R}_2 - (\vec{A} - \vec{B}) \right| < \sqrt{ \frac{a+b}{ab} \ln\left(\frac{\sqrt{2}\pi^{5/4} |C_\mu C_\nu|}{p\sqrt{p + 2\alpha_{min}} \delta_{pair-R}}\right) }
\label{eq:pbc_eri_pair_cutoff_implementation}
\end{align}

Similar to molecular calculations described in Ref \citenum{terachem_f} we also apply the Cauchy-Schwarz upper bound $\left| (\mu\nu^{local}|\mu\nu^{local}) \right|^{1/2}$ to the local pairs, according to the Cauchy-Schwarz inequality:
\begin{align}
\left| (\mu\nu^{local}|\lambda\sigma^{local}) \right| \leq \left| (\mu\nu^{local}|\mu\nu^{local}) \right|^{1/2} \left| (\lambda\sigma^{local}|\lambda\sigma^{local}) \right|^{1/2}
\label{eq:local_cauchy_schwarz_bound}
\end{align}
The integrals in this equation are all molecular integrals without lattice summation. So an additional cutoff threshold $\delta_{pair-Cauchy-Schwarz}$ and an additional upper bound relationship is defined:
\begin{align}
\left| (\mu\nu^{local}|\mu\nu^{local}) \right|^{1/2} = \left| 2^{1/4} C_{\mu}C_{\nu} e^{-\frac{ab}{a+b}\left| (\vec{A} - \vec{B}) - \vec{R}_2 \right|^2} \left(\frac{\pi}{p}\right)^{5/4} \right| > \delta_{pair-Cauchy-Schwarz}
\label{eq:pbc_eri_cauchy_schwarz_cutoff_definition}
\end{align}
Or equivalently:
\begin{align}
\left| \vec{R}_2 - (\vec{A} - \vec{B}) \right| < \sqrt{ \frac{a+b}{ab} \ln\left(\frac{2^{1/4} \pi^{5/4} |C_\mu C_\nu|}{p^{5/4} \delta_{pair-Cauchy-Schwarz}}\right) }
\label{eq:pbc_eri_cauchy_schwarz_cutoff_implementation}
\end{align}

Equation \ref{eq:pbc_eri_pair_cutoff_implementation} and \ref{eq:pbc_eri_cauchy_schwarz_cutoff_implementation} have different dependencies on exponent $p$, and we include a local pair only if both equations are satisfied. The thresholds can be different, but in practice they ought to be on the same order of magnitude. The $\ln$ term in either equation can result in a negative value if $|C_\mu C_\nu|$ is small or the threshold is big, in that case the whole crystalline pGTO pair is discarded. Every electron repulsion integral whose bra or ket pair is discarded is considered zero.

For the reciprocal space summation over $\vec{K}_1$, we consider two cases: if $\vec{k}_\mu = \vec{k}_\nu$, then the summand is bounded by
\begin{align}
\left| \frac{1}{|\vec{K}_1|^2} e^{-\frac{1}{4}(\frac{1}{p} + \frac{1}{q} + \frac{1}{\omega^2})|\vec{K}_1|^2} e^{i \vec{K}_1 \cdot \left(\vec{Q}(\vec{P}_3) - \vec{P}(\vec{P}_2)\right)} \right| \leq \frac{1}{|\vec{K}_{min}|^2} e^{-\frac{1}{4}(\frac{1}{p} + \frac{1}{q} + \frac{1}{\omega^2})|\vec{K}_1|^2}
\label{eq:pbc_eri_reciprocal_cutoff_1}
\end{align}
We define an overall prefactor $C_{2e} = C_\mu C_\nu C_\lambda C_\sigma \frac{2\pi^{5/2}}{pq\sqrt{p+q}} e^{-\frac{ab}{a+b} \left| \vec{A} - \vec{B}(\vec{R}_2) \right|^2} e^{-\frac{cd}{c+d} \left| \vec{C} - \vec{D}(\vec{R}_3) \right|^2}$ and use the same cutoff threshold $\delta_{Ewald-K}$ as in the nuclear attraction integral, and the truncation condition for $\vec{K}_1$ is:
\begin{align}
\left| C_{2e} \frac{2\pi^{3/2}}{V_{cell}} \sqrt{\frac{1}{p} + \frac{1}{q}} \frac{1}{|\vec{K}_{min}|^2} e^{-\frac{1}{4}(\frac{1}{p} + \frac{1}{q} + \frac{1}{\omega^2})|\vec{K}_1|^2} \right| > \delta_{Ewald-K}
\label{eq:pbc_eri_reciprocal_cutoff_definition}
\end{align}
Or equivalently:
\begin{align}
|\vec{K}_1| < \sqrt{ \frac{4}{\frac{1}{p} + \frac{1}{q} + \frac{1}{\omega^2}} \ln\left(\frac{2\pi^{3/2} |C_{2e}|}{V_{cell} |\vec{K}_{min}|^2 \delta_{Ewald-K}} \sqrt{\frac{1}{p} + \frac{1}{q}}\right) }
\label{eq:pbc_eri_reciprocal_cutoff_implementation}
\end{align}

Another case is if $\vec{k}_\mu \neq \vec{k}_\nu$. The upper bound condition in this case is:
\begin{align}
&\quad \left| \frac{1}{|\vec{K}_1  - (\vec{k}_\mu - \vec{k}_\nu)|^2} e^{-\frac{1}{4}(\frac{1}{p} + \frac{1}{q} + \frac{1}{\omega^2})|\vec{K}_1 - (\vec{k}_\mu - \vec{k}_\nu)|^2} e^{i \left(\vec{K}_1 - (\vec{k}_\mu - \vec{k}_\nu)\right) \cdot \left(\vec{Q}(\vec{P}_3) - \vec{P}(\vec{P}_2)\right)} \right| \notag \\
&\leq \frac{1}{|\vec{k}_\mu - \vec{k}_\nu|^2} e^{-\frac{1}{4}(\frac{1}{p} + \frac{1}{q} + \frac{1}{\omega^2})|\vec{K}_1 - (\vec{k}_\mu - \vec{k}_\nu)|^2}
\label{eq:pbc_eri_reciprocal_cutoff_2}
\end{align}
which results in a larger bounding sphere:
\begin{align}
|\vec{K}_1 - (\vec{k}_\mu - \vec{k}_\nu)| < \sqrt{ \frac{4}{\frac{1}{p} + \frac{1}{q} + \frac{1}{\omega^2}} \ln\left(\frac{2\pi^{3/2} |C_{2e}|}{V_{cell} |\vec{k}_\mu - \vec{k}_\nu|^2 \delta_{Ewald-K}} \sqrt{\frac{1}{p} + \frac{1}{q}}\right) }
\label{eq:pbc_eri_reciprocal_cutoff_k_sample}
\end{align}
The $\vec{k}_\mu \neq \vec{k}_\nu$ case is shown for completeness, but not implemented in TeraChem yet as we only support $\Gamma$-point calculations so far.

For the real space summation over $\vec{R}_1$, we follow the same procedure as in the nuclear attraction integral shown in equation \ref{eq:pbc_v1e_real_cutoff_1}, use the same cutoff threshold $\delta_{Ewald-R}$, and obtain the following inequality:
\begin{align}
\left| C_{2e} \left(1 - \frac{\omega\sqrt{p+q}}{\sqrt{ pq + \omega^2(p+q) }} \right) e^{-\frac{1}{\frac{1}{p} + \frac{1}{q} + \frac{1}{\omega^2}} \left|\vec{R}_1 - (\vec{P}(\vec{R}_2) - \vec{Q}(\vec{R}_3))\right|^2} \right| > \delta_{Ewald-R}
\label{eq:pbc_eri_real_cutoff_definition}
\end{align}
Or equivalently:
\begin{align}
\left|\vec{R}_1 - (\vec{P}(\vec{R}_2) - \vec{Q}(\vec{R}_3))\right| < \sqrt{ \left(\frac{1}{p} + \frac{1}{q} + \frac{1}{\omega^2}\right) \ln\left(\frac{|C_{2e}|}{\delta_{Ewald-R}} \left(1 - \frac{\omega\sqrt{p+q}}{\sqrt{ pq + \omega^2(p+q) }} \right) \right) }
\label{eq:pbc_eri_real_cutoff_implementation}
\end{align}

For the general case electron repulsion integral, we thus have:
\begin{align}
(\mu\nu|\lambda\sigma)^{diverge} &= \delta_{\vec{k}_\mu + \vec{k}_\lambda, \vec{k}_\nu + \vec{k}_\sigma} C_\mu C_\nu C_\lambda C_\sigma \frac{2\pi^{5/2}}{pq\sqrt{p+q}} \sum_{\vec{R}_2 \in \mathbb{Z}^3} e^{i \vec{k}_\mu \cdot \vec{R}_2} \sum_{\vec{R}_3 \in \mathbb{Z}^3} e^{i \vec{k}_\lambda \cdot \vec{R}_3} \notag \\
    & \left( \prod_{\tau \in \{x,y,z\}} \sum^{i_\tau + j_\tau}_{t_\tau = 0} E^{i_\tau, j_\tau}_{t_\tau, \tau}(A_\tau, B_\tau(R_{2,\tau}), p) \left( \frac{\partial}{\partial P_\tau} \right)^{t_\tau} \right) \notag \\
    & \left( \prod_{\tau \in \{x,y,z\}} \sum^{\kappa_\tau + l_\tau}_{s_\tau = 0} E^{k_\tau, l_\tau}_{s_\tau, \tau}(C_\tau, D_\tau(R_{3,\tau}), q) \left( \frac{\partial}{\partial Q_\tau} \right)^{s_\tau} \right) \notag \\
    &\quad \sum_{\vec{R}_1 \in \mathbb{Z}^3} e^{i (\vec{k}_\mu - \vec{k}_\nu) \cdot \vec{R}_1} F_0\left( \frac{pq}{p+q} \left| \vec{P}(\vec{R}_2) - \vec{Q}(\vec{R}_3) - \vec{R}_1 \right|^2 \right)
\label{eq:pbc_eri_general_to_hermite}
\end{align}

After applying the Ewald summation to the $\vec{R}_1$ summation, and simplifying the derivatives, we obtain the programmable expression of the general case electron repulsion integral:
\begin{align}
(\mu\nu|\lambda\sigma) &= \delta_{\vec{k}_\mu + \vec{k}_\lambda, \vec{k}_\nu + \vec{k}_\sigma} C_\mu C_\nu C_\lambda C_\sigma \frac{2\pi^{5/2}}{pq\sqrt{p+q}} \sum_{\vec{R}_2 \in \mathbb{Z}^3} e^{i \vec{k}_\mu \cdot \vec{R}_2} \sum_{\vec{R}_3 \in \mathbb{Z}^3} e^{i \vec{k}_\lambda \cdot \vec{R}_3} \notag \\
    &\quad \left( \frac{2\pi^{3/2}}{V_{cell}} \sqrt{\frac{1}{p} + \frac{1}{q}} \sum_{\vec{K}_1 \in \mathbb{Z}^3, \vec{K}_1 \neq \vec{0} \text{ if } \vec{k}_\mu = \vec{k}_\nu} \frac{1}{|\vec{K}_1  - (\vec{k}_\mu - \vec{k}_\nu)|^2} e^{-\frac{1}{4}(\frac{1}{p} + \frac{1}{q} + \frac{1}{\omega^2})|\vec{K}_1 - (\vec{k}_\mu - \vec{k}_\nu)|^2} \right. \notag \\
    &\qquad \qquad \sum^{i_x + j_x}_{t_x = 0} E^{i_x, j_x}_{t_x, x}(A_x, B_x(R_{2,x}), p) \sum^{i_y + j_y}_{t_y = 0} E^{i_y, j_y}_{t_y, y}(A_y, B_y(R_{2,y}), p) \sum^{i_z + j_z}_{t_z = 0} E^{i_z, j_z}_{t_z, z}(A_z, B_z(R_{2,z}), p) \notag \\
    &\qquad \qquad \sum^{\kappa_x + l_x}_{s_x = 0} E^{\kappa_x, l_x}_{s_x, x}(C_x, D_x(R_{3,x}), q) \sum^{\kappa_y + l_y}_{s_y = 0} E^{\kappa_y, l_y}_{s_y, y}(C_y, D_y(R_{3,y}), q) \sum^{\kappa_z + l_z}_{s_z = 0} E^{\kappa_z, l_z}_{s_z, z}(C_z, D_z(R_{3,z}), q) \notag \\
    &\qquad \qquad (-1)^{t_x+t_y+t_z} \ i^{t_x+t_y+t_z + s_x+s_y+s_z} e^{i \left(\vec{K}_1 - (\vec{k}_\mu - \vec{k}_\nu)\right) \cdot \left(\vec{Q}(\vec{P}_3) - \vec{P}(\vec{P}_2)\right)} \notag \\
    &\qquad \qquad \left(K_{1,x} - (k_{\mu,x} - k_{\nu,x})\right)^{t_x + s_x} \left(K_{1,y} - (k_{\mu,y} - k_{\nu,y})\right)^{t_y + s_y} \left(K_{1,z} - (k_{\mu,z} - k_{\nu,z})\right)^{t_z + s_z} \notag \\
    &\qquad + \sum_{\vec{R}_1 \in \mathbb{Z}^3} e^{i (\vec{k}_\mu - \vec{k}_\nu) \cdot \vec{R}_1} \notag \\
    &\qquad \qquad \sum^{i_x + j_x}_{t_x = 0} E^{i_x, j_x}_{t_x, x}(A_x, B_x(R_{2,x}), p) \sum^{i_y + j_y}_{t_y = 0} E^{i_y, j_y}_{t_y, y}(A_y, B_y(R_{2,y}), p) \sum^{i_z + j_z}_{t_z = 0} E^{i_z, j_z}_{t_z, z}(A_z, B_z(R_{2,z}), p) \notag \\
    &\qquad \qquad \sum^{\kappa_x + l_x}_{s_x = 0} E^{\kappa_x, l_x}_{s_x, x}(C_x, D_x(R_{3,x}), q) \sum^{\kappa_y + l_y}_{s_y = 0} E^{\kappa_y, l_y}_{s_y, y}(C_y, D_y(R_{3,y}), q) \sum^{\kappa_z + l_z}_{s_z = 0} E^{\kappa_z, l_z}_{s_z, z}(C_z, D_z(R_{3,z}), q) \notag \\
    &\qquad \qquad \left( R_{t_x,t_y,t_z}^0\left( \frac{1}{\frac{1}{p} + \frac{1}{q}}, \vec{Q}(\vec{R}_3) - \vec{P}(\vec{R}_2) + \vec{R}_1 \right) \right. \notag \\
    &\qquad \qquad \quad \left. - \frac{\omega\sqrt{p+q}}{\sqrt{ pq + \omega^2(p+q) }} R_{t_x,t_y,t_z}^0\left( \frac{1}{\frac{1}{p} + \frac{1}{q} + \frac{1}{\omega^2}}, \vec{Q}(\vec{R}_3) - \vec{P}(\vec{R}_2) + \vec{R}_1 \right) \right) \notag \\
    &\qquad - \delta_{\vec{k}_\mu, \vec{k}_\nu} \frac{2\pi^{3/2}}{V_{cell}} \sqrt{\frac{1}{p} + \frac{1}{q}} \frac{1}{4\omega^2} \notag \\
    &\qquad \qquad E^{i_x, j_x}_{0, x}(A_x, B_x(R_{2,x}), p) E^{i_y, j_y}_{0, y}(A_y, B_y(R_{2,y}), p) E^{i_z, j_z}_{0, z}(A_z, B_z(R_{2,z}), p) \notag \\
    &\qquad \qquad E^{\kappa_x, l_x}_{0, x}(C_x, D_x(R_{3,x}), q) E^{\kappa_y, l_y}_{0, y}(C_y, D_y(R_{3,y}), q) E^{\kappa_z, l_z}_{0, z}(C_z, D_z(R_{3,z}), q) \Biggl. \Biggr)
\label{eq:pbc_eri_general_final}
\end{align}

This equation is the most general form of an electron repulsion integral. As we will see in the next two sections, we only need to consider the two cases where either $\vec{k}_\mu = \vec{k}_\nu$ and $\vec{k}_\lambda = \vec{k}_\sigma$, or $\vec{k}_\mu = \vec{k}_\sigma$ and $\vec{k}_\nu = \vec{k}_\lambda$. Furthermore, within $\Gamma$-point approximation, we can limit our implementation to the $\vec{k}_\mu = \vec{k}_\nu = \vec{k}_\lambda = \vec{k}_\sigma = \vec{0}$ case, and so this equation can be simplified accordingly.

\subsection{Coulomb and exchange matrix}

Similar to the non-periodic case, in most HF or DFT implementations, we do not need to explicitly evaluate any individual $(\mu\nu|\lambda\sigma)$ elements, but instead, we construct the Coulomb and HF exchange matrices. The Coulomb matrix $\mathbf{J}^{\vec{k}}$ for a reciprocal vector $\vec{k}$ is defined as
\begin{align}
J_{\mu\nu}^{\vec{k}} = \frac{1}{V_{B.Z.}} \iiint_{B.Z.} d\vec{k}' \sum_{\lambda\sigma}^{n_{AO}} \left(\mu_{\vec{k}} \nu_{\vec{k}}\middle|\sigma_{\vec{k}'}\lambda_{\vec{k}'}\right) \mathcal{D}_{\lambda\sigma}^{\vec{k}'}
\label{eq:pbc_j_definition_integral}
\end{align}
where $B.Z.$ refers to the first Brillouin zone, and $V_{B.Z.} = \dfrac{8\pi^3}{V_{cell}}$ is the volume of the first Brillouin zone. $\bm{\mathcal{D}}^{\vec{k}'}$ is the density matrix for the reciprocal vector $\vec{k}'$, defined as
\begin{align}
\mathcal{D}^{\vec{k}'} = \sum_{i}^{n_{MO}} n_i^{occ} C_{\mu i}^{\vec{k}'} (C_{\nu i}^{\vec{k}'})^*
\label{eq:pbc_density_matrix_definition}
\end{align}
In this definition $n_i^{occ}$ is the occupation number of the $i$-th molecular orbital.

In practice the integral in equation \ref{eq:pbc_j_definition_integral} is performed numerically, called $\vec{k}$ sampling, where only a finite set of $\vec{k}'$ gridpoints within the first Brillouin zone is included, each with weight $\omega_{\vec{k}'}$:
\begin{align}
J_{\mu\nu}^{\vec{k}} \approx \sum_{\vec{k}' \in B.Z.} \omega_{\vec{k}'} \sum_{\lambda\sigma}^{n_{AO}} \left(\mu_{\vec{k}} \nu_{\vec{k}}\middle|\sigma_{\vec{k}'}\lambda_{\vec{k}'}\right) \mathcal{D}_{\lambda\sigma}^{\vec{k}'}
\label{eq:pbc_j_definition_k_sample}
\end{align}

The HF exchange matrix $\mathbf{K}^{\vec{k}}$ sampled at $\vec{k}$ is similarly defined as:
\begin{align}
K_{\mu\nu}^{\vec{k}} &= \frac{1}{V_{B.Z.}} \iiint_{B.Z.} d\vec{k}' \sum_{\lambda\sigma}^{n_{AO}} \left(\mu_{\vec{k}} \lambda_{\vec{k}'} \middle|\sigma_{\vec{k}'} \nu_{\vec{k}}\right) \mathcal{D}_{\lambda\sigma}^{\vec{k}'} \notag \\
    &\approx \sum_{\vec{k}' \in B.Z.} \omega_{\vec{k}'} \sum_{\lambda\sigma}^{n_{AO}} \left(\mu_{\vec{k}} \lambda_{\vec{k}'} \middle|\sigma_{\vec{k}'} \nu_{\vec{k}}\right) \mathcal{D}_{\lambda\sigma}^{\vec{k}'}
\label{eq:pbc_k_definition}
\end{align}

In the next section we will address how $k$ sampling is implemented in TeraChem.

\subsection{$\Gamma$-point approximation}

In a solid system where the unit cell is large, the reciprocal space unit cell is thus small, and the amount of $\vec{k}$ points necessary to sample the first Brillouin zone is also small. The extreme case is the $\Gamma$-point approximation: the only sample point of the reciprocal space is the $\Gamma$-point, where $\vec{k} = \vec{0}$ and $\omega_{\vec{0}} = 1$. When the unit cell of a system is not large, the $\Gamma$-point approximation is still valid if a supercell is constructed. By repeating the unit cell in real space, we effectively sample the Brillouin zone uniformly. \cite{gamma_point}

The advantage of $\Gamma$-point approximation in terms of implementation is very significant. First, under the $\Gamma$-point approximation, all phase factors in the basis functions and integral equations become unity, and as a result all basis functions and integrals become real-valued. We can improve the run time efficiency by avoiding complex value arithmetic. Second, since we only need to sample one $\vec{k}$ point, we only need to construct and store one density matrix or Fock matrix. This will allow us to reuse the program architecture for molecular calculations. In order to support $\Gamma$-point periodic HF or DFT calculation, we only swap the integrals from non-periodic integrals to the corresponding $\Gamma$-point periodic integrals. The $\Gamma$-point periodic integrals can be obtained by simply setting $\vec{k}_\mu = \vec{k}_\nu = \vec{k}_\lambda = \vec{k}_\sigma = 0$ in the equations in the preceding sections.

However, it is important to note that our general approach is not limited to the $\Gamma$-point approximation. The $\vec{k}$ sampling can be readily supported by transitioning to complex arithmetic and establishing an infrastructure to handle multiple density and Fock matrices, which will be explored in future.

\subsection{Exchange-correlation integral}

In order to support local exchange-correlation functionals within the Generalized Gradient Approximation (GGA), we need to evaluate the electron density $\rho(\vec{r})$ and its gradient $\vec{\nabla}\rho(\vec{r})$ at gridpoints $\{\vec{r}\}$:
\begin{align}
\rho(\vec{r}) &= \frac{1}{V_{B.Z.}} \iiint_{B.Z.} d\vec{k} \sum_{\mu\nu}^{n_{AO}} \mathcal{D}_{\mu\nu}^{\vec{k}} \mu_{\vec{k}}^*(\vec{r}; \vec{i},a,\vec{A}) \nu_{\vec{k}}(\vec{r}; \vec{j},b,\vec{B}) \notag \\
    &\approx \sum_{\vec{k} \in B.Z.} \omega_{\vec{k}} \sum_{\mu\nu}^{n_{AO}} \mathcal{D}_{\mu\nu}^{\vec{k}} \mu_{\vec{k}}^*(\vec{r}; \vec{i},a,\vec{A}) \nu_{\vec{k}}(\vec{r}; \vec{j},b,\vec{B}) \label{eq:pbc_xc_density_0} \\
\vec{\nabla} \rho(\vec{r}) &= \frac{1}{V_{B.Z.}} \iiint_{B.Z.} d\vec{k} \sum_{\mu\nu}^{n_{AO}} \mathcal{D}_{\mu\nu}^{\vec{k}} \nabla_{\vec{r}} \left( \mu_{\vec{k}}^*(\vec{r}; \vec{i},a,\vec{A}) \nu_{\vec{k}}(\vec{r}; \vec{j},b,\vec{B}) \right) \notag \\
    &\approx \sum_{\vec{k} \in B.Z.} \omega_{\vec{k}} \sum_{\mu\nu}^{n_{AO}} \mathcal{D}_{\mu\nu}^{\vec{k}} \nabla_{\vec{r}} \left( \mu_{\vec{k}}^*(\vec{r}; \vec{i},a,\vec{A}) \nu_{\vec{k}}(\vec{r}; \vec{j},b,\vec{B}) \right) \label{eq:pbc_xc_density_1}
\end{align}

These can be expressed in terms of local pGTOs as:
\begin{align}
\rho(\vec{r}) &= \sum_{\vec{k} \in B.Z.} \omega_{\vec{k}} \sum_{\mu\nu}^{n_{AO}} \mathcal{D}_{\mu\nu}^{\vec{k}} \sum_{\vec{R}_1 \in \mathbb{Z}^3} \sum_{\vec{R}_2 \in \mathbb{Z}^3} e^{i \vec{k} \cdot \vec{R}_2} \mu^{local}(\vec{r} - \vec{R}_1; \vec{i},a,\vec{A}) \nu^{local}(\vec{r} - \vec{R}_1 - \vec{R}_2; \vec{j},b,\vec{B}) \label{eq:pbc_xc_density_local_0} \\
\vec{\nabla} \rho(\vec{r}) &= \sum_{\vec{k} \in B.Z.} \omega_{\vec{k}} \sum_{\mu\nu}^{n_{AO}} \mathcal{D}_{\mu\nu}^{\vec{k}} \sum_{\vec{R}_1 \in \mathbb{Z}^3} \sum_{\vec{R}_2 \in \mathbb{Z}^3} e^{i \vec{k} \cdot \vec{R}_2} \nabla_{\vec{r}} \left( \mu^{local}(\vec{r} - \vec{R}_1; \vec{i},a,\vec{A}) \nu^{local}(\vec{r} - \vec{R}_1 - \vec{R}_2; \vec{j},b,\vec{B}) \right) \label{eq:pbc_xc_density_local_1}
\end{align}

Unlike all the other classes of integrals discussed above, we are not able to remove a lattice summation using equation \ref{eq:integral_translation_invariance}. It is still possible to use the local pGTO pair structure (equation \ref{eq:local_pair_definition}), by treating the $\vec{R}_1$ lattice vector summation as summing over periodic images of the gridpoint $\vec{r}$.

The electron density and its gradient on each gridpoint are used to evaluate the exchange-correlation potentials $\dfrac{\partial \varepsilon_{XC}}{\partial \rho} (\rho(\vec{r}), \vec{\nabla} \rho(\vec{r}) )$ and $\dfrac{\partial \varepsilon_{XC}}{\partial (\vec{\nabla} \rho)}( \rho(\vec{r}), \vec{\nabla} \rho(\vec{r}) )$ at each gridpoint.

The local exchange-correlation integrals in GGA are defined as:
\begin{align}
V_{\mu\nu}^{XC, \vec{k}, 0} &= \iiint_\infty d\vec{r} \ \frac{\partial \varepsilon_{XC}}{\partial \rho}( \rho(\vec{r}), \vec{\nabla} \rho(\vec{r}) ) \mu_{\vec{k}}^*(\vec{r}; \vec{i},a,\vec{A}) \nu_{\vec{k}}(\vec{r}; \vec{j},b,\vec{B}) \label{eq:pbc_xc_integral_definition_0} \\
V_{\mu\nu}^{XC, \vec{k}, 1} &= \iiint_\infty d\vec{r} \ \frac{\partial \varepsilon_{XC}}{\partial (\vec{\nabla} \rho)}( \rho(\vec{r}), \vec{\nabla} \rho(\vec{r}) ) \cdot \vec{\nabla}_{\vec{r}} \left( \mu_{\vec{k}}^*(\vec{r}; \vec{i},a,\vec{A}) \nu_{\vec{k}}(\vec{r}; \vec{j},b,\vec{B}) \right)\label{eq:pbc_xc_integral_definition_1}
\end{align}
which can be simplified to local pGTO form:
\begin{align}
V_{\mu\nu}^{XC, \vec{k}, 0} &= \sum_{\vec{R}_2 \in \mathbb{Z}^3} e^{i \vec{k} \cdot \vec{R}_2} \iiint_\infty d\vec{r} \ \frac{\partial \varepsilon_{XC}}{\partial \rho}( \rho(\vec{r}), \vec{\nabla} \rho(\vec{r}) ) \mu^{local}(\vec{r}; \vec{i},a,\vec{A}) \nu^{local}(\vec{r} - \vec{R}_2; \vec{j},b,\vec{B}) \label{eq:pbc_xc_integral_local_0} \\
V_{\mu\nu}^{XC, \vec{k}, 1} &= \sum_{\vec{R}_2 \in \mathbb{Z}^3} e^{i \vec{k} \cdot \vec{R}_2} \iiint_\infty d\vec{r} \ \frac{\partial \varepsilon_{XC}}{\partial (\vec{\nabla} \rho)}( \rho(\vec{r}), \vec{\nabla} \rho(\vec{r}) ) \cdot \vec{\nabla}_{\vec{r}} \left( \mu^{local}(\vec{r}; \vec{i},a,\vec{A}) \nu^{local}(\vec{r} - \vec{R}_2; \vec{j},b,\vec{B}) \right) \label{eq:pbc_xc_integral_local_1}
\end{align}

The exchange-correlation integral is generally evaluated numerically as a weighted sum over gridpoints in real space. The common ways of constructing grid includes atom-centered grids \cite{becke_weight_original, becke_weight_periodic, dft_integral, dft_standard_grid_1} and uniform grids \cite{cp2k_mixed_basis}. In either case, a gridpoint is defined with respect to a particular unit cell. In order to integrate over the whole space, we need to include the periodic images of the gridpoint as well. As a result, the numerical integration formula for the exchange-correlation integral over gridpoints $\vec{r}_g$ with weights $w_g$  is:
\begin{align}
V_{\mu\nu}^{XC, \vec{k}, 0} &\approx \sum_{\vec{R}_1 \in \mathbb{Z}^3} \sum_{\vec{R}_2 \in \mathbb{Z}^3} e^{i \vec{k} \cdot \vec{R}_2} \sum_g^{N_{grid}} w_g \frac{\partial \varepsilon_{XC}}{\partial \rho}( \rho(\vec{r}_g), \vec{\nabla} \rho(\vec{r}_g) ) \notag \\
    &\hspace{120pt} \mu^{local}(\vec{r}_g - \vec{R}_1; \vec{i},a,\vec{A}) \nu^{local}(\vec{r}_g - \vec{R}_1 - \vec{R}_2; \vec{j},b,\vec{B}) \label{eq:pbc_xc_grid_sum_0} \\
V_{\mu\nu}^{XC, \vec{k}, 1} &\approx \sum_{\vec{R}_1 \in \mathbb{Z}^3} \sum_{\vec{R}_2 \in \mathbb{Z}^3} e^{i \vec{k} \cdot \vec{R}_2} \sum_g^{N_{grid}} w_g \frac{\partial \varepsilon_{XC}}{\partial (\vec{\nabla} \rho)}( \rho(\vec{r}_g), \vec{\nabla} \rho(\vec{r}_g) ) \notag \\
    &\hspace{120pt} \cdot \vec{\nabla}_{\vec{r}} \left( \mu^{local}(\vec{r}_g - \vec{R}_1; \vec{i},a,\vec{A}) \nu^{local}(\vec{r}_g - \vec{R}_1 - \vec{R}_2; \vec{j},b,\vec{B}) \right) \label{eq:pbc_xc_grid_sum_1}
\end{align}

In the equation above we apply the fact that the electron density $\rho(\vec{r})$ and its gradient $\vec{\nabla} \rho(\vec{r})$ are all periodic functions and are invariant to lattice vector translation, and thus the exchange-correlation potentials $\dfrac{\partial \varepsilon_{XC}}{\partial \rho} (\rho(\vec{r}), \vec{\nabla} \rho(\vec{r}) )$ and $\dfrac{\partial \varepsilon_{XC}}{\partial (\vec{\nabla} \rho)}( \rho(\vec{r}), \vec{\nabla} \rho(\vec{r}) )$, which depend only on the electron density and its gradient, are also periodic.

For both the electron density evaluation (equation \ref{eq:pbc_xc_density_local_0} and \ref{eq:pbc_xc_density_local_1}) and exchange-correlation integral evaluation (equation \ref{eq:pbc_xc_grid_sum_0} and \ref{eq:pbc_xc_grid_sum_1}), the two lattice summations over $\vec{R}_1$ and $\vec{R}_2$ both decay as fast as Gaussians, so in order to truncate the lattice summation, we reuse the pair lattice summation cutoff scheme for overlap integral (equation \ref{eq:pbc_overlap_cutoff_implementation}) and the corresponding threshold $\delta_{pair-R}$.

\subsection{Becke weights}

While evaluating the exchange-correlation integral, an integration grid in real space is required. The inner-shell electron density varies rapidly in space, and would therefore require prohibitively large uniform grids \cite{periodic_gaussian_review}. As a result, we choose to construct atom-centered grids for numerical integration, which are effective for all-electron molecular calculations.

For each atom located at $\vec{A}$ (in any lattice image), we construct a spherical grid of points around $\vec{A}$, with quadrature position offset $\vec{r}_{g, A}^{quadrature}$ and quadrature weight $w_{g, A}^{quadrature}$. The quadrature position offset and weight are predefined for each element, and tabulated ahead of time. The actual gridpoint location is $\vec{r}_g = \vec{A} + \vec{r}_{g, A}^{quadrature}$. In order to partition each gridpoint to individual atoms in a smooth way, we compute the Becke weight of each gridpoint with respect to its origin atom $A$:
\begin{align}
\omega^{Becke}_{g, A} = \frac{P(\vec{A}, \vec{r}_g)}{ \sum_{B}^{N_{atom}} \prod_{\vec{R}_3 \in \mathbb{Z}^3} P(\vec{B} + \vec{R}_3, \vec{r}_g) }
\label{eq:becke_weight_definition}
\end{align}
where the vicinity function $P(\vec{A}, \vec{r})$ describes if $\vec{r}$ is in the vicinity of atom $A$ located at $\vec{A}$ compared to other atoms or images of atoms. $P(\vec{A}, \vec{r})$ can be expressed as
\begin{align}
P(\vec{A}, \vec{r}) = \left( \prod_{B \neq A}^{N_{atom}} \prod_{\vec{R}_1 \in \mathbb{Z}^3} s(\vec{A}, \vec{B} + \vec{R}_1, \vec{r}) \right) \cdot \left( \prod_{\vec{R}_2 \in \mathbb{Z}^3, \vec{R}_2 \neq \vec{0}} s(\vec{A}, \vec{A} + \vec{R}_2, \vec{r}) \right)
\label{eq:becke_vicinity_definition}
\end{align}
Here the switching function $s(\vec{A}, \vec{B}, \vec{r})$ is a smooth function varying from 1 to 0 as $\vec{r}$ moves from $A$ to $B$. It is usually implemented in terms of $u(\vec{A}, \vec{B}, \vec{r})$:
\begin{align}
u(\vec{A}, \vec{B}, \vec{r}) = \frac{|\vec{r} - \vec{A}| - |\vec{r} - \vec{B}|}{|\vec{A} - \vec{B}|}
\label{eq:becke_mu_definition}
\end{align}
And the functional form Becke proposed for the switching function is \cite{becke_weight_original}:
\begin{align}
s(\vec{A}, \vec{B}, \vec{r}) = \frac{1}{2} \left(1 - f\left( f\left( f\left( u(\vec{A}, \vec{B}, \vec{r}) \right) \right) \right) \right) \qquad f(x) = \frac{1}{2}x(3-x^2)
\label{eq:becke_switch_definition}
\end{align}

The first term of equation \ref{eq:becke_vicinity_definition} accounts for if the gridpoint is close to any atoms other than atom $A$ or their images. The second term accounts for if the gridpoint is close to any lattice images of $\vec{A}$ that is not the same image as $\vec{A}$. A gridpoint will get zero weight if it is in the vicinity of a different atom $B$ or a different lattice image $\vec{A} + \vec{R}$ from the source atom at $\vec{A}$.

The cutoff sphere for the three products in equation \ref{eq:becke_weight_definition} and \ref{eq:becke_vicinity_definition} is not obvious from any inequality. For each gridpoint, we set the center of the cutoff sphere at its origin atom location $\vec{A}$, and the radius of the cutoff sphere as 17 Bohr (about 9 Angstrom) following a previous study \cite{becke_weight_periodic}.

The procedure for obtaining the Becke weight of a gridpoint $\vec{r}_g$ is: we first compute the vicinity function $P(\vec{A}, \vec{r}_g)$ (the numerator of equation \ref{eq:becke_weight_definition}), where $\vec{A}$ is the origin atom location of $\vec{r}_g$. In order to do so, we loop over all other atoms $B$. We find the lattice image of $B$ that is closest to $\vec{A}$ in distance. We then construct a bounding sphere for the lattice product over $\vec{R}_1$, and compute all switching functions $s(\vec{A}, \vec{B} + \vec{R}_1, \vec{r})$. This will give us the first term in equation \ref{eq:becke_vicinity_definition}. We similarly construct a bounding sphere for $\vec{R}_2$ product, and compute the second term in equation \ref{eq:becke_vicinity_definition}. If the vicinity function $P(\vec{A}, \vec{r}_g)$ is close to zero (below a threshold $\delta_{Becke}$), then the Becke weight of $\vec{r}_g$ is 0 and the subsequent computation is skipped. Otherwise, we go through all other atoms and their lattice images within the bounding sphere, as well as the lattice images of $\vec{A}$ within the bounding sphere, for the denominator of equation \ref{eq:becke_weight_definition}. During each vicinity function evaluation, whenever a switching function produces a near-zero value (less than $\delta_{Becke}$), the vicinity function is considered zero and subsequent iterations will be dropped.

The total weight of a gridpoint is the product of its quadrature weight and its Becke weight with respect to the origin atom:
\begin{align}
\omega_g = \omega^{quadrature}_{g, A} \omega^{Becke}_{g, A}
\label{eq:total_weight_definition}
\end{align}

A way to assess the quality of the integration grid is to integrate a unit function ($f(\vec{r}) = 1$), which should integrate to the unit cell volume in ideal cases:
\begin{align}
\sum_{g}^{N_{grid}} \omega_g = V_{cell}
\label{eq:becke_grid_integrate_one}
\end{align}
In practice, for a reasonably large molecular grid, the summation of weights is usually greater than the cell volume by 0.01\% to 0.25\%.

\subsection{Nuclear repulsion energy}
The nuclear repulsion energy is formally divergent as well for a periodic system, and the divergent term needs to be manually removed via an Ewald strategy like the other electrostatic integrals previously described. The formula for nuclear repulsion energy is:
\begin{align}
E^{nuclear} &= \sum_{A = 1}^{N_{atom}} \sum_{B = A + 1}^{N_{atom}} q_A q_B \left( \frac{4\pi}{V_{cell}} \sum_{\vec{K} \in \mathbb{Z}^3, \vec{K} \neq \vec{0}} \frac{1}{|\vec{K}|^2} e^{-\frac{1}{4\omega^2} |\vec{K}|^2} \mathrm{cos}(\vec{K} \cdot (\vec{A} - \vec{B})) - \frac{4\pi}{V_{cell}} \frac{1}{4\omega^2} \right. \notag \\
    &\hspace{110pt} \left. + \sum_{\vec{R} \in \mathbb{Z}^3} \frac{\mathrm{erfc}\left( \omega |\vec{A} - \vec{B} + \vec{R}| \right)}{ |\vec{A} - \vec{B} + \vec{R}| } \right) \notag \\
    &\quad + \frac{1}{2} \sum_{A = 1}^{N_{atom}} q_A^2 \left( \frac{4\pi}{V_{cell}} \sum_{\vec{K} \in \mathbb{Z}^3, \vec{K} \neq \vec{0}} \frac{1}{|\vec{K}|^2} e^{-\frac{1}{4\omega^2} |\vec{K}|^2} - \frac{4\pi}{V_{cell}} \frac{1}{4\omega^2} + \sum_{\vec{R} \in \mathbb{Z}^3, \vec{R} \neq \vec{0}} \frac{\mathrm{erfc}\left( \omega |\vec{R}| \right)}{ |\vec{R}| } - \frac{2\omega}{\sqrt{\pi}} \right)
\label{eq:nuclear_repulsion_energy}
\end{align}
The term accounting for reciprocal space divergence removal ($\dfrac{4\pi}{V_{cell}} \dfrac{1}{4\omega^2}$) does not appear in the standard Ewald summation formula, and we need to include it because the total charge of all nuclei is not zero.

Since the complementary error function is bounded by:
\begin{align}
\mathrm{erfc}(x) < \frac{2}{\sqrt{\pi}} \frac{e^{-x^2}}{x}
\label{eq:erfc_upper_bound}
\end{align}
The cutoff radius $r_{bound}$ for the real space lattice summation can be obtained by solving the following equation numerically:
\begin{align}
\frac{2}{\sqrt{\pi}} \frac{e^{-\omega^2 r_{bound}^2}}{r_{bound}^2} = \delta_{Ewald-R}
\label{eq:nuclear_repulsion_energy_real_cutoff}
\end{align}

Similarly, the cutoff radius $k_{bound}$ for the reciprocal space lattice summation can be obtained by solving the following equation numerically:
\begin{align}
\frac{4\pi}{V_{cell}} \frac{e^{-\frac{1}{4\omega^2} k_{bound}^2}}{k_{bound}^2} = \delta_{Ewald-K}
\label{eq:nuclear_repulsion_energy_reciprocal_cutoff}
\end{align}

The bounding sphere for nuclear repulsion energy calculation depends only on the range separation $\omega$ parameter, and is obtained only once for a single point calculation. As a result, there is little concern about computational expense in the evaluating the bounding radii.


\section{Implementation\label{sec:implementation}}

\subsection{Data structure}

\begin{figure}[htb!]
    \centering
    \includegraphics[width=0.6\linewidth]{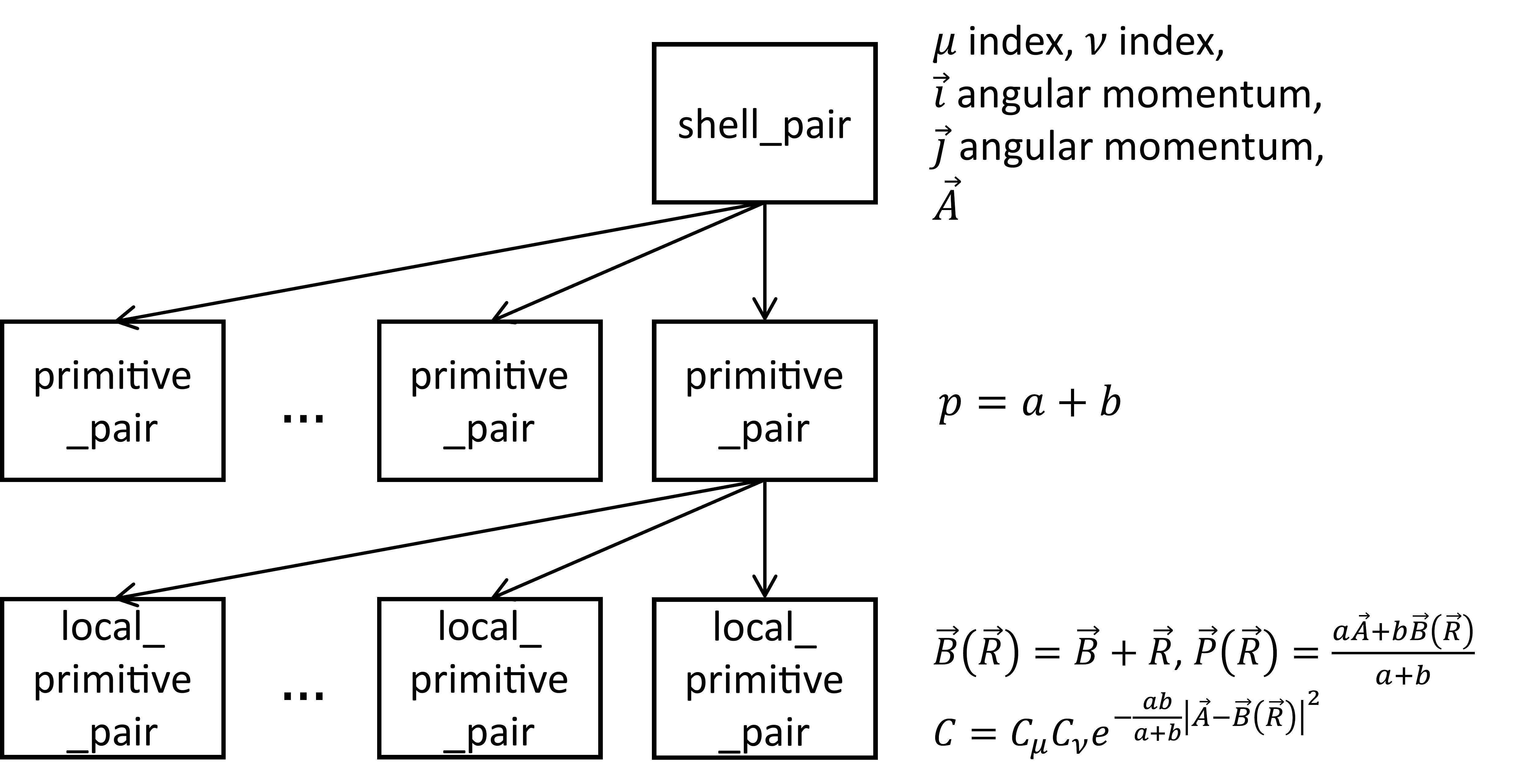}
\caption{The shell pair, pGTO pair and local pGTO pair data structure. The arrows show a inclusion relationship: a shell pair includes a list of pGTO pairs, and a pGTO pair includes a list of local pGTO pairs. On the right-hand side we show the values stored in each layer of the data structure.}
\label{fig:pair_data_structure}
\end{figure}

In order to support the pair lattice vector summation (the $\vec{R}$ summation in equation \ref{eq:pbc_overlap_formula} and \ref{eq:pbc_v1e_general_final}, the $\vec{R}_2$ and $\vec{R}_3$ summation in equation \ref{eq:pbc_eri_general_final}, and the $\vec{R}_2$ summation in equation \ref{eq:pbc_xc_density_local_0}, \ref{eq:pbc_xc_density_local_1}, \ref{eq:pbc_xc_integral_local_0} and \ref{eq:pbc_xc_integral_local_1}), we define a three-layer pair data structure, as shown in Fig. \ref{fig:pair_data_structure}. The atomic orbitals are grouped first into shells, where each shell contains atomic orbitals sharing the same total angular momentum as well as the same contracted Gaussian exponents and coefficients in equation \ref{eq:local_contracted_basis}, but having different angular momentum indices (the $p_x$, $p_y$ and $p_z$ orbitals, for example). We construct pairs of shells, and for each shell pair, we store the list of $\mu$ and $\nu$ indices, as well as the $\mu$ atom center location $\vec{A}$ (in any image). Since each contracted atomic orbital is composed of a list of primitive Gaussian functions, a shell pair contains a list of primitive shell pairs, each with a unique pair of exponents and contraction coefficients. Each primitive shell pair requires a list of local primitive shell pairs with lattice vector $\vec{R}$. A local primitive shell pair stores the offset $\nu$ center location $\vec{B}(\vec{R})$, the offset pair center location $\vec{P}(\vec{R})$, and a overall prefactor.

The pair data structure is constructed when the atom positions and basis set are read in, before any integrals are calculated. For each primitive shell pair, we construct a bounding sphere according to equation \ref{eq:pbc_eri_pair_cutoff_implementation} and equation \ref{eq:pbc_eri_cauchy_schwarz_cutoff_implementation}, iterate through all lattice vectors inside the bounding sphere, and compute the prefactor of the corresponding local primitive shell pair. We check if the prefactor is above the thresholds $\delta_{pair-R}$ and $\delta_{pair-Cauchy-Schwarz}$, and if not, the primitive shell pair is discarded. If the primitive shell pair has no local primitive shell pairs in its list, the whole primitive shell pair is discarded, and similarly an empty shell pair is discarded completely, which happens when the two atoms are far away even after accounting for lattice images. If the primitive shell pair is not empty, we sort its list of local primitive shell pairs by the absolute value of prefactors. We also sort the list of primitive shell pairs within a shell pair by the largest absolute value of prefactor, and then sort the list of all shell pairs in the same way. The sorted three-layer approach for pairs allows us to apply different truncation thresholds efficiently, for example, the switch-over threshold between single and double precision in a mixed precision calculation.

The three-layer data structure is defined in terms of shells of orbitals. Each primitive shell is expanded into pGTOs in a kernel call, which we will describe in the next section.

\subsection{GPU acceleration scheme}

The GPU acceleration algorithm for integrals under periodic boundary conditions in TeraChem is similar to the molecular integral implementation \cite{terachem_f}. Here we briefly describe the algorithm, focusing on the treatment of the Ewald summation.

The overlap and kinetic energy integrals are computed on the CPU as they are quit inexpensive. 
The nuclear attraction integrals are evaluated on the GPU in the following way: Each GPU thread obtains a local primitive shell pair, loops through all the point charges, and computes an Ewald summation in each iteration. We split the Ewald summation calculation into two GPU kernel functions, one for real space and one for reciprocal space. The Ewald summation result (also summed over all point charges according to $V_{\mu\nu} = \sum_C^{N_{point-charge}} V_{\mu\nu C}$) is copied back to CPU, and the summation over local pGTO pairs (over $\vec{R}$) is performed on CPU. If the angular momentum of $\mu$ or $\nu$ is non-zero, the $E^{i_\tau, j_\tau}_{t_\tau, \tau}$ and $R_{t_x,t_y,t_z}^0$ terms are generated in the same way as previously described for molecular integrals\cite{terachem_f}.  Each thread will handle all local pGTO pairs in a local primitive shell pair. For example, a PD primitive shell pair will have 18 pGTO pairs, and thus in every lattice vector summation iteration, all 18 elements of the nuclear attraction integral ($V_{p_xd_{xx}}$, $V_{p_xd_{xy}}$, ..., $V_{p_zd_{zz}}$) will be computed and accumulated into 18 individual variables.

The Coulomb matrix $\mathbf{J}$ construction is simplified by a periodic version of the ``J-engine'' \cite{j_engine_hgp, j_prepostprocess_original, j_prepostprocess_family_basis_set, terachem_gpu_2}. The  density matrix  in the Hermite Gaussian basis under the $\Gamma$-point approximation is: 

\begin{align}
\mathcal{D}_{s_xs_ys_z}(\vec{R}_3) &= \sum\limits_{\substack{\lambda(\vec{r}; \vec{\kappa},c,\vec{C}),\\ \sigma(\vec{r}; \vec{l},d,\vec{D})}}^{n_{AO}} E^{\kappa_x, l_x}_{s_x, x}(C_x, D_x(R_{3,x}), q) E^{\kappa_y, l_y}_{s_y, y}(C_y, D_y(R_{3,y}), q) E^{\kappa_z, l_z}_{s_z, z}(C_z, D_z(R_{3,z}), q) \notag \\
    &\hspace{60pt} \mathcal{D}_{ \lambda(\vec{r}; \vec{\kappa},c,\vec{C}), \sigma(\vec{r}; \vec{l},d,\vec{D}) }
\label{eq:pbc_j_split_1}
\end{align}

By combining the electron repulsion integral formula (equation \ref{eq:pbc_eri_general_final}), the definition of $\mathbf{J}$ (equation \ref{eq:pbc_j_definition_k_sample}), and reordering the summations, the $\mathbf{J}$ matrix elements under $\Gamma$-point approximation are obtained as:
\begin{align}
J_{\mu\nu} &= \sum_{\vec{R}_2 \in \mathbb{Z}^3} \sum_{\vec{R}_3 \in \mathbb{Z}^3} J_{\mu\nu}(\vec{R}_2, \vec{R}_3)
\label{eq:pbc_j_split_4}\\
J_{ \mu(\vec{r};\vec{i},a,\vec{A}), \nu(\vec{r};\vec{j},b,\vec{B}) }(\vec{R}_2, \vec{R}_3) &= \sum^{i_x + j_x}_{t_x = 0} E^{i_x, j_x}_{t_x, x}(A_x, B_x(R_{2,x}), p) \sum^{i_y + j_y}_{t_y = 0} E^{i_y, j_y}_{t_y, y}(A_y, B_y(R_{2,y}), p) \notag \\
    &\quad \sum^{i_z + j_z}_{t_z = 0} E^{i_z, j_z}_{t_z, z}(A_z, B_z(R_{2,z}), p) J_{t_xt_yt_z}(\vec{R}_2,\vec{R}_3)
\label{eq:pbc_j_split_3}
\end{align}
where the Hermite Gaussian basis $J_{t_xt_yt_z}(\vec{R}_2,\vec{R}_3)$ terms are:
\begin{align}
J_{t_xt_yt_z}(\vec{R}_2,\vec{R}_3) &= C_\mu C_\nu C_\lambda C_\sigma \frac{2\pi^{5/2}}{pq\sqrt{p+q}} \sum^{\kappa_x + l_x}_{s_x = 0} \sum^{\kappa_y + l_y}_{s_y = 0} \sum^{\kappa_z + l_z}_{s_z = 0} \mathcal{D}_{s_xs_ys_z}(\vec{R}_3) \notag \\
    &\quad \left( \frac{2\pi^{3/2}}{V_{cell}} \sqrt{\frac{1}{p} + \frac{1}{q}} \sum_{\vec{K}_1 \in \mathbb{Z}^3, \vec{K}_1 \neq \vec{0}} \frac{1}{|\vec{K}_1|^2} e^{-\frac{1}{4}(\frac{1}{p} + \frac{1}{q} + \frac{1}{\omega^2})|\vec{K}_1|^2} \right. \notag \\
    &\qquad \qquad (-1)^{t_x+t_y+t_z} \ i^{t_x+t_y+t_z + s_x+s_y+s_z} e^{i \vec{K}_1 \cdot \left(\vec{Q}(\vec{P}_3) - \vec{P}(\vec{P}_2)\right)} \left.K_{1,x}\right.^{t_x + s_x} \left.K_{1,y}\right.^{t_y + s_y} \left.K_{1,z}\right.^{t_z + s_z} \notag \\
    &\qquad + \sum_{\vec{R}_1 \in \mathbb{Z}^3} \left( R_{t_x,t_y,t_z}^0\left( \frac{1}{\frac{1}{p} + \frac{1}{q}}, \vec{Q}(\vec{R}_3) - \vec{P}(\vec{R}_2) + \vec{R}_1 \right) \right. \notag \\
    &\hspace{80pt} \left. - \frac{\omega\sqrt{p+q}}{\sqrt{ pq + \omega^2(p+q) }} R_{t_x,t_y,t_z}^0\left( \frac{1}{\frac{1}{p} + \frac{1}{q} + \frac{1}{\omega^2}}, \vec{Q}(\vec{R}_3) - \vec{P}(\vec{R}_2) + \vec{R}_1 \right) \right) \notag \\
    &\qquad - \frac{2\pi^{3/2}}{V_{cell}} \sqrt{\frac{1}{p} + \frac{1}{q}} \frac{1}{4\omega^2} \delta_{t_x,0} \delta_{t_y,0} \delta_{t_z,0} \delta_{s_x,0} \delta_{s_y,0} \delta_{s_z,0} \Biggl. \Biggr)
\label{eq:pbc_j_split_2}
\end{align}

We perform the density matrix Cartesian Gaussian to Hermite Gaussian transformation (equation \ref{eq:pbc_j_split_1}), the $\mathbf{J}$ matrix Hermite Gaussian to Cartesian Gaussian transformation (equation \ref{eq:pbc_j_split_3}), as well as the $\vec{R}_2$ and $\vec{R}_3$ summation (equation \ref{eq:pbc_j_split_4}) on CPU. The Hermite $\mathbf{J}$ matrix evaluation (equation \ref{eq:pbc_j_split_2}) is performed on GPU. Similar to the nuclear attraction integral implementation, we have two separate GPU kernels for the Ewald summation, one for real space and another for reciprocal space. Each GPU thread will map to a $\mu\nu$ primitive shell pair, and iterate through the $\lambda\sigma$ primitive shell pairs. Each thread will keep a copy of all $\mathbf{J}$ matrix elements corresponding to $\mu\nu$ pair in the Hermite Gaussian basis, and update their values in each Ewald summation iteration. We point the interested reader to Ref \citenum{terachem_f} (which describes the implementation for molecular systems) for greater details.

Due to the index mismatch between the bra ket pair and input ($\bm{\mathcal{D}}$) output ($\mathbf{K}$) pair, the GPU implementation of HF exchange matrix $\mathbf{K}$ matrix construction is more complicated, and is described in greater detail in Ref \citenum{terachem_f} for isolated molecules. We reuse most of the molecular algorithm, with two major changes: an Ewald summation is performed in place of an electron repulsion integral element evaluation, and every GPU kernel function is split into two, one for real space Ewald summation and one for reciprocal space.

For a DFT calculation, both density evaluation at gridpoints and exchange-correlation integral computation entail $O(N^3)$ formal computation cost and two lattice summations, and thus benefit from GPU acceleration. For the density evaluation at gridpoints, we assign a gridpoint to each thread, which then iterates through all local pGTO pairs. In every iteration, we construct a bounding sphere for $\vec{R}_1$ in equation \ref{eq:pbc_xc_density_local_0} and \ref{eq:pbc_xc_density_local_1}, loop through the lattice vectors inside the bounding sphere, and compute the summation for the density and its gradient. For the exchange-correlation integral computation, we assign a local primitive shell pair to each thread, which then iterates through all gridpoints. In every iteration, another lattice vector summation is performed over $\vec{R}_1$ in equation \ref{eq:pbc_xc_integral_local_0} and \ref{eq:pbc_xc_integral_local_1}. In the process of iterating through local pGTO pairs and gridpoints, we first group these items into boxes. The boxes are built as cubes that evenly encompasses a unit cell, whose dimensions are between $2\times2\times2$ Bohr$^3$ and $4\times4\times4$ Bohr$^3$. An item is considered inside a box if its center location or any lattice image of its center location falls within the box, and any item is contained in exactly one box. If a thread finds an item in the box too far away from its own gridpoint or local pair center ($|\vec{r}_g - \vec{R}_1 - \vec{P}(\vec{R}_2)|^2 > 50 \ \mathrm{Bohr}^2$ and $p|\vec{r}_g - \vec{R}_1 - \vec{P}(\vec{R}_2)|^2 > 100$), then the whole box is discarded.

The Becke weight evaluation for each gridpoint is also performed on GPU. Each thread will fetch one gridpoint, and compute its Becke weight with respect to its origin atom. Different from all the integral kernels, the Becke weight kernel function contains two nested lattice vector summation loops, and a hardcoded cutoff radius (17 Bohr, as mentioned earlier). Although Becke weights are evaluated only once at the beginning of a single point calculation, for a large system it requires compute time of nearly one SCF iteration, and further acceleration would be helpful.

\subsection{Lattice vector summation}

In most of the integral procedures, we need to find all lattice vectors $\vec{R}$ within a bounding sphere:
\begin{align}
\left| \vec{R} - \vec{R}_{offset} \right| < R_{cutoff}
\label{eq:bounding_sphere}
\end{align}
The center $\vec{R}_{offset}$ and radius $R_{cutoff}$ of the bounding sphere varies with different types of integrals, but they all have this general form.

We find all lattice vectors by first constructing a bounding parallelepiped that is guaranteed to enclose all lattices points that are within the bounding sphere. The bounding parallelepiped is the supercell that is slightly larger than a bounding cube with center $\vec{A} - \vec{B}$ and edge length $2R_{cutoff}$. In order to obtain such a parallelepiped, we define the primitive cell vectors as $\vec{R}_a$, $\vec{R}_b$ and $\vec{R}_c$ and the matrix $\mathbf{R} = \left( \vec{R}_a, \vec{R}_b, \vec{R}_c \right)$. 
The corners of the bounding cube are given by $\vec{P} = (A_x-B_x \pm R_{cutoff}, A_y-B_y \pm R_{cutoff}, A_z-B_z \pm R_{cutoff})$, and we compute the fractional lattice index $\vec{f}= \mathbf{R}^{-1} \vec{P}$ of these corner points.
Two sets of integer lattice indices are defined by taking the ceiling and floor function along each lattice vector:
\begin{align}
\vec{n}_{ceil} = (\lceil f_{1} \rceil, \lceil f_{2} \rceil, \lceil f_{3} \rceil) \label{eq:fractional_index_to_ceil_index} \\
\vec{n}_{floor} = (\lfloor f_{1} \rfloor, \lfloor f_{2} \rfloor, \lfloor f_{3} \rfloor) \label{eq:fractional_index_to_floor_index}
\end{align}
We then gather the minimum and maximum integer lattice indices for all corner points of the bounding cube. The lattice vectors corresponding to these minimum and maximum lattice indices define the bounding parallelepiped. Lastly we iterate through the lattice indices from minimum to maximum along each of the three directions, and check if the corresponding lattice vector is within the bounding sphere. We drop the lattice vectors within the bounding parallelepiped and outside the bounding sphere.

It is usually the case that $\vec{R}_{offset}$ can be in any lattice image. For example, when constructing the crystalline pGTO pairs, $\vec{R}_{offset} = \vec{A} - \vec{B}$, and the two atom center locations $\vec{A}$ and $\vec{B}$ can each be in any lattice image. In order to guarantee numerical stability, we look for the $\vec{R}_{offset} + \vec{R}$ with minimal norm value, and replace $\vec{R}_{offset}$ with $\left(\vec{R}_{offset} + \underset{\vec{R} \in  \mathbb{Z}^3}{\text{arg min}} \left| \vec{R}_{offset} + \vec{R} \right|\right)$ before constructing the bounding cube. The $\vec{R}_{offset} + \vec{R}$ with minimal norm value is obtained in the following way: we obtain the fractional lattice index of $\vec{R}_{offset}$ with $\vec{f}= \mathbf{R}^{-1} \vec{R}_{offset}$. We then find the integer lattice index closest to the fractional lattice index:
\begin{align}
\vec{n}_{round} = (\mathrm{round}(f_{1}), \mathrm{round}(f_{2}), \mathrm{round}(f_{3}))
\label{eq:fractional_index_to_round_index}
\end{align}
and subtract the corresponding lattice vector from $\vec{R}_{offset}$:
\begin{align}
\vec{R}_{offset} + \underset{\vec{R} \in  \mathbb{Z}^3}{\text{arg min}} \left| \vec{R}_{offset} + \vec{R} \right| = \vec{R}_{offset} - \mathbf{R} \vec{n}_{round}
\label{eq:lattice_image_min_norm}
\end{align}
This step is not strictly necessary but makes the program more robust against edge cases.

\subsection{Selection of the range separation parameter parameter}

The Ewald summation procedure involves the use of a range separation parameter $\omega$, which determines the size of the bounding sphere for both the real and reciprocal space lattice vector summation. An optimal choice of $\omega$ is thus key for computational efficiency, although the final results should be effectively independent of $\omega$. If we assume the minimal exponent value of all pGTOs is large ($p >> \omega^2$), and ignore all the prefactor terms (approximating the $\ln$ terms as 1), then the reciprocal space bounding sphere radius is $2\omega$ (in equations \ref{eq:pbc_v1e_reciprocal_cutoff_implementation} and \ref{eq:pbc_eri_reciprocal_cutoff_implementation}). The volume of reciprocal space bounding sphere is thus $\dfrac{32}{3}\pi \omega^3$, and the number of lattice vectors inside the bounding sphere is approximately:
\begin{align}
n_K = \frac{\frac{32}{3}\pi \omega^3}{V_{reciprocal-cell}} = \frac{\frac{32}{3}\pi \omega^3}{\frac{8\pi^3}{V_{cell}}} = \frac{4}{3\pi^2} \omega^3 V_{cell}
\label{eq:n_lattice_estimation_reciprocal}
\end{align}

Similarly, the real space bounding sphere radius is $\frac{1}{\omega}$ (in equations \ref{eq:pbc_v1e_real_cutoff_implementation} and \ref{eq:pbc_eri_real_cutoff_implementation}). Therefore, the volume of real space bounding sphere is $\dfrac{4}{3}\pi \dfrac{1}{\omega^3}$, and the number of lattice vectors inside the bounding sphere is approximately:
\begin{align}
n_R = \frac{4}{3}\pi \frac{1}{\omega^3 V_{cell}}
\label{eq:n_lattice_estimation_real}
\end{align}

\begin{figure}[htb!]
    \begin{minipage}{0.48\linewidth}
        \centering
    \includegraphics[width=\linewidth]{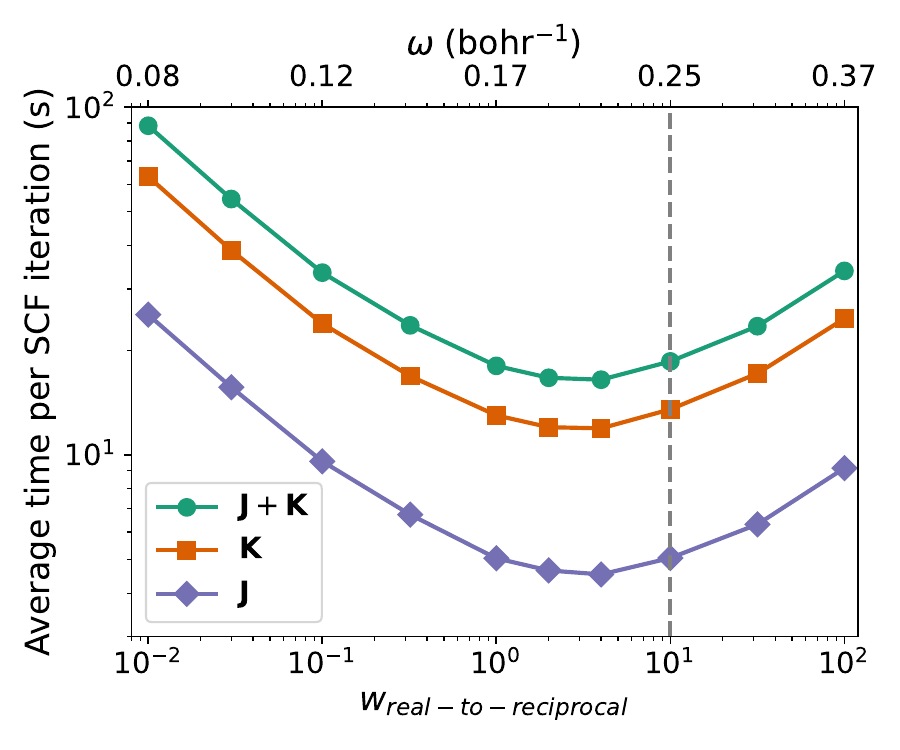}
    \subcaption{Silicon}
    \end{minipage}
        \begin{minipage}{0.48\linewidth}
        \centering
    \includegraphics[width=\linewidth]{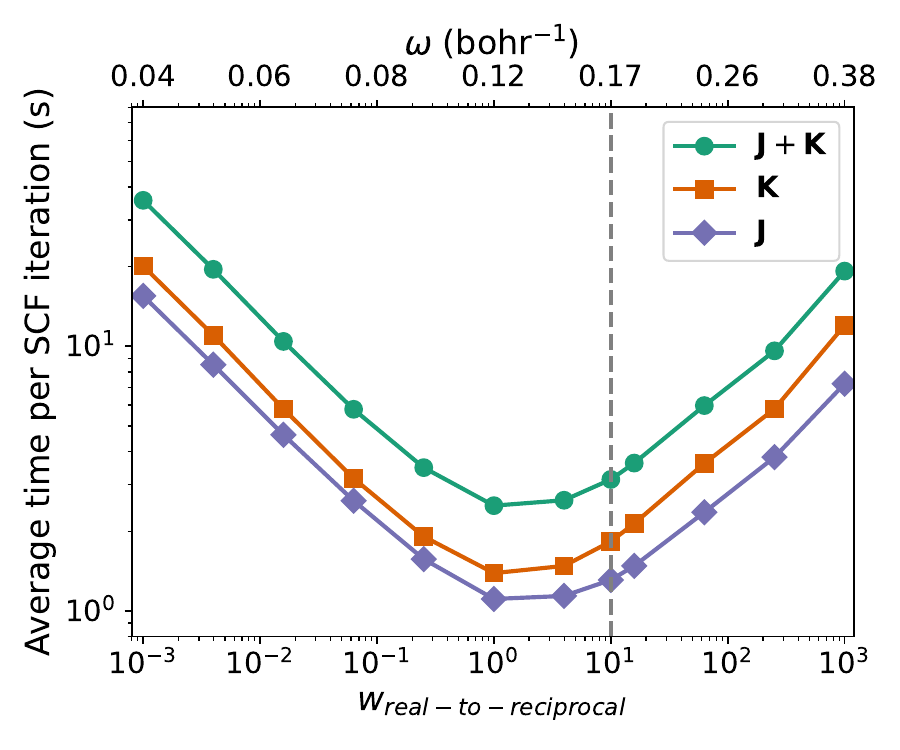}
    \subcaption{Benzene}
    \end{minipage}
\caption{Average time per SCF iteration for $\mathbf{J},\mathbf{K}$ and $\mathbf{J}+\mathbf{K}$ (averaged over all cycles) for HF/def2-SVP all-electron calculations on crystalline silicon and benzene on one NVIDIA A100 GPU, with varying real-to-reciprocal space weights and corresponding $\omega$ values. The chosen $w_{real-to-reciprocal} = 10$ is shown as a dashed line, and is quite close to the optimum for both systems. Note that both axes are on log scales.}
\label{fig:omega_scan}
\end{figure}

Integral evaluations in real space involve Boys function evaluation, which is more expensive than integral evaluation in reciprocal space. 
We therefore define the optimal $\omega$ as the value that minimizes the weighted sum of the number of real and reciprocal space lattice vectors within the respective bounding spheres:
\begin{align}
\omega &= \underset{\omega}{\text{arg min}} (n_K + w_{real-to-reciprocal} n_R) = \left( \frac{w_{real-to-reciprocal} \pi^3}{V_{cell}^2} \right)^{1/6}
\label{eq:omega_implementation}
\end{align}
We set the default weight $w_{real-to-reciprocal} = 10$, which means there will be 10 times more reciprocal space lattice vectors in the corresponding summation than real space vectors. The value of the weight is not rigorously chosen, and a more careful optimization of $w_{real-to-reciprocal}$ will be carried out in future. However, exploratory calculations on crystalline silicon and benzene (shown in Figure \ref{fig:omega_scan}) suggest that the chosen $w_{real-to-reciprocal} = 10$ is likely close to optimal for a broad range of systems, albeit potentially slightly too large. They also indicate that the same $\omega$ should minimize computation time for both $\mathbf{J}$ and $\mathbf{K}$. We also allow the user to adjust the $\omega$ parameter if they so desire. However, an improper choice of $\omega$ will lead to a steep increase in run time, as shown in Fig. \ref{fig:omega_scan}.

\subsection{Precision}
TeraChem supports calculations with single, double, and mixed precision for both molecular and (now) periodic systems. The mixed precision scheme involves evaluating integrals with a density weighted Schwartz upper bound above a threshold (the TeraChem default being $10^{-5}$) in double precision while the rest are evaluated in single precision\cite{terachem_dynamic_precision}. This approach thus provides a balance between computational cost and numerical accuracy. All TeraChem calculations reported in this work therefore use the aforementioned mixed precision scheme for computing ERIs, unless specified otherwise. 

\subsection{Thresholds}

In the theory section, we introduced four cutoff thresholds to truncate the corresponding lattice vector summation: $\delta_{pair-R}$ and $\delta_{pair-Cauchy-Schwarz}$ for truncating the real space lattice vector summation while computing pGTO pairs, $\delta_{Ewald-R}$ for truncating the real space lattice vector summation in an Ewald summation, and $\delta_{Ewald-K}$ for truncating the reciprocal space lattice vector summation in an Ewald summation. We set the default value for all four thresholds to $10^{-14}$, and allow the users to modify them if they so desire.

\begin{figure}[htb!]
    \begin{minipage}{0.48\linewidth}
        \centering
    \includegraphics[width=\linewidth]{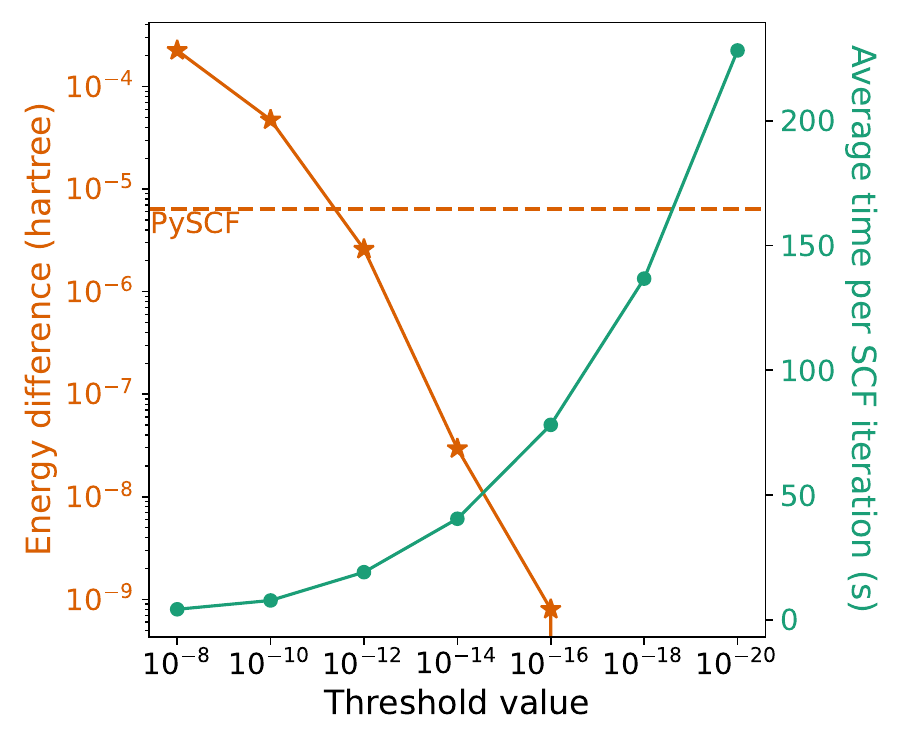}
    \subcaption{Silicon}
    \end{minipage}
        \begin{minipage}{0.48\linewidth}
        \centering
    \includegraphics[width=\linewidth]{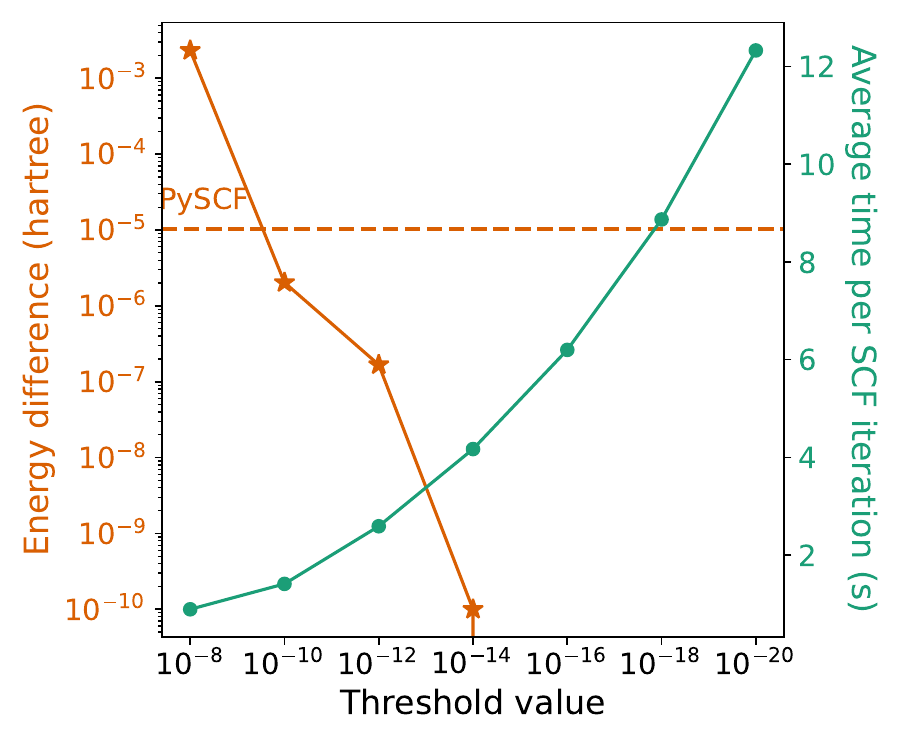}
    \subcaption{Benzene}
    \end{minipage}
\caption{The SCF energy convergence and average SCF iteration time with all-electron HF/def2-SVP for crystalline silicon and benzene on a single NVIDIA A100 GPU, with double precision and varying lattice summation cutoff thresholds. The reference energy is the TeraChem calculation result with all thresholds set to $10^{-20}$. We also show the energy computed by PySCF with Gaussian and plane wave mixed density fitting\cite{pyscf_mixed_density_fitting}  (precision parameter set to $10^{-14}$), as a dashed orange line. Note that the axes for the threshold value and the energy difference are on a log scale, while the axis for time is on a linear scale.}
\label{fig:threshold_double_hf}
\end{figure}

\begin{figure}[htb!]
    \begin{minipage}{0.48\linewidth}
        \centering
    \includegraphics[width=\linewidth]{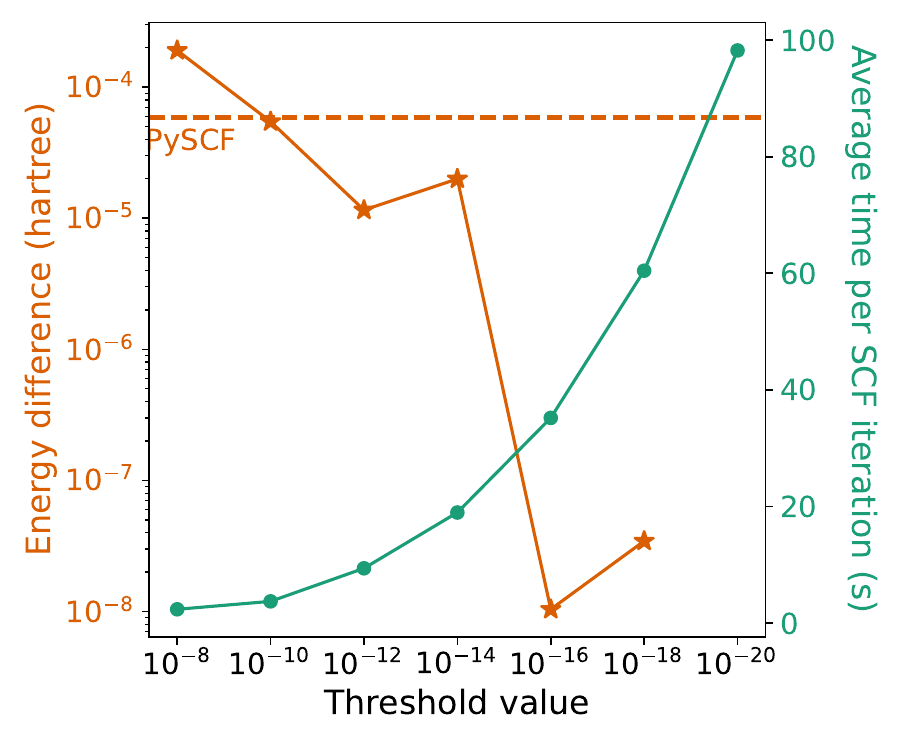}
    \subcaption{Silicon}
    \end{minipage}
        \begin{minipage}{0.48\linewidth}
        \centering
    \includegraphics[width=\linewidth]{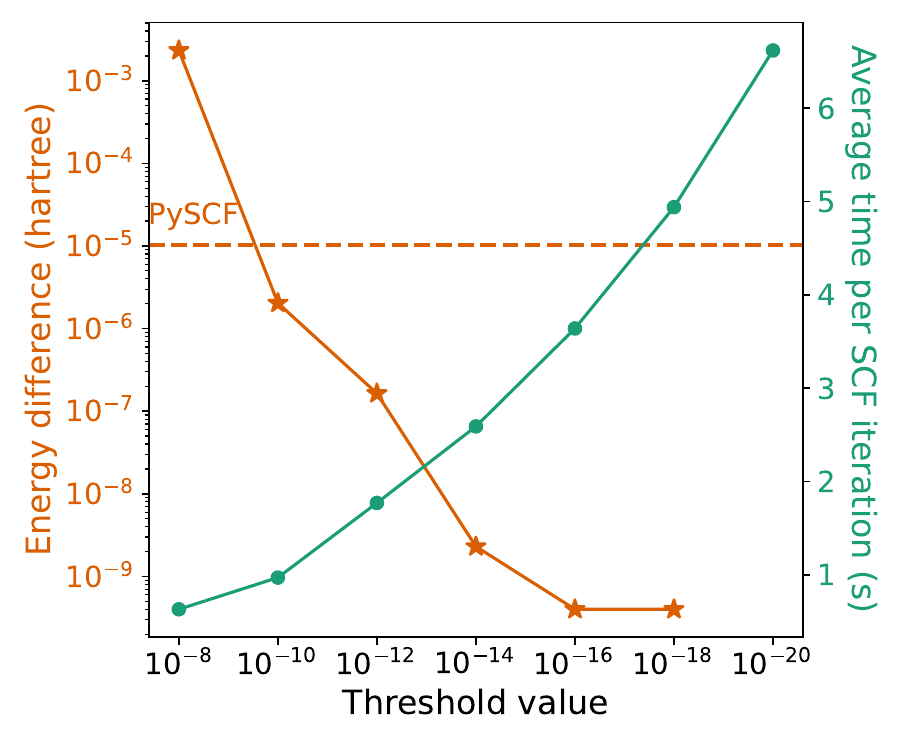}
    \subcaption{Benzene}
    \end{minipage}
\caption{The SCF energy convergence and average SCF iteration time with all-electron HF/def2-SVP for crystalline silicon and benzene on a single NVIDIA A100 GPU, with mixed precision and varying lattice summation cutoff thresholds. The reference energy is the TeraChem calculation result with all thresholds set to $10^{-20}$. We also show the energy computed by PySCF with Gaussian and plane wave mixed density fitting\cite{pyscf_mixed_density_fitting}  (precision parameter set to $10^{-14}$), as a dashed orange line. Note that the axes for the threshold value and the energy difference are on a log scale, while the axis for time is on a linear scale.}
\label{fig:threshold_mixed_hf}
\end{figure}

In Fig. \ref{fig:threshold_double_hf} we show the absolute error and run time as a function of the thresholds (we took $\delta_{pair-R} = \delta_{pair-Cauchy-Schwarz} = \delta_{Ewald-R} = \delta_{Ewald-K}$) for silicon and benzene with HF/def2-SVP\cite{def2_basis} and double precision. It is clear that the HF energy converges with respect to the thresholds at around $\delta = 10^{-14}$. The behavior with mixed precision (shown in Fig. \ref{fig:threshold_mixed_hf}) is a little less straightforward for the case of silicon as the energy convergence is not monotonic. Nonetheless, the mixed precision calculations appear to offer a sufficient level of accuracy for $\delta = 10^{-14}$, for roughly half the computational cost of pure double precision calculations. A tighter threshold than the TeraChem default of $10^{-5}$ (which would ideally provide double precision accuracy when $\delta = 10^{-14}$) for switching between single and double precision may offer better numerical performance but has not been explored in this work.

\section{Performance\label{sec:performance}}

We explore the performance of our code by examining the run
times and scaling vs increasing unit cell sizes for HF and DFT. We also compare the behavior of our periodic code to molecular cluster calculations, and examine parallelism over multiple GPUs.

\begin{figure}[htb!]
\begin{minipage}{0.49\textwidth}
    \includegraphics[width=\linewidth]{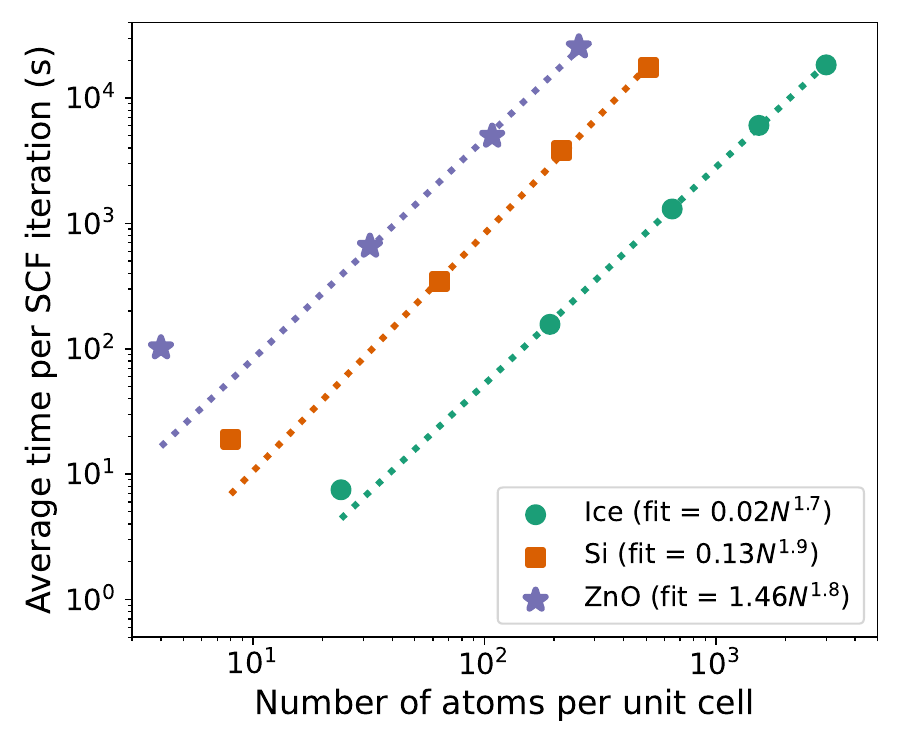}
    \subcaption{HF}
\end{minipage}
\begin{minipage}{0.49\textwidth}
    \includegraphics[width=\linewidth]{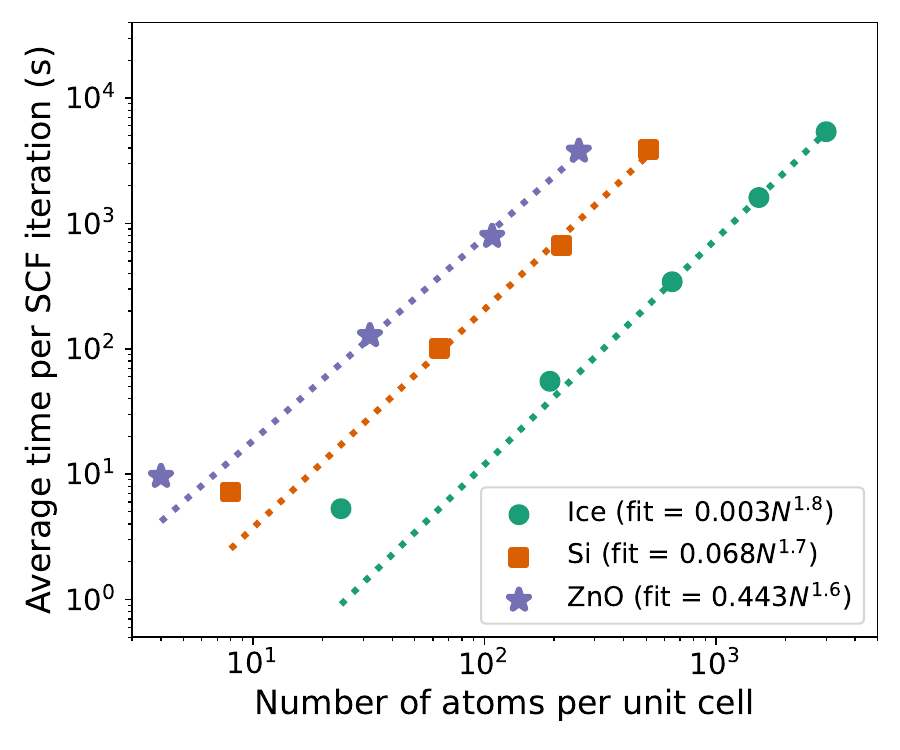}
    \subcaption{PBE}
\end{minipage}
\caption{Run time per SCF iteration (averaged over all cycles) for increasingly large unit cells of ice XI, silicon and wurtzite zinc oxide, on one NVIDIA A100 GPU, with HF (left) and PBE (right). The associated power law fits (via fitting to the three largest unit cell sizes) is also shown as a dotted line and the fit parameters are given in the legend. The calculations are all-electron (i.e. pseudopotential free) and utilize the def2-SVP basis (with the $f$ type function on Zn removed). Note that both axes are on log scales.}
\label{fig:material_performance}
\end{figure}

\subsection{Run times and Scaling}

We explored the performance of our periodic implementation through all-electron def2-SVP basis calculations on increasingly large unit cells of ice XI\cite{ice_xi_structure} (the global minimum polymorph), silicon\cite{si_structure}, and ZnO\cite{zno_structure} (wurtzite structure). The unit cells for all three materials were constructed as supercells of the primitive unit cell by uniform expansion along all three spatial dimensions. The number of atoms/molecules in the supercell therefore grow cubically with the chosen (integer) scaling parameter. The $f$ type function on Zn in the def2-SVP basis was removed, as our present implementation can only utilize $s,p,$ and $d$ type orbitals. 

The average time per SCF cycle for these three materials are shown in Fig. \ref{fig:material_performance} and the numerical values are provided in the supporting information. The HF calculations are bottlenecked by the computation of $\mathbf{K}$, and as a result are $3-6\times$ slower than the corresponding local DFT (PBE\cite{pbe_functional}) calculations, per iteration. The average time per SCF cycle for the hybrid PBE0\cite{pbe0_functional} functional was also quite similar to HF, indicating computation of $\mathbf{K}$ is the most expensive step for hybrid DFT calculations. Nonetheless, the cost of hybrid DFT or HF calculations with our implementation is not prohibitively high compared to local DFT calculations. We also note that the PBE calculations was bottlenecked by computation of $\mathbf{J}$, and not the evaluation of local DFT exchange-correlation or matrix linear algebra. 

The most computationally demanding calculation shown in Fig \ref{fig:material_performance} is HF on the $4\times4\times4$ supercell of ZnO, which has 256 atoms total (5248 basis functions with def2-SVP, as the $f$ function on Zn was removed) and requires $7.1$ h per SCF iteration on one NVIDIA A100 GPU. In contrast, the $5\times5\times5$ supercell of ice XI has 3000 atoms (25000 basis functions with def2-SVP) and only requires $5.1$ h per HF iteration with the same hardware. The greater spatial density of atoms in ionic ZnO (and covalently bonded silicon) contributes to the much larger computational cost of periodic calculations than lower density molecular crystals like ice, as many more basis functions lie within the bounding sphere. We also note that the computational cost reported in Fig. \ref{fig:material_performance} shows  near quadratic scaling with system size, ranging from $O(N^{1.6})$ for ZnO/PBE to $O(N^{1.9})$ for Si/HF. We show that $\mathbf{J}$ and $\mathbf{K}$ individually have near quadratic scaling in the Supporting Information.
Our periodic calculations are thus not constrained by the cubic scaling linear algebra steps for these systems.

\begin{table}[htb!]
\begin{tabular}{r|r|rrr|rrr}
\hline
\multicolumn{2}{c|}{System Size} &\multicolumn{3}{c|}{Periodic Boundaries }                                                & \multicolumn{3}{c}{Molecular Cluster}                                               \\ \hline
Number of  & Number of &HF &PBE & PBE0 & HF & PBE & PBE0 \\ 
molecules & basis functions & & &  &  &  &  \\ \hline 
27     & 3240                                   & 283.6                 & 72.8                    & 289.0                    & 3.1                    & 4.8                     & 6.1                      \\
64     & 7680                                   & 1135.8                 & 302.6                   & 1127.7                   & 19.3                   & 17.3                    & 25.8                     \\
125    & 15000                                   & 3395.4                 & 1051.3                   & 2987.6                   & 76.9                   & 68.8                    & 99.2                     \\
216    & 25920                                   & 7933.9                 & 3043.4                    & 7574.0                    & 299.2                   & 260.1                   &  358.9                    \\\hline
\end{tabular}
\caption{Average time per SCF iteration (in s) for all-electron calculations on benzene supercells of varying sizes, both using periodic boundary conditions and treating the supercell as an isolated molecular cluster. The calculations use the def2-SVP basis and are run on a single NVIDIA A100 GPU.}
\label{tab:timings_benzene}
\end{table}

\begin{table}[htb!]
\begin{tabular}{r|r|rrr|rrr}
\hline
\multicolumn{2}{c|}{System Size} &\multicolumn{3}{c|}{Periodic Boundaries }                                                & \multicolumn{3}{c}{Molecular Cluster}                                               \\ \hline
Number of  & Number of &HF &PBE & PBE0 & HF & PBE & PBE0 \\ 
molecules & basis functions & & &  &  &  &  \\ \hline 
32 & 736 & 310.6  & 92.9  & 286.9   & 0.5  & 0.8 & 1.2  \\
108 & 2484 & 2704.1  & 585.0 & 2613.7 & 8.5  & 6.3 & 12.9 \\
256 & 5888 & 15441.6 &   3027.1     &   11905.2      & 64.1 &  29.3    &   84.8   \\ \hline
\end{tabular}
\caption{Average time per SCF iteration (in s) for all-electron calculations on crystalline LiF unit cells of varying sizes, both using periodic boundary conditions and treating the unit cell as an isolated molecular cluster. The calculations use a modified version of the def2-SVP basis (where the most diffuse s function is removed from Li to avoid linear dependencies) and are run on a single NVIDIA A100 GPU.}
\label{tab:timings_lif}
\end{table}

\subsection{Comparison to Molecular Cluster Calculations}

The differences between periodic boundary calculations of an unit cell and corresponding molecular calculations (i.e. treating the unit cell as a gas phase cluster) are interesting to consider. We analyze this through studies on increasingly large unit cells of benzene\cite{benzene_structure} and LiF\cite{lif_structure}, which are reported in Tables \ref{tab:timings_benzene} and \ref{tab:timings_lif}, respectively. The unit cells for both materials were constructed through uniform expansion of the primitive unit cell along all three spatial dimensions. Similar calculations on covalent crystals like silicon are not possible due to `dangling bonds' on surface atoms. 

Periodic HF/PBE0 calculations are slower by a factor of 20-50 for benzene and by $\sim 200$ for LiF, relative to the corresponding molecular cluster. The slowdown partly originates from the summation over periodic images, which is not applicable for the non-periodic molecular cluster calculation. Another potential contributing factor for the slowdown originates from the branching of the GPU threads. When performing the Ewald summation for electron repulsion integrals on the GPU, each thread will have a different local pGTO pair, and thus a different pair exponent $p$, and therefore likely a different size of bounding sphere. Each GPU thread will need to iterate through different number of lattice vectors, and due to the strong synchronization design of a warp of threads \cite{cuda}, a thread with less lattice vectors has to wait for the thread with more lattice vectors, which damages the parallel efficiency. The algorithm we have proposed effectively saves most amount of memory, because we never expand the lattice vectors in an Ewald summation on the memory space, but it suffers from the branching issue. A more efficient algorithm might be achieved by two separate GPU kernels: one kernel setting up the lattice vectors in the summation and store it on GPU memory, and another kernel computes the Ewald term for each lattice vector in the summation. This aspect will be investigated further in future work.

\begin{figure}[htb!]
    \centering
    \includegraphics[width=0.48\linewidth]{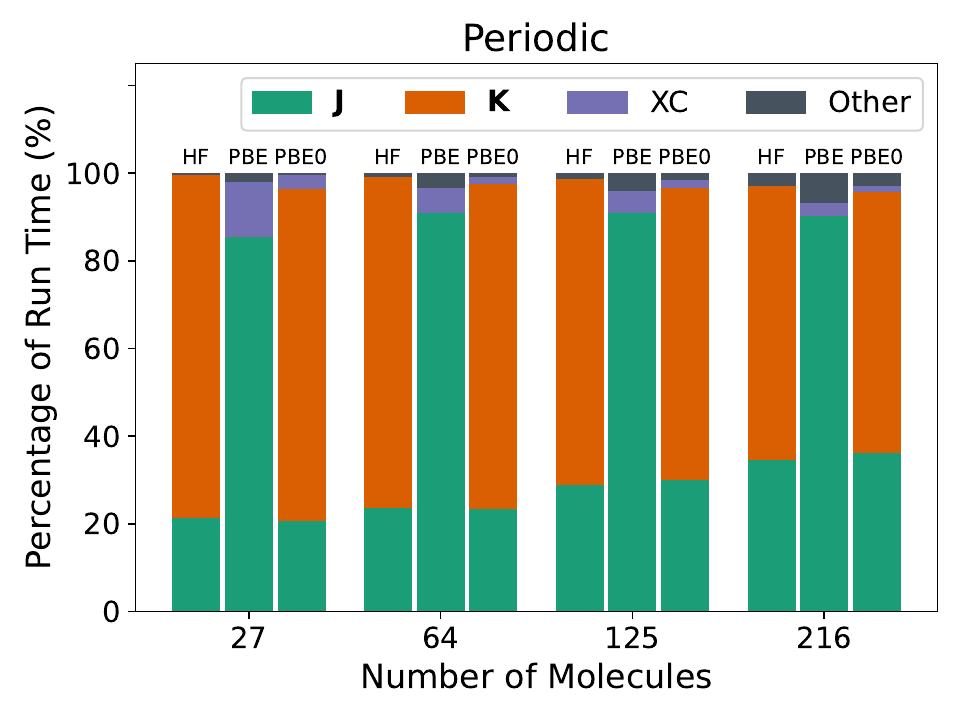}
    \includegraphics[width=0.48\linewidth]{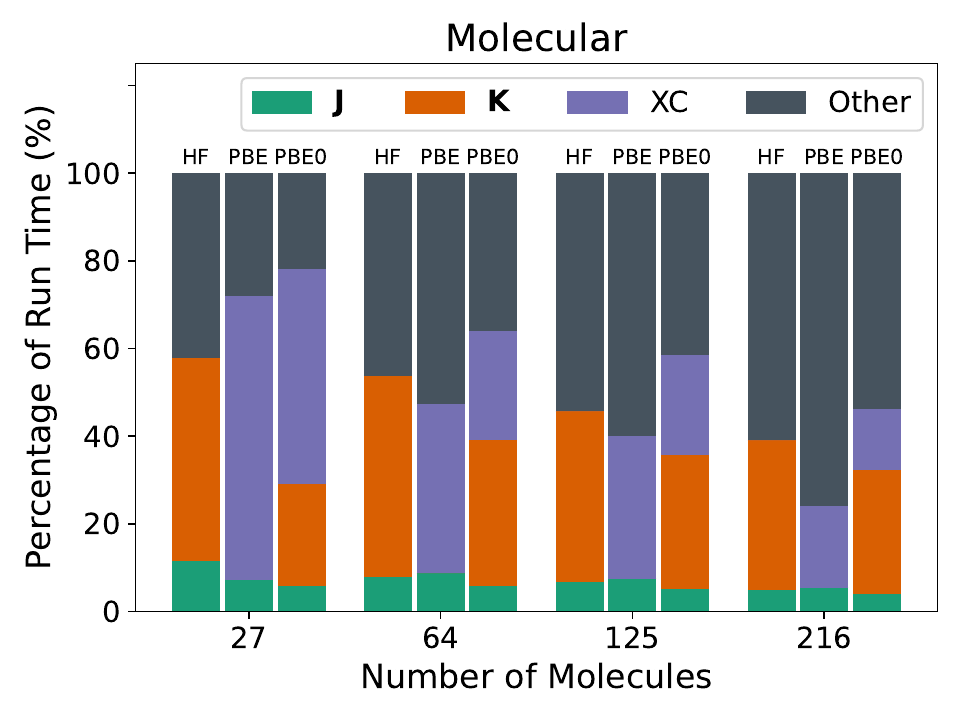}
\caption{SCF iteration run time component distribution (averaged over all cycles) for benzene with different unit cell sizes, using periodic boundary conditions (left) and as isolated molecular clusters (right). The ``Other" bar includes linear algebra operations like Fock matrix diagonalization. The average time per SCF cycle is reported in Table \ref{tab:timings_benzene}.}
\label{fig:benzene_component}
\end{figure}

\begin{figure}[htb!]
    \centering
    \includegraphics[width=0.48\linewidth]{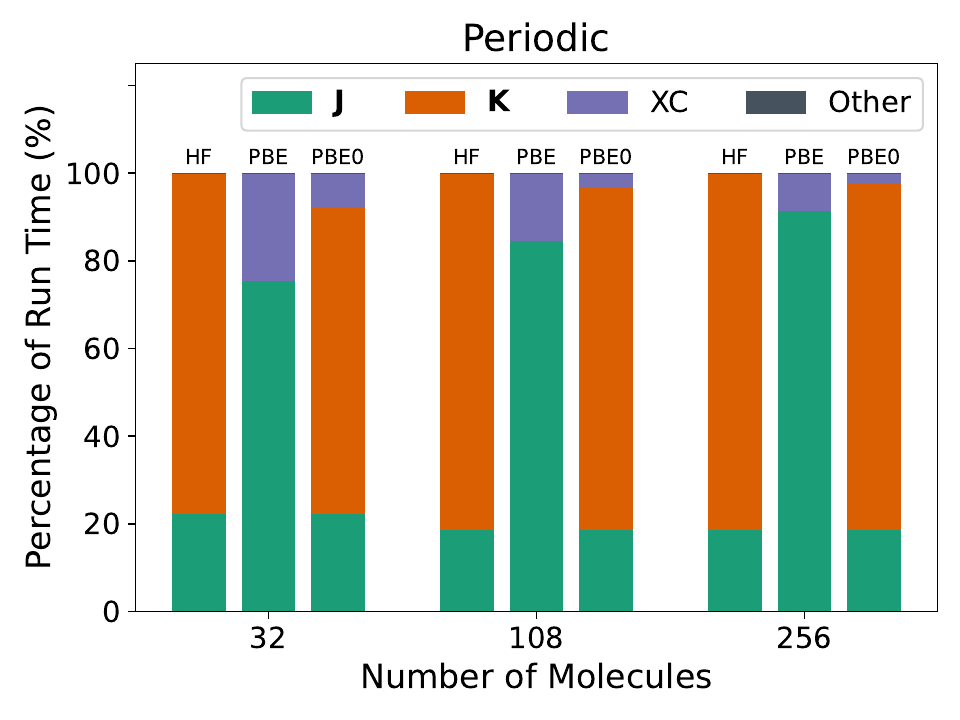}
    \includegraphics[width=0.48\linewidth]{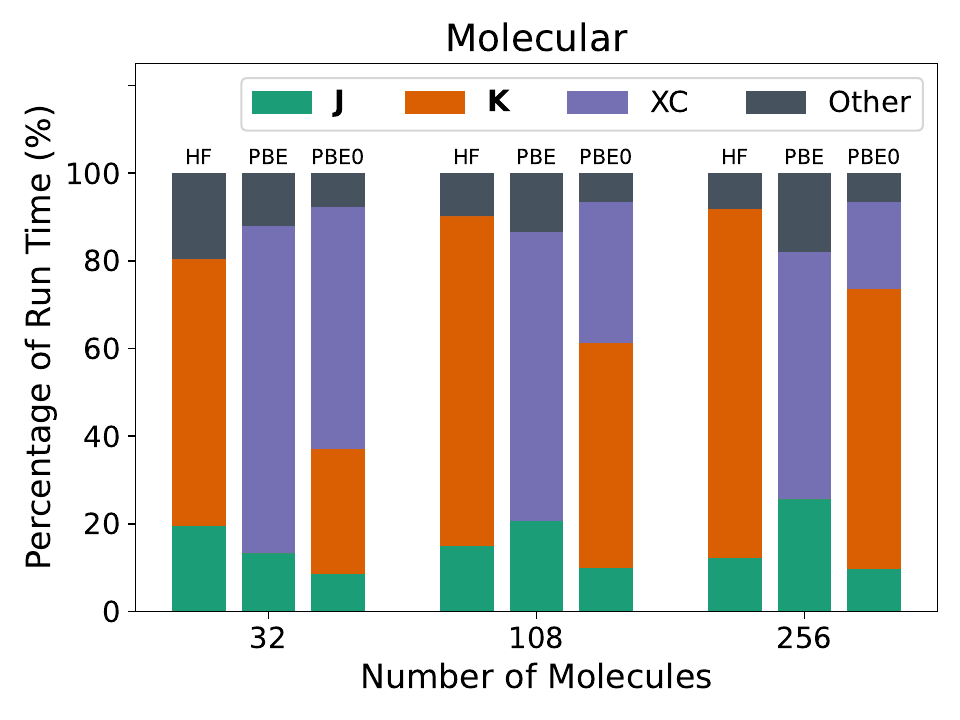}
\caption{SCF iteration run time component distribution (averaged over all cycles) for LiF with different unit cell sizes, using with periodic boundary conditions (left) and as isolated molecular clusters (right). The ``Other" bar includes linear algebra operations like Fock matrix diagonalization. The average time per SCF cycle is reported in Table \ref{tab:timings_lif}.}
\label{fig:lif_component}
\end{figure}

Figs. \ref{fig:benzene_component} and \ref{fig:lif_component} shows how much various components of the full Fock matrix contribute to the total SCF cycle run time for benzene and LiF, respectively. In these periodic calculations, the quadratic scaling $\mathbf{J}$ and $\mathbf{K}$ matrix construction dominates the compute time of SCF iterations. Computation of local exchange-correlation, as well as linear algebra operations like the cubic scaling matrix diagonalization and multiplication operations (included in the ``Other" bar) take a much smaller amount of time in comparison. In contrast, the corresponding molecular cluster calculations are often bottlenecked by computation of local exchange-correlation or linear algebra (especially for benzene, as shown in Fig. \ref{fig:benzene_component}), despite using a very similar procedure for GPU accelerated integral computation\cite{terachem_f}.  This analysis also reveals that the Coulomb matrix construction constitutes a significantly greater fraction of the total run time for periodic HF/PBE0 calculations of benzene as compared to isolated molecular cluster calculations, likely because the classical Coulomb interaction is more long-ranged than HF exchange or local exchange-correlation interaction. However, computation of $\mathbf{K}$ remains the main computational bottleneck for both benzene and LiF (as well as for other calculations shown in the supporting information), and will be the main target for future optimization.

\subsection{Performance over multiple GPUs}
\begin{figure}[htb!]
    \centering
    \begin{minipage}{0.49\textwidth}
    \includegraphics[width=\linewidth]{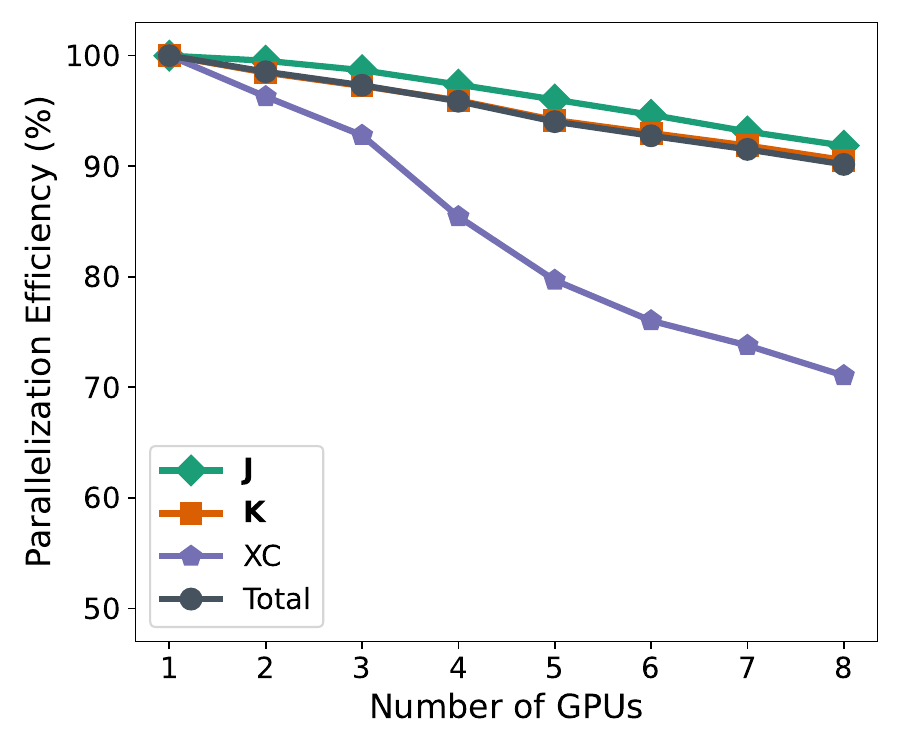}
    \subcaption{Wurtzite ZnO $3\times3\times3$ supercell (108 atoms).}
    \end{minipage}
    \begin{minipage}{0.49\textwidth}
    \includegraphics[width=\linewidth]{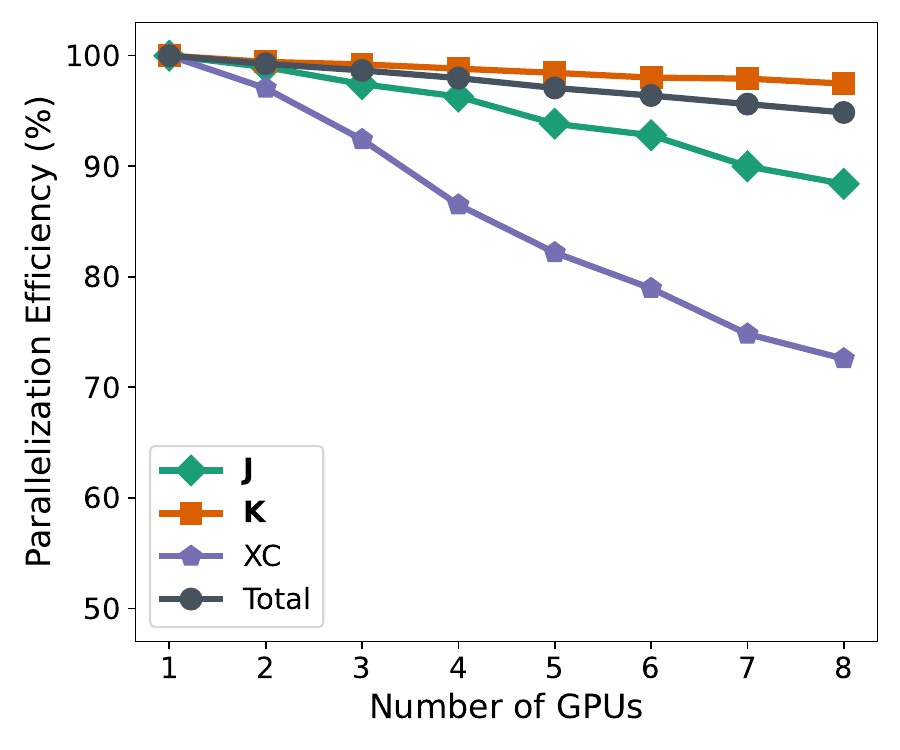}
    \subcaption{Si $3\times3\times3$ supercell (216 atoms).}
    \end{minipage}
    \begin{minipage}{0.49\textwidth}
    \includegraphics[width=\linewidth]{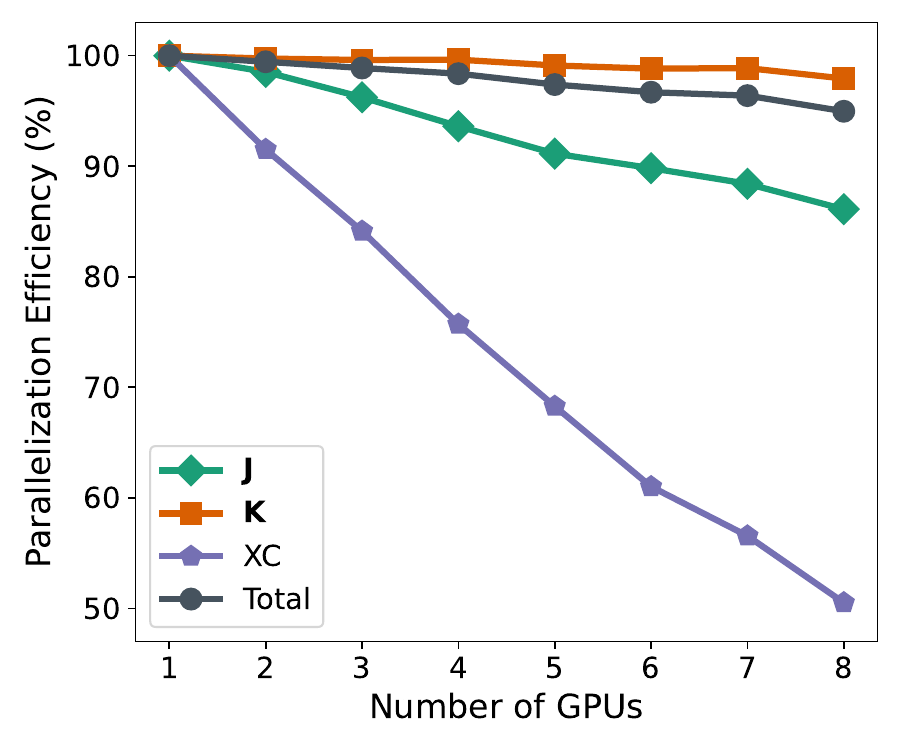}
    \subcaption{MOF-5 primitive unit cell (424 atoms).}
    \end{minipage}
    \begin{minipage}{0.49\textwidth}
    \includegraphics[width=\linewidth]{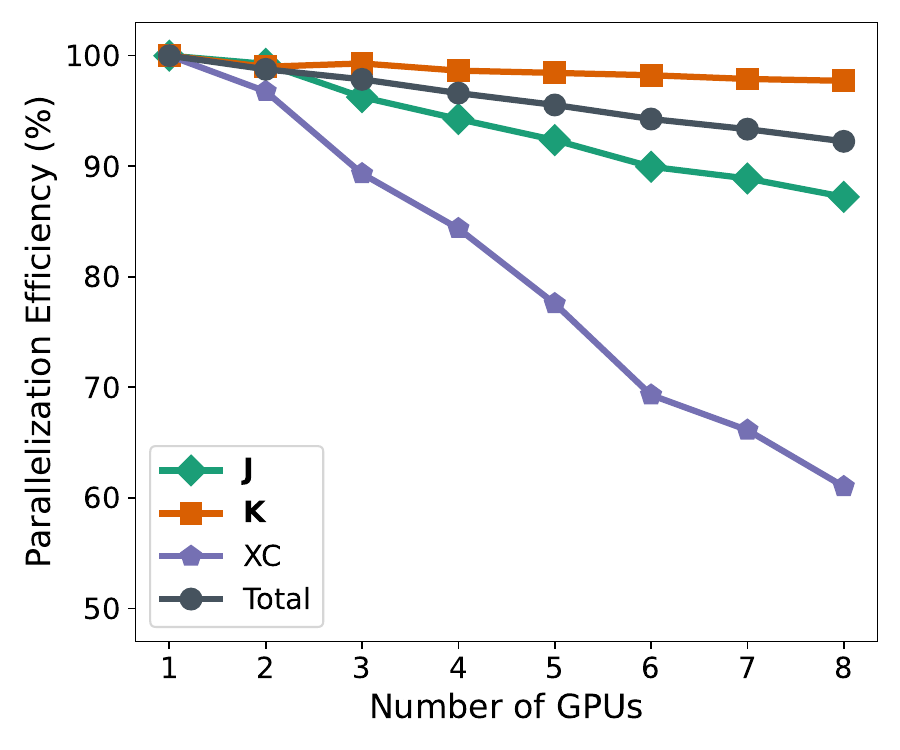}
    \subcaption{Ice XI $4\times4\times4$ supercell (1536 atoms).}
    \end{minipage}
\caption{Parallelization efficiency vs number of NVIDIA A100 GPUs for all-electron PBE0/def2-SVP calculations. The average times per SCF cycle on one GPU are 5071 s (ZnO),  4232 s (Si), 3429 s (MOF-5), and 6652 s (Ice). The parallelization efficiency for n GPUs is defined as $\dfrac{\textrm{run time (n GPUs)}}{\textrm{run time (1 GPU)}}\times \dfrac{100\%}{\textrm{n}}$. The $f$ type function on Zn in the def2-SVP basis was removed.}
\label{fig:parallelization_efficiency}
\end{figure}

We examined how well our implementation parallelizes over multiple NVIDIA A100 GPUs through all-electron PBE0/def2-SVP calculations on four systems: the $3\times3\times3$ supercells of wurtzite ZnO and silicon, the primitive unit cell of MOF-5\cite{mof5_structure} and the $4\times4\times4$ supercell of Ice XI. The resulting parallelization efficiencies for 1-8 NVIDIA A100 GPUs is shown in Fig \ref{fig:parallelization_efficiency}. The total run time parallelizes quite well, with the lowest efficiency being 90\% over 8 GPUS for the ZnO supercell, which still represents a $7.2\times$ speedup over a single GPU. The good parallelization performance is largely a consequence of $\mathbf{K}$ computation being both the most demanding step and very parallelizable in practice. Evaluation of $\mathbf{J}$ is slightly less parallelizable (with the lowest efficiency being 86\% for the MOF-5 unit cell with 8 GPUs) than $\mathbf{K}$, but is still fairly scaleable. The local exchange-correlation computation does not parallelize quite as well, only a attaining 4$\times$ speedup for the MOF-5 unit cell with 8 GPUs (i.e. 50\% parallelization efficiency). However, evaluation of this local potential is too computationally inexpensive to significantly affect total parallelization efficiency, which remains quite high ($>90\%$) over 8 GPUs for all four studied systems.

We could not examine parallelization performance with more than 8 GPUs as that is the maximum number available to us on one compute node. We also note that lower parallelization efficiencies are possible for smaller unit cell sizes, and are indeed observed for the $2\times2\times2$ supercells of silicon and wurtzite ZnO (as reported in the supporting information).   However, the $>90\%$ parallelization efficiency shown in Fig. \ref{fig:parallelization_efficiency} is adequate for good performance on modern HPC systems for sizeable systems like metal organic frameworks. In particular, individual GPU nodes of the NERSC Perlmutter supercomputer have 4 NVIDIA A100 cards each\cite{nersc}. This is a regime where we have $>95\%$ parallelization efficiency in total runtime for all four systems shown in Fig. \ref{fig:parallelization_efficiency}, indicating that our periodic calculations would be readily scaleable on NERSC. Indeed, we obtain virtually identical run times between NERSC and our local cluster for the $3\times3\times3$ supercell of Si, using 1-4 NVIDIA A100 GPUs (as shown in the Supporting Information).  

\subsection{Comparison to CPU based PySCF}

\begin{table}[htb!]
\begin{tabular}{c|c|l|rr|rr}
\hline System                               & Number of                       & Method & \multicolumn{2}{c|}{TeraChem} & \multicolumn{2}{c}{PySCF} \\
                            & atoms                   &  & \multicolumn{2}{c|}{} & \multicolumn{2}{c}{} \\ \hline 
                                    &                                   &        & SCF     & Initialization     & SCF    & Initialization   \\ \hline 
\multirow{2}{*}{Benzene} & \multirow{2}{*}{96} & HF     & 39      & 19                 & 56     & 8224             \\
                                                          &                                                           & PBE    & 13      & 20                 & 34     & 7955             \\
\multirow{2}{*}{Ice XI} & \multirow{2}{*}{192} & HF     & 154     & 27                 & 408    & 21752            \\
                                                          &                                                           & PBE    & 54      & 39                 & 186    & 21933            \\
\multirow{2}{*}{Silicon}      & \multirow{2}{*}{64}      & HF     & 340     & 65                 & 238    & 11136            \\
                                                          &                                                           & PBE    & 100     & 65                 & 88     & 11001   \\\hline         
\end{tabular}
\caption{Comparison between TeraChem and PySCF run times (in s) for $2\times2\times2$ supercells of benzene, ice XI and Si, with the def2-SVP basis. Both the average time per SCF cycle and the ``initialization time" (which represents all time not spent on SCF iterations) are reported, with the latter being substantial for PySCF due to density fitting. All calculations (HF and PBE, TeraChem and PySCF) for a given system were run on the same GPU node in the NERSC Perlmutter supercomputer\cite{nersc}, where each GPU node has 4 NVIDIA A100 GPU cards and an associated AMD EPYC 7763 CPU with 128 cores. The TeraChem calculations utilize 1 NVIDIA A100 GPU, while PySCF calculations use 32 CPU threads.  }
\label{tab:pyscf}
\end{table}

We also compared the performance of our implementation against the CPU based PySCF package\cite{pyscf_recent}, which also utilizes Gaussian type orbitals for periodic systems. PySCF utilizes density fitting for periodic systems\cite{pyscf_mixed_density_fitting,pyscf_range_separated_gaussian_density_fitting}, where the cost of the most time-consuming auxiliary integral construction step scales cubically vs system size. As a result, our quadratic scaling implementation is guaranteed to be better for large unit cells but might be less optimal for smaller systems. We therefore only carried out a few tests on NERSC, where the GPU nodes have an associated AMD EPYC 7763 CPU with 128 cores. Our preliminary testing revealed that the fastest PySCF performance is obtained from range-separated density fitting \cite{pyscf_range_separated_gaussian_density_fitting} with 32 CPU threads. A higher number of threads lead to no appreciable further speedup and at times caused the wall time to increase. 

Table \ref{tab:pyscf} reports the TeraChem and PySCF run times for all-electron HF and PBE calculations on $2\times2\times2$ supercells of benzene, ice XI and silicon, with the def2-SVP basis. The PySCF run times are dominated by the density fitting auxiliary integral construction step, and not by the SCF iterations themselves. In contrast, initialization constitutes a small fraction of the total cost of TeraChem calculations.  For a $2\times2\times 2$ supercell of benzene (8 molecules), the average time per HF/def2-SVP iteration with TeraChem on one NVIDIA A100 GPU is 39 s. For PySCF with 32 threads of a AMD EPYC 7763, initialization alone takes 8224 s due to density fitting, while subsequent SCF steps take 56 s on average. As a result, the total TeraChem run time of 317 s is $\sim$ 25 times smaller than the 8671 s wall time needed for PySCF, and the relative speedup is only expected to increase for larger systems due to difference in scaling behavior. Similar behavior is observed for PBE/def2-SVP, where TeraChem requires 13 s per SCF iteration, while PySCF takes 34 s per iteration following 7955 s for initializiation.
In this regard, we note that we used the default grid for PySCF and the  number of electrons per unit cell estimated via integrating the density over the grid utilized for local exchange correlation is 336.0002 for both TeraChem and PySCF (vs 336 in reality), indicating that PySCF is not being unfairly burdened with a too large grid.

We also observe that the average time for SCF iterations is significantly larger with PySCF than TeraChem for the $2\times2\times 2$ supercell of ice XI (64 molecules), on top of very large initialization times exceeding $21000$ s. For the smaller, but more dense $2\times2\times 2$ supercell of silicon, the individual PySCF SCF iterations using 32 CPU threads are faster than TeraChem with 1 NVIDIA A100 GPU, for both HF and PBE. However, the large ($\sim 11000$ s) initialization time leads to a longer total wall time for the PySCF calculations, in comparison to TeraChem.  
Work is currently underway to incorporate density fitting into TeraChem, which would permit a more `apples to apples' comparison in future. 

\section{Conclusion\label{ref:conclusion}}

We have presented a GPU-accelerated implementation of $\Gamma$-point Hamiltonian integral evaluation with Gaussian basis sets in the TeraChem software package, which utilizes Ewald summation for computation of electrostatic interactions. We have designed truncation schemes for speeding up summations over lattice vectors in both the real and reciprocal space, and demonstrate how adjustment of truncation thresholds or the range-separation parameter for computation of electrostatic integrals affect performance. Our code is capable of computing HF exchange efficiently, permitting HF and hybrid DFT periodic calculations for large unit cells with hundreds or thousands of atoms at an affordable computational cost. The average time per SCF iteration for our implementation grows quadratically with system size, similar to TeraChem's performance for molecular calculations. Our code furthermore parallelizes well over 8 or fewer GPUs, with efficiencies exceeding 90\% ( $\ge 95\%$ for 4 GPUs or less).

We nonetheless note that there remains considerable scope for realizing further improvements in performance, particularly through the development of more efficient GPU kernels for HF exchange and Coulomb matrix construction. We also currently do not perform particle mesh Ewald \cite{particle_mesh_ewald, particle_mesh_ewald_force, particle_mesh_ewald_gpu_1, particle_mesh_ewald_gpu_2} or other Fast Fourier Transform (FFT) based approaches for reciprocal space summation, and instead directly sum over reciprocal space vectors with different bounding spheres for every relevant electrostatic integral. The availability of efficient GPU-accelerated FFT implementations \cite{cufft} however indicate that the compute cost for reciprocal space summation can be further reduced. Developmental efforts along these directions will be pursued in future. We are also exploring the use of density fitting for smaller unit cells.

Potential future improvements aside, our current implementation of periodic Hamiltonian integrals nonetheless enables efficient, large-scale calculations on solid state systems with the GPU resources available on modern supercomputing clusters. The implementation of analytical forces to enable \textit{ab initio} dynamics simulations is presently ongoing.
This work therefore paves the way for many potential applications utilizing large unit-cell hybrid DFT calculations, including better modeling of defects\cite{point_defect_review}, discovery of heterogenous catalysis based pathways\cite{heterogenous_cat_review_2,heterogenous_cat_review}, and simulation of electrochemical processes like corrosion\cite{corrosion}. We are also exploring the use of correlated wavefunction methods like coupled cluster singles and doubles (CCSD)\cite{terachem_ccsd_gpu} and complete active space self-consistent field\cite{terachem_casscf} to simulate the ground and excited states of extended materials, with the objective of performing noadiabatic molecular dynamics simulations.

\section*{Acknowledgments} 
This research was financially supported by the Office of Naval Research (N00014-21-1-2151). D.H. is a Stanford Science Fellow. P.A.U. acknowledges support by the National Science Foundation MPS-Ascend Postdoctoral Research Fellowship, under Grant No. 2213324. This work used computational resources of the National Energy Research Scientific Computing Center (NERSC), a U.S. Department of Energy Office of Science User Facility located at Lawrence Berkeley National Laboratory, operated under Contract No. DE-AC02-05CH11231 using NERSC award BES-ERCAP0028744.

\section*{Supporting Information}
\noindent PDF: Kinetic energy integral formulas, Scaling for $\mathbf{J}$ and $\mathbf{K}$ construction, Comparison between our local cluster and NERSC. 
\newline 
XLXS: Raw timings for all the systems reported. 
\newline ZIP: TeraChem and PySCF output files, unit cell geometries, analysis scripts. 
\bibliography{references}

\providecommand{\latin}[1]{#1}
\makeatletter
\providecommand{\doi}
  {\begingroup\let\do\@makeother\dospecials
  \catcode`\{=1 \catcode`\}=2 \doi@aux}
\providecommand{\doi@aux}[1]{\endgroup\texttt{#1}}
\makeatother
\providecommand*\mcitethebibliography{\thebibliography}
\csname @ifundefined\endcsname{endmcitethebibliography}  {\let\endmcitethebibliography\endthebibliography}{}
\begin{mcitethebibliography}{137}
\providecommand*\natexlab[1]{#1}
\providecommand*\mciteSetBstSublistMode[1]{}
\providecommand*\mciteSetBstMaxWidthForm[2]{}
\providecommand*\mciteBstWouldAddEndPuncttrue
  {\def\EndOfBibitem{\unskip.}}
\providecommand*\mciteBstWouldAddEndPunctfalse
  {\let\EndOfBibitem\relax}
\providecommand*\mciteSetBstMidEndSepPunct[3]{}
\providecommand*\mciteSetBstSublistLabelBeginEnd[3]{}
\providecommand*\EndOfBibitem{}
\mciteSetBstSublistMode{f}
\mciteSetBstMaxWidthForm{subitem}{(\alph{mcitesubitemcount})}
\mciteSetBstSublistLabelBeginEnd
  {\mcitemaxwidthsubitemform\space}
  {\relax}
  {\relax}

\bibitem[Kohn and Sham(1965)Kohn, and Sham]{ksdft}
Kohn,~W.; Sham,~L.~J. {Self-consistent equations including exchange and correlation effects}. \emph{Phys. Rev.} \textbf{1965}, \emph{140}, A1133--A1138\relax
\mciteBstWouldAddEndPuncttrue
\mciteSetBstMidEndSepPunct{\mcitedefaultmidpunct}
{\mcitedefaultendpunct}{\mcitedefaultseppunct}\relax
\EndOfBibitem
\bibitem[Hasnip \latin{et~al.}(2014)Hasnip, Refson, Probert, Yates, Clark, and Pickard]{review_dft_solid}
Hasnip,~P.~J.; Refson,~K.; Probert,~M.~I.; Yates,~J.~R.; Clark,~S.~J.; Pickard,~C.~J. Density functional theory in the solid state. \emph{Philosophical Transactions of the Royal Society A: Mathematical, Physical and Engineering Sciences} \textbf{2014}, \emph{372}, 20130270\relax
\mciteBstWouldAddEndPuncttrue
\mciteSetBstMidEndSepPunct{\mcitedefaultmidpunct}
{\mcitedefaultendpunct}{\mcitedefaultseppunct}\relax
\EndOfBibitem
\bibitem[Jain \latin{et~al.}(2016)Jain, Shin, and Persson]{dft_energy_review}
Jain,~A.; Shin,~Y.; Persson,~K.~A. Computational predictions of energy materials using density functional theory. \emph{Nature Reviews Materials} \textbf{2016}, \emph{1}, 1--13\relax
\mciteBstWouldAddEndPuncttrue
\mciteSetBstMidEndSepPunct{\mcitedefaultmidpunct}
{\mcitedefaultendpunct}{\mcitedefaultseppunct}\relax
\EndOfBibitem
\bibitem[Perdew \latin{et~al.}(1996)Perdew, Burke, and Ernzerhof]{pbe_functional}
Perdew,~J.~P.; Burke,~K.; Ernzerhof,~M. {Generalized gradient approximation made simple}. \emph{Phys. Rev. Lett.} \textbf{1996}, \emph{77}, 3865--3868\relax
\mciteBstWouldAddEndPuncttrue
\mciteSetBstMidEndSepPunct{\mcitedefaultmidpunct}
{\mcitedefaultendpunct}{\mcitedefaultseppunct}\relax
\EndOfBibitem
\bibitem[Sun \latin{et~al.}(2015)Sun, Ruzsinszky, and Perdew]{SCAN_functional}
Sun,~J.; Ruzsinszky,~A.; Perdew,~J.~P. {Strongly Constrained and Appropriately Normed Semilocal Density Functional}. \emph{Phys. Rev. Lett.} \textbf{2015}, \emph{115}, 036402\relax
\mciteBstWouldAddEndPuncttrue
\mciteSetBstMidEndSepPunct{\mcitedefaultmidpunct}
{\mcitedefaultendpunct}{\mcitedefaultseppunct}\relax
\EndOfBibitem
\bibitem[Becke(1993)]{becke_hybrid}
Becke,~A.~D. {Density-functional thermochemistry. III. The role of exact exchange}. \emph{J. Chem. Phys.} \textbf{1993}, \emph{98}, 5648--5652\relax
\mciteBstWouldAddEndPuncttrue
\mciteSetBstMidEndSepPunct{\mcitedefaultmidpunct}
{\mcitedefaultendpunct}{\mcitedefaultseppunct}\relax
\EndOfBibitem
\bibitem[Adamo and Barone(1999)Adamo, and Barone]{pbe0_functional}
Adamo,~C.; Barone,~V. Toward reliable density functional methods without adjustable parameters: The PBE0 model. \emph{The Journal of chemical physics} \textbf{1999}, \emph{110}, 6158--6170\relax
\mciteBstWouldAddEndPuncttrue
\mciteSetBstMidEndSepPunct{\mcitedefaultmidpunct}
{\mcitedefaultendpunct}{\mcitedefaultseppunct}\relax
\EndOfBibitem
\bibitem[Krukau \latin{et~al.}(2006)Krukau, Vydrov, Izmaylov, and Scuseria]{hse06_functional}
Krukau,~A.~V.; Vydrov,~O.~A.; Izmaylov,~A.~F.; Scuseria,~G.~E. Influence of the exchange screening parameter on the performance of screened hybrid functionals. \emph{The Journal of Chemical Physics} \textbf{2006}, \emph{125}\relax
\mciteBstWouldAddEndPuncttrue
\mciteSetBstMidEndSepPunct{\mcitedefaultmidpunct}
{\mcitedefaultendpunct}{\mcitedefaultseppunct}\relax
\EndOfBibitem
\bibitem[Hirata \latin{et~al.}(2001)Hirata, Grabowski, Tobita, and Bartlett]{cc_mp2_1d}
Hirata,~S.; Grabowski,~I.; Tobita,~M.; Bartlett,~R.~J. Highly accurate treatment of electron correlation in polymers: coupled-cluster and many-body perturbation theories. \emph{Chemical Physics Letters} \textbf{2001}, \emph{345}, 475--480\relax
\mciteBstWouldAddEndPuncttrue
\mciteSetBstMidEndSepPunct{\mcitedefaultmidpunct}
{\mcitedefaultendpunct}{\mcitedefaultseppunct}\relax
\EndOfBibitem
\bibitem[Hirata \latin{et~al.}(2004)Hirata, Podeszwa, Tobita, and Bartlett]{ccsd_pbc_1}
Hirata,~S.; Podeszwa,~R.; Tobita,~M.; Bartlett,~R.~J. Coupled-cluster singles and doubles for extended systems. \emph{The Journal of Chemical Physics} \textbf{2004}, \emph{120}, 2581--2592\relax
\mciteBstWouldAddEndPuncttrue
\mciteSetBstMidEndSepPunct{\mcitedefaultmidpunct}
{\mcitedefaultendpunct}{\mcitedefaultseppunct}\relax
\EndOfBibitem
\bibitem[McClain \latin{et~al.}(2017)McClain, Sun, Chan, and Berkelbach]{ccsd_pbc_2}
McClain,~J.; Sun,~Q.; Chan,~G. K.-L.; Berkelbach,~T.~C. Gaussian-based coupled-cluster theory for the ground-state and band structure of solids. \emph{Journal of Chemical Theory and Computation} \textbf{2017}, \emph{13}, 1209--1218\relax
\mciteBstWouldAddEndPuncttrue
\mciteSetBstMidEndSepPunct{\mcitedefaultmidpunct}
{\mcitedefaultendpunct}{\mcitedefaultseppunct}\relax
\EndOfBibitem
\bibitem[Katagiri(2005)]{eom_ccsd_pbc}
Katagiri,~H. Equation-of-motion coupled-cluster study on exciton states of polyethylene with periodic boundary condition. \emph{The Journal of Chemical Physics} \textbf{2005}, \emph{122}\relax
\mciteBstWouldAddEndPuncttrue
\mciteSetBstMidEndSepPunct{\mcitedefaultmidpunct}
{\mcitedefaultendpunct}{\mcitedefaultseppunct}\relax
\EndOfBibitem
\bibitem[Neufeld \latin{et~al.}(2022)Neufeld, Ye, and Berkelbach]{cc_metal}
Neufeld,~V.~A.; Ye,~H.-Z.; Berkelbach,~T.~C. Ground-state properties of metallic solids from ab initio coupled-cluster theory. \emph{The Journal of Physical Chemistry Letters} \textbf{2022}, \emph{13}, 7497--7503\relax
\mciteBstWouldAddEndPuncttrue
\mciteSetBstMidEndSepPunct{\mcitedefaultmidpunct}
{\mcitedefaultendpunct}{\mcitedefaultseppunct}\relax
\EndOfBibitem
\bibitem[Ye and Berkelbach(2024)Ye, and Berkelbach]{local_periodic_cc}
Ye,~H.-Z.; Berkelbach,~T.~C. Periodic Local Coupled-Cluster Theory for Insulators and Metals. \emph{J. Chem. Theory Comput.} \textbf{2024}, \relax
\mciteBstWouldAddEndPunctfalse
\mciteSetBstMidEndSepPunct{\mcitedefaultmidpunct}
{}{\mcitedefaultseppunct}\relax
\EndOfBibitem
\bibitem[Bloch(1929)]{bloch_original}
Bloch,~F. {\"U}ber die quantenmechanik der elektronen in kristallgittern. \emph{Zeitschrift f{\"u}r physik} \textbf{1929}, \emph{52}, 555--600\relax
\mciteBstWouldAddEndPuncttrue
\mciteSetBstMidEndSepPunct{\mcitedefaultmidpunct}
{\mcitedefaultendpunct}{\mcitedefaultseppunct}\relax
\EndOfBibitem
\bibitem[Bylaska(2017)]{plane_wave_dft}
Bylaska,~E.~J. Plane-wave DFT methods for chemistry. In \emph{Annual Reports in Computational Chemistry}; Elsevier, 2017; Vol.~13; pp 185--228\relax
\mciteBstWouldAddEndPuncttrue
\mciteSetBstMidEndSepPunct{\mcitedefaultmidpunct}
{\mcitedefaultendpunct}{\mcitedefaultseppunct}\relax
\EndOfBibitem
\bibitem[Giannozzi \latin{et~al.}(2009)Giannozzi, Baroni, Bonini, Calandra, Car, Cavazzoni, Ceresoli, Chiarotti, Cococcioni, Dabo, \latin{et~al.} others]{quantum_espresso_original}
Giannozzi,~P.; Baroni,~S.; Bonini,~N.; Calandra,~M.; Car,~R.; Cavazzoni,~C.; Ceresoli,~D.; Chiarotti,~G.~L.; Cococcioni,~M.; Dabo,~I. \latin{et~al.}  QUANTUM ESPRESSO: a modular and open-source software project for quantum simulations of materials. \emph{Journal of Physics: Condensed Matter} \textbf{2009}, \emph{21}, 395502\relax
\mciteBstWouldAddEndPuncttrue
\mciteSetBstMidEndSepPunct{\mcitedefaultmidpunct}
{\mcitedefaultendpunct}{\mcitedefaultseppunct}\relax
\EndOfBibitem
\bibitem[Giannozzi \latin{et~al.}(2017)Giannozzi, Andreussi, Brumme, Bunau, Nardelli, Calandra, Car, Cavazzoni, Ceresoli, Cococcioni, \latin{et~al.} others]{quantum_espresso_recent}
Giannozzi,~P.; Andreussi,~O.; Brumme,~T.; Bunau,~O.; Nardelli,~M.~B.; Calandra,~M.; Car,~R.; Cavazzoni,~C.; Ceresoli,~D.; Cococcioni,~M. \latin{et~al.}  Advanced capabilities for materials modelling with Quantum ESPRESSO. \emph{Journal of Physics: Condensed Matter} \textbf{2017}, \emph{29}, 465901\relax
\mciteBstWouldAddEndPuncttrue
\mciteSetBstMidEndSepPunct{\mcitedefaultmidpunct}
{\mcitedefaultendpunct}{\mcitedefaultseppunct}\relax
\EndOfBibitem
\bibitem[Carnimeo \latin{et~al.}(2019)Carnimeo, Baroni, and Giannozzi]{quantum_espresso_plane_wave_exchange}
Carnimeo,~I.; Baroni,~S.; Giannozzi,~P. Fast hybrid density-functional computations using plane-wave basis sets. \emph{Electronic Structure} \textbf{2019}, \emph{1}, 015009\relax
\mciteBstWouldAddEndPuncttrue
\mciteSetBstMidEndSepPunct{\mcitedefaultmidpunct}
{\mcitedefaultendpunct}{\mcitedefaultseppunct}\relax
\EndOfBibitem
\bibitem[Kresse and Hafner(1993)Kresse, and Hafner]{vasp_1}
Kresse,~G.; Hafner,~J. Ab initio molecular dynamics for liquid metals. \emph{Physical Review B} \textbf{1993}, \emph{47}, 558\relax
\mciteBstWouldAddEndPuncttrue
\mciteSetBstMidEndSepPunct{\mcitedefaultmidpunct}
{\mcitedefaultendpunct}{\mcitedefaultseppunct}\relax
\EndOfBibitem
\bibitem[Kresse and Furthm{\"u}ller(1996)Kresse, and Furthm{\"u}ller]{vasp_2}
Kresse,~G.; Furthm{\"u}ller,~J. Efficiency of ab-initio total energy calculations for metals and semiconductors using a plane-wave basis set. \emph{Computational Materials Science} \textbf{1996}, \emph{6}, 15--50\relax
\mciteBstWouldAddEndPuncttrue
\mciteSetBstMidEndSepPunct{\mcitedefaultmidpunct}
{\mcitedefaultendpunct}{\mcitedefaultseppunct}\relax
\EndOfBibitem
\bibitem[Kresse and Furthm{\"u}ller(1996)Kresse, and Furthm{\"u}ller]{vasp_3}
Kresse,~G.; Furthm{\"u}ller,~J. Efficient iterative schemes for ab initio total-energy calculations using a plane-wave basis set. \emph{Physical Review B} \textbf{1996}, \emph{54}, 11169\relax
\mciteBstWouldAddEndPuncttrue
\mciteSetBstMidEndSepPunct{\mcitedefaultmidpunct}
{\mcitedefaultendpunct}{\mcitedefaultseppunct}\relax
\EndOfBibitem
\bibitem[Hafner(2008)]{vasp_recent}
Hafner,~J. Ab-initio simulations of materials using VASP: Density-functional theory and beyond. \emph{Journal of Computational Chemistry} \textbf{2008}, \emph{29}, 2044--2078\relax
\mciteBstWouldAddEndPuncttrue
\mciteSetBstMidEndSepPunct{\mcitedefaultmidpunct}
{\mcitedefaultendpunct}{\mcitedefaultseppunct}\relax
\EndOfBibitem
\bibitem[Gonze \latin{et~al.}(2002)Gonze, Beuken, Caracas, Detraux, Fuchs, Rignanese, Sindic, Verstraete, Zerah, Jollet, \latin{et~al.} others]{abinit_original}
Gonze,~X.; Beuken,~J.-M.; Caracas,~R.; Detraux,~F.; Fuchs,~M.; Rignanese,~G.-M.; Sindic,~L.; Verstraete,~M.; Zerah,~G.; Jollet,~F. \latin{et~al.}  First-principles computation of material properties: the ABINIT software project. \emph{Computational Materials Science} \textbf{2002}, \emph{25}, 478--492\relax
\mciteBstWouldAddEndPuncttrue
\mciteSetBstMidEndSepPunct{\mcitedefaultmidpunct}
{\mcitedefaultendpunct}{\mcitedefaultseppunct}\relax
\EndOfBibitem
\bibitem[Gonze \latin{et~al.}(2020)Gonze, Amadon, Antonius, Arnardi, Baguet, Beuken, Bieder, Bottin, Bouchet, Bousquet, \latin{et~al.} others]{abinit_recent}
Gonze,~X.; Amadon,~B.; Antonius,~G.; Arnardi,~F.; Baguet,~L.; Beuken,~J.-M.; Bieder,~J.; Bottin,~F.; Bouchet,~J.; Bousquet,~E. \latin{et~al.}  The ABINIT project: Impact, environment and recent developments. \emph{Computer Physics Communications} \textbf{2020}, \emph{248}, 107042\relax
\mciteBstWouldAddEndPuncttrue
\mciteSetBstMidEndSepPunct{\mcitedefaultmidpunct}
{\mcitedefaultendpunct}{\mcitedefaultseppunct}\relax
\EndOfBibitem
\bibitem[Enkovaara \latin{et~al.}(2010)Enkovaara, Rostgaard, Mortensen, Chen, Du{\l}ak, Ferrighi, Gavnholt, Glinsvad, Haikola, Hansen, \latin{et~al.} others]{gpaw_2010}
Enkovaara,~J.; Rostgaard,~C.; Mortensen,~J.~J.; Chen,~J.; Du{\l}ak,~M.; Ferrighi,~L.; Gavnholt,~J.; Glinsvad,~C.; Haikola,~V.; Hansen,~H. \latin{et~al.}  Electronic structure calculations with GPAW: a real-space implementation of the projector augmented-wave method. \emph{Journal of Physics: Condensed Matter} \textbf{2010}, \emph{22}, 253202\relax
\mciteBstWouldAddEndPuncttrue
\mciteSetBstMidEndSepPunct{\mcitedefaultmidpunct}
{\mcitedefaultendpunct}{\mcitedefaultseppunct}\relax
\EndOfBibitem
\bibitem[Mortensen \latin{et~al.}(2024)Mortensen, Larsen, Kuisma, Ivanov, Taghizadeh, Peterson, Haldar, Dohn, Sch{\"a}fer, J{\'o}nsson, \latin{et~al.} others]{gpaw_recent}
Mortensen,~J.~J.; Larsen,~A.~H.; Kuisma,~M.; Ivanov,~A.~V.; Taghizadeh,~A.; Peterson,~A.; Haldar,~A.; Dohn,~A.~O.; Sch{\"a}fer,~C.; J{\'o}nsson,~E.~{\"O}. \latin{et~al.}  GPAW: An open Python package for electronic structure calculations. \emph{The Journal of Chemical Physics} \textbf{2024}, \emph{160}\relax
\mciteBstWouldAddEndPuncttrue
\mciteSetBstMidEndSepPunct{\mcitedefaultmidpunct}
{\mcitedefaultendpunct}{\mcitedefaultseppunct}\relax
\EndOfBibitem
\bibitem[Blum \latin{et~al.}(2009)Blum, Gehrke, Hanke, Havu, Havu, Ren, Reuter, and Scheffler]{fhi_aims_2009}
Blum,~V.; Gehrke,~R.; Hanke,~F.; Havu,~P.; Havu,~V.; Ren,~X.; Reuter,~K.; Scheffler,~M. Ab initio molecular simulations with numeric atom-centered orbitals. \emph{Computer Physics Communications} \textbf{2009}, \emph{180}, 2175--2196\relax
\mciteBstWouldAddEndPuncttrue
\mciteSetBstMidEndSepPunct{\mcitedefaultmidpunct}
{\mcitedefaultendpunct}{\mcitedefaultseppunct}\relax
\EndOfBibitem
\bibitem[Goedecker \latin{et~al.}(1996)Goedecker, Teter, and Hutter]{pseudopotential_GTH_1}
Goedecker,~S.; Teter,~M.; Hutter,~J. Separable dual-space Gaussian pseudopotentials. \emph{Physical Review B} \textbf{1996}, \emph{54}, 1703\relax
\mciteBstWouldAddEndPuncttrue
\mciteSetBstMidEndSepPunct{\mcitedefaultmidpunct}
{\mcitedefaultendpunct}{\mcitedefaultseppunct}\relax
\EndOfBibitem
\bibitem[Hartwigsen \latin{et~al.}(1998)Hartwigsen, G{\oe}decker, and Hutter]{pseudopotential_GTH_2}
Hartwigsen,~C.; G{\oe}decker,~S.; Hutter,~J. Relativistic separable dual-space Gaussian pseudopotentials from H to Rn. \emph{Physical Review B} \textbf{1998}, \emph{58}, 3641\relax
\mciteBstWouldAddEndPuncttrue
\mciteSetBstMidEndSepPunct{\mcitedefaultmidpunct}
{\mcitedefaultendpunct}{\mcitedefaultseppunct}\relax
\EndOfBibitem
\bibitem[Krack(2005)]{pseudopotential_GTH_3}
Krack,~M. Pseudopotentials for H to Kr optimized for gradient-corrected exchange-correlation functionals. \emph{Theoretical Chemistry Accounts} \textbf{2005}, \emph{114}, 145--152\relax
\mciteBstWouldAddEndPuncttrue
\mciteSetBstMidEndSepPunct{\mcitedefaultmidpunct}
{\mcitedefaultendpunct}{\mcitedefaultseppunct}\relax
\EndOfBibitem
\bibitem[Hamann \latin{et~al.}(1979)Hamann, Schl{\"u}ter, and Chiang]{pseudopotential_norm_conserving_1}
Hamann,~D.; Schl{\"u}ter,~M.; Chiang,~C. Norm-conserving pseudopotentials. \emph{Physical Review Letters} \textbf{1979}, \emph{43}, 1494\relax
\mciteBstWouldAddEndPuncttrue
\mciteSetBstMidEndSepPunct{\mcitedefaultmidpunct}
{\mcitedefaultendpunct}{\mcitedefaultseppunct}\relax
\EndOfBibitem
\bibitem[Pickett(1989)]{pseudopotential_norm_conserving_2}
Pickett,~W.~E. Pseudopotential methods in condensed matter applications. \emph{Computer Physics Reports} \textbf{1989}, \emph{9}, 115--197\relax
\mciteBstWouldAddEndPuncttrue
\mciteSetBstMidEndSepPunct{\mcitedefaultmidpunct}
{\mcitedefaultendpunct}{\mcitedefaultseppunct}\relax
\EndOfBibitem
\bibitem[Troullier and Martins(1991)Troullier, and Martins]{pseudopotential_norm_conserving_3}
Troullier,~N.; Martins,~J.~L. Efficient pseudopotentials for plane-wave calculations. \emph{Physical Review B} \textbf{1991}, \emph{43}, 1993\relax
\mciteBstWouldAddEndPuncttrue
\mciteSetBstMidEndSepPunct{\mcitedefaultmidpunct}
{\mcitedefaultendpunct}{\mcitedefaultseppunct}\relax
\EndOfBibitem
\bibitem[Vanderbilt(1990)]{vanderbilt_soft_pseudopotential}
Vanderbilt,~D. Soft self-consistent pseudopotentials in a generalized eigenvalue formalism. \emph{Physical review B} \textbf{1990}, \emph{41}, 7892\relax
\mciteBstWouldAddEndPuncttrue
\mciteSetBstMidEndSepPunct{\mcitedefaultmidpunct}
{\mcitedefaultendpunct}{\mcitedefaultseppunct}\relax
\EndOfBibitem
\bibitem[Slater(1937)]{augmented_plane_wave_original}
Slater,~J.~C. Wave functions in a periodic potential. \emph{Physical Review} \textbf{1937}, \emph{51}, 846\relax
\mciteBstWouldAddEndPuncttrue
\mciteSetBstMidEndSepPunct{\mcitedefaultmidpunct}
{\mcitedefaultendpunct}{\mcitedefaultseppunct}\relax
\EndOfBibitem
\bibitem[Andersen(1975)]{linear_augmented_plane_wave_original}
Andersen,~O.~K. Linear methods in band theory. \emph{Physical Review B} \textbf{1975}, \emph{12}, 3060\relax
\mciteBstWouldAddEndPuncttrue
\mciteSetBstMidEndSepPunct{\mcitedefaultmidpunct}
{\mcitedefaultendpunct}{\mcitedefaultseppunct}\relax
\EndOfBibitem
\bibitem[Bl{\"o}chl(1994)]{projector_augmented_plane_wave_original}
Bl{\"o}chl,~P.~E. Projector augmented-wave method. \emph{Physical Review B} \textbf{1994}, \emph{50}, 17953\relax
\mciteBstWouldAddEndPuncttrue
\mciteSetBstMidEndSepPunct{\mcitedefaultmidpunct}
{\mcitedefaultendpunct}{\mcitedefaultseppunct}\relax
\EndOfBibitem
\bibitem[Kresse and Joubert(1999)Kresse, and Joubert]{projector_augmented_plane_wave_2}
Kresse,~G.; Joubert,~D. From ultrasoft pseudopotentials to the projector augmented-wave method. \emph{Physical Review B} \textbf{1999}, \emph{59}, 1758\relax
\mciteBstWouldAddEndPuncttrue
\mciteSetBstMidEndSepPunct{\mcitedefaultmidpunct}
{\mcitedefaultendpunct}{\mcitedefaultseppunct}\relax
\EndOfBibitem
\bibitem[Mortensen \latin{et~al.}(2005)Mortensen, Hansen, and Jacobsen]{projector_augmented_plane_wave_3}
Mortensen,~J.~J.; Hansen,~L.~B.; Jacobsen,~K.~W. Real-space grid implementation of the projector augmented wave method. \emph{Physical Review B} \textbf{2005}, \emph{71}, 035109\relax
\mciteBstWouldAddEndPuncttrue
\mciteSetBstMidEndSepPunct{\mcitedefaultmidpunct}
{\mcitedefaultendpunct}{\mcitedefaultseppunct}\relax
\EndOfBibitem
\bibitem[Bylaska \latin{et~al.}(2002)Bylaska, Valiev, Kawai, and Weare]{projector_augmented_plane_wave_nice_equation}
Bylaska,~E.~J.; Valiev,~M.; Kawai,~R.; Weare,~J.~H. Parallel implementation of the projector augmented plane wave method for charged systems. \emph{Computer Physics Communications} \textbf{2002}, \emph{143}, 11--28\relax
\mciteBstWouldAddEndPuncttrue
\mciteSetBstMidEndSepPunct{\mcitedefaultmidpunct}
{\mcitedefaultendpunct}{\mcitedefaultseppunct}\relax
\EndOfBibitem
\bibitem[Sanchez-Portal \latin{et~al.}(1995)Sanchez-Portal, Artacho, and Soler]{plane_wave_projection_to_gaussian_1}
Sanchez-Portal,~D.; Artacho,~E.; Soler,~J.~M. Projection of plane-wave calculations into atomic orbitals. \emph{Solid State Communications} \textbf{1995}, \emph{95}, 685--690\relax
\mciteBstWouldAddEndPuncttrue
\mciteSetBstMidEndSepPunct{\mcitedefaultmidpunct}
{\mcitedefaultendpunct}{\mcitedefaultseppunct}\relax
\EndOfBibitem
\bibitem[Boffi \latin{et~al.}(2016)Boffi, Jain, and Natan]{plane_wave_exchange_projection}
Boffi,~N.~M.; Jain,~M.; Natan,~A. Efficient computation of the hartree--fock exchange in real-space with projection operators. \emph{Journal of Chemical Theory and Computation} \textbf{2016}, \emph{12}, 3614--3622\relax
\mciteBstWouldAddEndPuncttrue
\mciteSetBstMidEndSepPunct{\mcitedefaultmidpunct}
{\mcitedefaultendpunct}{\mcitedefaultseppunct}\relax
\EndOfBibitem
\bibitem[Duchemin and Gygi(2010)Duchemin, and Gygi]{plane_wave_exchange_exact_2}
Duchemin,~I.; Gygi,~F. A scalable and accurate algorithm for the computation of Hartree--Fock exchange. \emph{Computer Physics Communications} \textbf{2010}, \emph{181}, 855--860\relax
\mciteBstWouldAddEndPuncttrue
\mciteSetBstMidEndSepPunct{\mcitedefaultmidpunct}
{\mcitedefaultendpunct}{\mcitedefaultseppunct}\relax
\EndOfBibitem
\bibitem[Rossomme \latin{et~al.}(2023)Rossomme, Cunha, Li, Chen, McIsaac, Head-Gordon, and Head-Gordon]{pseudopotential_error}
Rossomme,~E.; Cunha,~L.~A.; Li,~W.; Chen,~K.; McIsaac,~A.~R.; Head-Gordon,~T.; Head-Gordon,~M. The Good, the Bad, and the Ugly: Pseudopotential Inconsistency Errors in Molecular Applications of Density Functional Theory. \emph{Journal of Chemical Theory and Computation} \textbf{2023}, \emph{19}, 2827--2841\relax
\mciteBstWouldAddEndPuncttrue
\mciteSetBstMidEndSepPunct{\mcitedefaultmidpunct}
{\mcitedefaultendpunct}{\mcitedefaultseppunct}\relax
\EndOfBibitem
\bibitem[Sorouri \latin{et~al.}(2006)Sorouri, Foulkes, and Hine]{plane_wave_exchange_exact}
Sorouri,~A.; Foulkes,~W. M.~C.; Hine,~N.~D. Accurate and efficient method for the treatment of exchange in a plane-wave basis. \emph{The Journal of Chemical Physics} \textbf{2006}, \emph{124}\relax
\mciteBstWouldAddEndPuncttrue
\mciteSetBstMidEndSepPunct{\mcitedefaultmidpunct}
{\mcitedefaultendpunct}{\mcitedefaultseppunct}\relax
\EndOfBibitem
\bibitem[Mardirossian and Head-Gordon(2017)Mardirossian, and Head-Gordon]{molecular_dft_review}
Mardirossian,~N.; Head-Gordon,~M. Thirty years of density functional theory in computational chemistry: an overview and extensive assessment of 200 density functionals. \emph{Molecular physics} \textbf{2017}, \emph{115}, 2315--2372\relax
\mciteBstWouldAddEndPuncttrue
\mciteSetBstMidEndSepPunct{\mcitedefaultmidpunct}
{\mcitedefaultendpunct}{\mcitedefaultseppunct}\relax
\EndOfBibitem
\bibitem[Pisani and Dovesi(1980)Pisani, and Dovesi]{periodic_hf_1}
Pisani,~C.; Dovesi,~R. Exact-exchange Hartree--Fock calculations for periodic systems. I. Illustration of the method. \emph{International Journal of Quantum Chemistry} \textbf{1980}, \emph{17}, 501--516\relax
\mciteBstWouldAddEndPuncttrue
\mciteSetBstMidEndSepPunct{\mcitedefaultmidpunct}
{\mcitedefaultendpunct}{\mcitedefaultseppunct}\relax
\EndOfBibitem
\bibitem[Dovesi \latin{et~al.}(2018)Dovesi, Erba, Orlando, Zicovich-Wilson, Civalleri, Maschio, R{\'e}rat, Casassa, Baima, Salustro, \latin{et~al.} others]{crystal_17}
Dovesi,~R.; Erba,~A.; Orlando,~R.; Zicovich-Wilson,~C.~M.; Civalleri,~B.; Maschio,~L.; R{\'e}rat,~M.; Casassa,~S.; Baima,~J.; Salustro,~S. \latin{et~al.}  Quantum-mechanical condensed matter simulations with CRYSTAL. \emph{Wiley Interdisciplinary Reviews: Computational Molecular Science} \textbf{2018}, \emph{8}, e1360\relax
\mciteBstWouldAddEndPuncttrue
\mciteSetBstMidEndSepPunct{\mcitedefaultmidpunct}
{\mcitedefaultendpunct}{\mcitedefaultseppunct}\relax
\EndOfBibitem
\bibitem[Erba \latin{et~al.}(2022)Erba, Desmarais, Casassa, Civalleri, Don{\`a}, Bush, Searle, Maschio, Edith-Daga, Cossard, \latin{et~al.} others]{crystal_23}
Erba,~A.; Desmarais,~J.~K.; Casassa,~S.; Civalleri,~B.; Don{\`a},~L.; Bush,~I.~J.; Searle,~B.; Maschio,~L.; Edith-Daga,~L.; Cossard,~A. \latin{et~al.}  CRYSTAL23: a program for computational solid state physics and chemistry. \emph{Journal of Chemical Theory and Computation} \textbf{2022}, \emph{19}, 6891--6932\relax
\mciteBstWouldAddEndPuncttrue
\mciteSetBstMidEndSepPunct{\mcitedefaultmidpunct}
{\mcitedefaultendpunct}{\mcitedefaultseppunct}\relax
\EndOfBibitem
\bibitem[Sun \latin{et~al.}(2017)Sun, Berkelbach, McClain, and Chan]{pyscf_mixed_density_fitting}
Sun,~Q.; Berkelbach,~T.~C.; McClain,~J.~D.; Chan,~G.~K. Gaussian and plane-wave mixed density fitting for periodic systems. \emph{The Journal of chemical physics} \textbf{2017}, \emph{147}\relax
\mciteBstWouldAddEndPuncttrue
\mciteSetBstMidEndSepPunct{\mcitedefaultmidpunct}
{\mcitedefaultendpunct}{\mcitedefaultseppunct}\relax
\EndOfBibitem
\bibitem[Sun(2023)]{pyscf_periodic_new}
Sun,~Q. Various integral estimations and screening schemes for extended systems in PySCF. \emph{arXiv preprint arXiv:2302.11307} \textbf{2023}, \relax
\mciteBstWouldAddEndPunctfalse
\mciteSetBstMidEndSepPunct{\mcitedefaultmidpunct}
{}{\mcitedefaultseppunct}\relax
\EndOfBibitem
\bibitem[Hutter \latin{et~al.}(2014)Hutter, Iannuzzi, Schiffmann, and VandeVondele]{cp2k_recent}
Hutter,~J.; Iannuzzi,~M.; Schiffmann,~F.; VandeVondele,~J. cp2k: atomistic simulations of condensed matter systems. \emph{Wiley Interdisciplinary Reviews: Computational Molecular Science} \textbf{2014}, \emph{4}, 15--25\relax
\mciteBstWouldAddEndPuncttrue
\mciteSetBstMidEndSepPunct{\mcitedefaultmidpunct}
{\mcitedefaultendpunct}{\mcitedefaultseppunct}\relax
\EndOfBibitem
\bibitem[Lee \latin{et~al.}(2021)Lee, Feng, Cunha, Gonthier, Epifanovsky, and Head-Gordon]{qchem_periodic_1}
Lee,~J.; Feng,~X.; Cunha,~L.~A.; Gonthier,~J.~F.; Epifanovsky,~E.; Head-Gordon,~M. Approaching the basis set limit in Gaussian-orbital-based periodic calculations with transferability: Performance of pure density functionals for simple semiconductors. \emph{The Journal of Chemical Physics} \textbf{2021}, \emph{155}\relax
\mciteBstWouldAddEndPuncttrue
\mciteSetBstMidEndSepPunct{\mcitedefaultmidpunct}
{\mcitedefaultendpunct}{\mcitedefaultseppunct}\relax
\EndOfBibitem
\bibitem[Lee \latin{et~al.}(2022)Lee, Rettig, Feng, Epifanovsky, and Head-Gordon]{qchem_periodic_2}
Lee,~J.; Rettig,~A.; Feng,~X.; Epifanovsky,~E.; Head-Gordon,~M. Faster exact exchange for solids via occ-RI-K: Application to combinatorially optimized range-separated hybrid functionals for simple solids with pseudopotentials near the basis set limit. \emph{Journal of Chemical Theory and Computation} \textbf{2022}, \emph{18}, 7336--7349\relax
\mciteBstWouldAddEndPuncttrue
\mciteSetBstMidEndSepPunct{\mcitedefaultmidpunct}
{\mcitedefaultendpunct}{\mcitedefaultseppunct}\relax
\EndOfBibitem
\bibitem[Rettig \latin{et~al.}(2023)Rettig, Lee, and Head-Gordon]{qchem_periodic_3}
Rettig,~A.; Lee,~J.; Head-Gordon,~M. Even faster exact exchange for solids via tensor hypercontraction. \emph{Journal of Chemical Theory and Computation} \textbf{2023}, \emph{19}, 5773--5784\relax
\mciteBstWouldAddEndPuncttrue
\mciteSetBstMidEndSepPunct{\mcitedefaultmidpunct}
{\mcitedefaultendpunct}{\mcitedefaultseppunct}\relax
\EndOfBibitem
\bibitem[Wang \latin{et~al.}(2020)Wang, Lewis, and Valeev]{mpqc_real_density_fitting}
Wang,~X.; Lewis,~C.~A.; Valeev,~E.~F. Efficient evaluation of exact exchange for periodic systems via concentric atomic density fitting. \emph{The Journal of Chemical Physics} \textbf{2020}, \emph{153}\relax
\mciteBstWouldAddEndPuncttrue
\mciteSetBstMidEndSepPunct{\mcitedefaultmidpunct}
{\mcitedefaultendpunct}{\mcitedefaultseppunct}\relax
\EndOfBibitem
\bibitem[Sharma and Beylkin(2021)Sharma, and Beylkin]{periodic_integral_gamma}
Sharma,~S.; Beylkin,~G. Efficient evaluation of two-center Gaussian integrals in periodic systems. \emph{Journal of Chemical Theory and Computation} \textbf{2021}, \emph{17}, 3916--3922\relax
\mciteBstWouldAddEndPuncttrue
\mciteSetBstMidEndSepPunct{\mcitedefaultmidpunct}
{\mcitedefaultendpunct}{\mcitedefaultseppunct}\relax
\EndOfBibitem
\bibitem[Ewald(1921)]{ewald_original}
Ewald,~P.~P. Die Berechnung optischer und elektrostatischer Gitterpotentiale. \emph{Annalen der Physik} \textbf{1921}, \emph{369}, 253--287\relax
\mciteBstWouldAddEndPuncttrue
\mciteSetBstMidEndSepPunct{\mcitedefaultmidpunct}
{\mcitedefaultendpunct}{\mcitedefaultseppunct}\relax
\EndOfBibitem
\bibitem[VandeVondele \latin{et~al.}(2005)VandeVondele, Krack, Mohamed, Parrinello, Chassaing, and Hutter]{cp2k_mixed_basis}
VandeVondele,~J.; Krack,~M.; Mohamed,~F.; Parrinello,~M.; Chassaing,~T.; Hutter,~J. Quickstep: Fast and accurate density functional calculations using a mixed Gaussian and plane waves approach. \emph{Computer Physics Communications} \textbf{2005}, \emph{167}, 103--128\relax
\mciteBstWouldAddEndPuncttrue
\mciteSetBstMidEndSepPunct{\mcitedefaultmidpunct}
{\mcitedefaultendpunct}{\mcitedefaultseppunct}\relax
\EndOfBibitem
\bibitem[Ye and Berkelbach(2021)Ye, and Berkelbach]{pyscf_range_separated_gaussian_density_fitting}
Ye,~H.-Z.; Berkelbach,~T.~C. Fast periodic Gaussian density fitting by range separation. \emph{The Journal of Chemical Physics} \textbf{2021}, \emph{154}\relax
\mciteBstWouldAddEndPuncttrue
\mciteSetBstMidEndSepPunct{\mcitedefaultmidpunct}
{\mcitedefaultendpunct}{\mcitedefaultseppunct}\relax
\EndOfBibitem
\bibitem[Saunders \latin{et~al.}(1992)Saunders, Freyria-Fava, Dovesi, Salasco, and Roetti]{crystal_bipolar_expansion}
Saunders,~V.; Freyria-Fava,~C.; Dovesi,~R.; Salasco,~L.; Roetti,~C. On the electrostatic potential in crystalline systems where the charge density is expanded in Gaussian functions. \emph{Molecular Physics} \textbf{1992}, \emph{77}, 629--665\relax
\mciteBstWouldAddEndPuncttrue
\mciteSetBstMidEndSepPunct{\mcitedefaultmidpunct}
{\mcitedefaultendpunct}{\mcitedefaultseppunct}\relax
\EndOfBibitem
\bibitem[Challacombe \latin{et~al.}(1997)Challacombe, White, and Head-Gordon]{fast_multipole_method}
Challacombe,~M.; White,~C.; Head-Gordon,~M. Periodic boundary conditions and the fast multipole method. \emph{The Journal of Chemical Physics} \textbf{1997}, \emph{107}, 10131--10140\relax
\mciteBstWouldAddEndPuncttrue
\mciteSetBstMidEndSepPunct{\mcitedefaultmidpunct}
{\mcitedefaultendpunct}{\mcitedefaultseppunct}\relax
\EndOfBibitem
\bibitem[Kudin and Scuseria(1998)Kudin, and Scuseria]{very_fast_multipole_method}
Kudin,~K.~N.; Scuseria,~G.~E. A fast multipole algorithm for the efficient treatment of the Coulomb problem in electronic structure calculations of periodic systems with Gaussian orbitals. \emph{Chemical Physics Letters} \textbf{1998}, \emph{289}, 611--616\relax
\mciteBstWouldAddEndPuncttrue
\mciteSetBstMidEndSepPunct{\mcitedefaultmidpunct}
{\mcitedefaultendpunct}{\mcitedefaultseppunct}\relax
\EndOfBibitem
\bibitem[Tymczak and Challacombe(2005)Tymczak, and Challacombe]{fast_multipole_method_linear}
Tymczak,~C.; Challacombe,~M. Linear scaling computation of the Fock matrix. VII. Periodic density functional theory at the $\Gamma$ point. \emph{The Journal of Chemical Physics} \textbf{2005}, \emph{122}\relax
\mciteBstWouldAddEndPuncttrue
\mciteSetBstMidEndSepPunct{\mcitedefaultmidpunct}
{\mcitedefaultendpunct}{\mcitedefaultseppunct}\relax
\EndOfBibitem
\bibitem[Ufimtsev and Martinez(2008)Ufimtsev, and Martinez]{terachem_gpu_1}
Ufimtsev,~I.~S.; Martinez,~T.~J. Quantum chemistry on graphical processing units. 1. Strategies for two-electron integral evaluation. \emph{Journal of Chemical Theory and Computation} \textbf{2008}, \emph{4}, 222--231\relax
\mciteBstWouldAddEndPuncttrue
\mciteSetBstMidEndSepPunct{\mcitedefaultmidpunct}
{\mcitedefaultendpunct}{\mcitedefaultseppunct}\relax
\EndOfBibitem
\bibitem[Ufimtsev and Martinez(2009)Ufimtsev, and Martinez]{terachem_gpu_2}
Ufimtsev,~I.~S.; Martinez,~T.~J. Quantum chemistry on graphical processing units. 2. Direct self-consistent-field implementation. \emph{Journal of Chemical Theory and Computation} \textbf{2009}, \emph{5}, 1004--1015\relax
\mciteBstWouldAddEndPuncttrue
\mciteSetBstMidEndSepPunct{\mcitedefaultmidpunct}
{\mcitedefaultendpunct}{\mcitedefaultseppunct}\relax
\EndOfBibitem
\bibitem[Ufimtsev and Martinez(2009)Ufimtsev, and Martinez]{terachem_gpu_3}
Ufimtsev,~I.~S.; Martinez,~T.~J. Quantum chemistry on graphical processing units. 3. Analytical energy gradients, geometry optimization, and first principles molecular dynamics. \emph{Journal of Chemical Theory and Computation} \textbf{2009}, \emph{5}, 2619--2628\relax
\mciteBstWouldAddEndPuncttrue
\mciteSetBstMidEndSepPunct{\mcitedefaultmidpunct}
{\mcitedefaultendpunct}{\mcitedefaultseppunct}\relax
\EndOfBibitem
\bibitem[Titov \latin{et~al.}(2013)Titov, Ufimtsev, Luehr, and Martinez]{terachem_2013}
Titov,~A.~V.; Ufimtsev,~I.~S.; Luehr,~N.; Martinez,~T.~J. Generating efficient quantum chemistry codes for novel architectures. \emph{Journal of Chemical Theory and Computation} \textbf{2013}, \emph{9}, 213--221\relax
\mciteBstWouldAddEndPuncttrue
\mciteSetBstMidEndSepPunct{\mcitedefaultmidpunct}
{\mcitedefaultendpunct}{\mcitedefaultseppunct}\relax
\EndOfBibitem
\bibitem[Seritan \latin{et~al.}(2021)Seritan, Bannwarth, Fales, Hohenstein, Isborn, Kokkila-Schumacher, Li, Liu, Luehr, Snyder~Jr, \latin{et~al.} others]{terachem_2021}
Seritan,~S.; Bannwarth,~C.; Fales,~B.~S.; Hohenstein,~E.~G.; Isborn,~C.~M.; Kokkila-Schumacher,~S.~I.; Li,~X.; Liu,~F.; Luehr,~N.; Snyder~Jr,~J.~W. \latin{et~al.}  TeraChem: A graphical processing unit-accelerated electronic structure package for large-scale ab initio molecular dynamics. \emph{Wiley Interdisciplinary Reviews: Computational Molecular Science} \textbf{2021}, \emph{11}, e1494\relax
\mciteBstWouldAddEndPuncttrue
\mciteSetBstMidEndSepPunct{\mcitedefaultmidpunct}
{\mcitedefaultendpunct}{\mcitedefaultseppunct}\relax
\EndOfBibitem
\bibitem[Wang \latin{et~al.}(2024)Wang, Hait, Johnson, Fajen, Guerrero, and Mart{\'\i}nez]{terachem_f}
Wang,~Y.; Hait,~D.; Johnson,~K.~G.; Fajen,~O.~J.; Guerrero,~R.~D.; Mart{\'\i}nez,~T.~J. Extending GPU-Accelerated Gaussian Integrals in the TeraChem Software Package to f Type Orbitals: Implementation and Applications. \emph{arXiv preprint arXiv:2406.14920} \textbf{2024}, \relax
\mciteBstWouldAddEndPunctfalse
\mciteSetBstMidEndSepPunct{\mcitedefaultmidpunct}
{}{\mcitedefaultseppunct}\relax
\EndOfBibitem
\bibitem[Tornai \latin{et~al.}(2019)Tornai, Ladj{\'a}nszki, R{\'a}k, Kis, and Cserey]{brianqc}
Tornai,~G.~J.; Ladj{\'a}nszki,~I.; R{\'a}k,~{\'A}.; Kis,~G.; Cserey,~G. Calculation of quantum chemical two-electron integrals by applying compiler technology on GPU. \emph{Journal of Chemical Theory and Computation} \textbf{2019}, \emph{15}, 5319--5331\relax
\mciteBstWouldAddEndPuncttrue
\mciteSetBstMidEndSepPunct{\mcitedefaultmidpunct}
{\mcitedefaultendpunct}{\mcitedefaultseppunct}\relax
\EndOfBibitem
\bibitem[Yasuda(2008)]{gaussian_program}
Yasuda,~K. Two-electron integral evaluation on the graphics processor unit. \emph{Journal of Computational Chemistry} \textbf{2008}, \emph{29}, 334--342\relax
\mciteBstWouldAddEndPuncttrue
\mciteSetBstMidEndSepPunct{\mcitedefaultmidpunct}
{\mcitedefaultendpunct}{\mcitedefaultseppunct}\relax
\EndOfBibitem
\bibitem[Barca \latin{et~al.}(2021)Barca, Alkan, Galvez-Vallejo, Poole, Rendell, and Gordon]{gamess_spd}
Barca,~G.~M.; Alkan,~M.; Galvez-Vallejo,~J.~L.; Poole,~D.~L.; Rendell,~A.~P.; Gordon,~M.~S. Faster self-consistent field (SCF) calculations on GPU clusters. \emph{Journal of Chemical Theory and Computation} \textbf{2021}, \emph{17}, 7486--7503\relax
\mciteBstWouldAddEndPuncttrue
\mciteSetBstMidEndSepPunct{\mcitedefaultmidpunct}
{\mcitedefaultendpunct}{\mcitedefaultseppunct}\relax
\EndOfBibitem
\bibitem[Kussmann and Ochsenfeld(2017)Kussmann, and Ochsenfeld]{fermions_original}
Kussmann,~J.; Ochsenfeld,~C. Hybrid CPU/GPU integral engine for strong-scaling ab initio methods. \emph{Journal of Chemical Theory and Computation} \textbf{2017}, \emph{13}, 3153--3159\relax
\mciteBstWouldAddEndPuncttrue
\mciteSetBstMidEndSepPunct{\mcitedefaultmidpunct}
{\mcitedefaultendpunct}{\mcitedefaultseppunct}\relax
\EndOfBibitem
\bibitem[Miao and Merz~Jr(2013)Miao, and Merz~Jr]{quick}
Miao,~Y.; Merz~Jr,~K.~M. Acceleration of electron repulsion integral evaluation on graphics processing units via use of recurrence relations. \emph{Journal of Chemical Theory and Computation} \textbf{2013}, \emph{9}, 965--976\relax
\mciteBstWouldAddEndPuncttrue
\mciteSetBstMidEndSepPunct{\mcitedefaultmidpunct}
{\mcitedefaultendpunct}{\mcitedefaultseppunct}\relax
\EndOfBibitem
\bibitem[Asadchev and Valeev(2023)Asadchev, and Valeev]{libintx_up_to_iiii}
Asadchev,~A.; Valeev,~E.~F. High-performance evaluation of high angular momentum 4-center Gaussian integrals on modern accelerated processors. \emph{The Journal of Physical Chemistry A} \textbf{2023}, \emph{127}, 10889--10895\relax
\mciteBstWouldAddEndPuncttrue
\mciteSetBstMidEndSepPunct{\mcitedefaultmidpunct}
{\mcitedefaultendpunct}{\mcitedefaultseppunct}\relax
\EndOfBibitem
\bibitem[Wu \latin{et~al.}(2024)Wu, Sun, Pu, Zheng, Ma, Yan, Yu, Wu, Huo, Li, \latin{et~al.} others]{pyscf_gpu}
Wu,~X.; Sun,~Q.; Pu,~Z.; Zheng,~T.; Ma,~W.; Yan,~W.; Yu,~X.; Wu,~Z.; Huo,~M.; Li,~X. \latin{et~al.}  Python-Based Quantum Chemistry Calculations with GPU Acceleration. \emph{arXiv preprint arXiv:2404.09452} \textbf{2024}, \relax
\mciteBstWouldAddEndPunctfalse
\mciteSetBstMidEndSepPunct{\mcitedefaultmidpunct}
{}{\mcitedefaultseppunct}\relax
\EndOfBibitem
\bibitem[Fernandes \latin{et~al.}(2015)Fernandes, Renison, and Naidoo]{quantum_supercharger_library}
Fernandes,~K.~D.; Renison,~C.~A.; Naidoo,~K.~J. Quantum supercharger library: Hyper-parallelism of the H artree--F ock method. \emph{Journal of Computational Chemistry} \textbf{2015}, \emph{36}, 1399--1409\relax
\mciteBstWouldAddEndPuncttrue
\mciteSetBstMidEndSepPunct{\mcitedefaultmidpunct}
{\mcitedefaultendpunct}{\mcitedefaultseppunct}\relax
\EndOfBibitem
\bibitem[Holzer(2020)]{turbomole}
Holzer,~C. An improved seminumerical Coulomb and exchange algorithm for properties and excited states in modern density functional theory. \emph{The Journal of Chemical Physics} \textbf{2020}, \emph{153}\relax
\mciteBstWouldAddEndPuncttrue
\mciteSetBstMidEndSepPunct{\mcitedefaultmidpunct}
{\mcitedefaultendpunct}{\mcitedefaultseppunct}\relax
\EndOfBibitem
\bibitem[Qi \latin{et~al.}(2023)Qi, Zhang, and Yang]{wuhan_electronic_structure_package}
Qi,~J.; Zhang,~Y.; Yang,~M. A hybrid CPU/GPU method for Hartree--Fock self-consistent-field calculation. \emph{The Journal of Chemical Physics} \textbf{2023}, \emph{159}\relax
\mciteBstWouldAddEndPuncttrue
\mciteSetBstMidEndSepPunct{\mcitedefaultmidpunct}
{\mcitedefaultendpunct}{\mcitedefaultseppunct}\relax
\EndOfBibitem
\bibitem[Fales \latin{et~al.}(2020)Fales, Curtis, Johnson, Lahana, Seritan, Wang, Weir, Mart{\'\i}nez, and Hohenstein]{terachem_ccsd_gpu}
Fales,~B.~S.; Curtis,~E.~R.; Johnson,~K.~G.; Lahana,~D.; Seritan,~S.; Wang,~Y.; Weir,~H.; Mart{\'\i}nez,~T.~J.; Hohenstein,~E.~G. Performance of coupled-cluster singles and doubles on modern stream processing architectures. \emph{Journal of Chemical Theory and Computation} \textbf{2020}, \emph{16}, 4021--4028\relax
\mciteBstWouldAddEndPuncttrue
\mciteSetBstMidEndSepPunct{\mcitedefaultmidpunct}
{\mcitedefaultendpunct}{\mcitedefaultseppunct}\relax
\EndOfBibitem
\bibitem[Parrish \latin{et~al.}(2019)Parrish, Zhao, Hohenstein, and Mart{\'\i}nez]{terachem_rank_reduce_cc_1}
Parrish,~R.~M.; Zhao,~Y.; Hohenstein,~E.~G.; Mart{\'\i}nez,~T.~J. Rank reduced coupled cluster theory. I. Ground state energies and wavefunctions. \emph{The Journal of Chemical Physics} \textbf{2019}, \emph{150}\relax
\mciteBstWouldAddEndPuncttrue
\mciteSetBstMidEndSepPunct{\mcitedefaultmidpunct}
{\mcitedefaultendpunct}{\mcitedefaultseppunct}\relax
\EndOfBibitem
\bibitem[Hohenstein \latin{et~al.}(2019)Hohenstein, Zhao, Parrish, and Mart{\'\i}nez]{terachem_rank_reduce_cc_2}
Hohenstein,~E.~G.; Zhao,~Y.; Parrish,~R.~M.; Mart{\'\i}nez,~T.~J. Rank reduced coupled cluster theory. II. Equation-of-motion coupled-cluster singles and doubles. \emph{The Journal of Chemical Physics} \textbf{2019}, \emph{151}\relax
\mciteBstWouldAddEndPuncttrue
\mciteSetBstMidEndSepPunct{\mcitedefaultmidpunct}
{\mcitedefaultendpunct}{\mcitedefaultseppunct}\relax
\EndOfBibitem
\bibitem[Hohenstein \latin{et~al.}(2022)Hohenstein, Fales, Parrish, and Mart{\'\i}nez]{terachem_rank_reduce_cc_3}
Hohenstein,~E.~G.; Fales,~B.~S.; Parrish,~R.~M.; Mart{\'\i}nez,~T.~J. Rank-reduced coupled-cluster. III. Tensor hypercontraction of the doubles amplitudes. \emph{The Journal of Chemical Physics} \textbf{2022}, \emph{156}\relax
\mciteBstWouldAddEndPuncttrue
\mciteSetBstMidEndSepPunct{\mcitedefaultmidpunct}
{\mcitedefaultendpunct}{\mcitedefaultseppunct}\relax
\EndOfBibitem
\bibitem[Hohenstein and Mart{\'\i}nez(2021)Hohenstein, and Mart{\'\i}nez]{terachem_rank_reduce_ccsd}
Hohenstein,~E.~G.; Mart{\'\i}nez,~T.~J. GPU acceleration of rank-reduced coupled-cluster singles and doubles. \emph{The Journal of Chemical Physics} \textbf{2021}, \emph{155}\relax
\mciteBstWouldAddEndPuncttrue
\mciteSetBstMidEndSepPunct{\mcitedefaultmidpunct}
{\mcitedefaultendpunct}{\mcitedefaultseppunct}\relax
\EndOfBibitem
\bibitem[Snyder \latin{et~al.}(2017)Snyder, Fales, Hohenstein, Levine, and Mart{\'\i}nez]{terachem_casscf}
Snyder,~J.~W.; Fales,~B.~S.; Hohenstein,~E.~G.; Levine,~B.~G.; Mart{\'\i}nez,~T.~J. A direct-compatible formulation of the coupled perturbed complete active space self-consistent field equations on graphical processing units. \emph{The Journal of Chemical Physics} \textbf{2017}, \emph{146}\relax
\mciteBstWouldAddEndPuncttrue
\mciteSetBstMidEndSepPunct{\mcitedefaultmidpunct}
{\mcitedefaultendpunct}{\mcitedefaultseppunct}\relax
\EndOfBibitem
\bibitem[Slav{\'\i}{\v{c}}ek and Mart{\'\i}nez(2010)Slav{\'\i}{\v{c}}ek, and Mart{\'\i}nez]{terachem_fomo_casci}
Slav{\'\i}{\v{c}}ek,~P.; Mart{\'\i}nez,~T.~J. Ab initio floating occupation molecular orbital-complete active space configuration interaction: An efficient approximation to CASSCF. \emph{The Journal of Chemical Physics} \textbf{2010}, \emph{132}\relax
\mciteBstWouldAddEndPuncttrue
\mciteSetBstMidEndSepPunct{\mcitedefaultmidpunct}
{\mcitedefaultendpunct}{\mcitedefaultseppunct}\relax
\EndOfBibitem
\bibitem[Fales and Levine(2015)Fales, and Levine]{terachem_direct_ci_1}
Fales,~B.~S.; Levine,~B.~G. Nanoscale multireference quantum chemistry: Full configuration interaction on graphical processing units. \emph{Journal of Chemical Theory and Computation} \textbf{2015}, \emph{11}, 4708--4716\relax
\mciteBstWouldAddEndPuncttrue
\mciteSetBstMidEndSepPunct{\mcitedefaultmidpunct}
{\mcitedefaultendpunct}{\mcitedefaultseppunct}\relax
\EndOfBibitem
\bibitem[Fales and Mart{\'\i}nez(2020)Fales, and Mart{\'\i}nez]{terachem_direct_ci_2}
Fales,~B.~S.; Mart{\'\i}nez,~T.~J. Fast transformations between configuration state function and Slater determinant bases for direct configuration interaction. \emph{The Journal of Chemical Physics} \textbf{2020}, \emph{152}\relax
\mciteBstWouldAddEndPuncttrue
\mciteSetBstMidEndSepPunct{\mcitedefaultmidpunct}
{\mcitedefaultendpunct}{\mcitedefaultseppunct}\relax
\EndOfBibitem
\bibitem[Fales and Mart{\'\i}nez(2020)Fales, and Mart{\'\i}nez]{terachem_direct_ci_3}
Fales,~B.~S.; Mart{\'\i}nez,~T.~J. Efficient treatment of large active spaces through multi-GPU parallel implementation of direct configuration interaction. \emph{Journal of Chemical Theory and Computation} \textbf{2020}, \emph{16}, 1586--1596\relax
\mciteBstWouldAddEndPuncttrue
\mciteSetBstMidEndSepPunct{\mcitedefaultmidpunct}
{\mcitedefaultendpunct}{\mcitedefaultseppunct}\relax
\EndOfBibitem
\bibitem[Fales \latin{et~al.}(2018)Fales, Seritan, Settje, Levine, Koch, and Mart{\'\i}nez]{terachem_rank_reduced_ci}
Fales,~B.~S.; Seritan,~S.; Settje,~N.~F.; Levine,~B.~G.; Koch,~H.; Mart{\'\i}nez,~T.~J. Large-scale electron correlation calculations: Rank-reduced full configuration interaction. \emph{Journal of Chemical Theory and Computation} \textbf{2018}, \emph{14}, 4139--4150\relax
\mciteBstWouldAddEndPuncttrue
\mciteSetBstMidEndSepPunct{\mcitedefaultmidpunct}
{\mcitedefaultendpunct}{\mcitedefaultseppunct}\relax
\EndOfBibitem
\bibitem[Wolfowicz \latin{et~al.}(2021)Wolfowicz, Heremans, Anderson, Kanai, Seo, Gali, Galli, and Awschalom]{point_defect_review}
Wolfowicz,~G.; Heremans,~F.~J.; Anderson,~C.~P.; Kanai,~S.; Seo,~H.; Gali,~A.; Galli,~G.; Awschalom,~D.~D. Quantum guidelines for solid-state spin defects. \emph{Nature Reviews Materials} \textbf{2021}, \emph{6}, 906--925\relax
\mciteBstWouldAddEndPuncttrue
\mciteSetBstMidEndSepPunct{\mcitedefaultmidpunct}
{\mcitedefaultendpunct}{\mcitedefaultseppunct}\relax
\EndOfBibitem
\bibitem[Maintz \latin{et~al.}(2011)Maintz, Eck, and Dronskowski]{vasp_gpu_1}
Maintz,~S.; Eck,~B.; Dronskowski,~R. Speeding up plane-wave electronic-structure calculations using graphics-processing units. \emph{Computer Physics Communications} \textbf{2011}, \emph{182}, 1421--1427\relax
\mciteBstWouldAddEndPuncttrue
\mciteSetBstMidEndSepPunct{\mcitedefaultmidpunct}
{\mcitedefaultendpunct}{\mcitedefaultseppunct}\relax
\EndOfBibitem
\bibitem[Maintz \latin{et~al.}(2012)Maintz, Eck, and Dronskowski]{vasp_gpu_2}
Maintz,~S.; Eck,~B.; Dronskowski,~R. cuVASP: A GPU-Accelerated Plane-Wave Electronic-Structure Code. High Performance Computing in Science and Engineering'11: Transactions of the High Performance Computing Center, Stuttgart (HLRS) 2011. 2012; pp 201--205\relax
\mciteBstWouldAddEndPuncttrue
\mciteSetBstMidEndSepPunct{\mcitedefaultmidpunct}
{\mcitedefaultendpunct}{\mcitedefaultseppunct}\relax
\EndOfBibitem
\bibitem[Hacene \latin{et~al.}(2012)Hacene, Anciaux-Sedrakian, Rozanska, Klahr, Guignon, and Fleurat-Lessard]{vasp_gpu_3}
Hacene,~M.; Anciaux-Sedrakian,~A.; Rozanska,~X.; Klahr,~D.; Guignon,~T.; Fleurat-Lessard,~P. Accelerating VASP electronic structure calculations using graphic processing units. \emph{Journal of Computational Chemistry} \textbf{2012}, \emph{33}, 2581--2589\relax
\mciteBstWouldAddEndPuncttrue
\mciteSetBstMidEndSepPunct{\mcitedefaultmidpunct}
{\mcitedefaultendpunct}{\mcitedefaultseppunct}\relax
\EndOfBibitem
\bibitem[Jia \latin{et~al.}(2013)Jia, Fu, Cao, Wang, Chi, Gao, and Wang]{pwmat_gpu_1}
Jia,~W.; Fu,~J.; Cao,~Z.; Wang,~L.; Chi,~X.; Gao,~W.; Wang,~L.-W. Fast plane wave density functional theory molecular dynamics calculations on multi-GPU machines. \emph{Journal of Computational Physics} \textbf{2013}, \emph{251}, 102--115\relax
\mciteBstWouldAddEndPuncttrue
\mciteSetBstMidEndSepPunct{\mcitedefaultmidpunct}
{\mcitedefaultendpunct}{\mcitedefaultseppunct}\relax
\EndOfBibitem
\bibitem[Jia \latin{et~al.}(2013)Jia, Cao, Wang, Fu, Chi, Gao, and Wang]{pwmat_gpu_2}
Jia,~W.; Cao,~Z.; Wang,~L.; Fu,~J.; Chi,~X.; Gao,~W.; Wang,~L.-W. The analysis of a plane wave pseudopotential density functional theory code on a GPU machine. \emph{Computer Physics Communications} \textbf{2013}, \emph{184}, 9--18\relax
\mciteBstWouldAddEndPuncttrue
\mciteSetBstMidEndSepPunct{\mcitedefaultmidpunct}
{\mcitedefaultendpunct}{\mcitedefaultseppunct}\relax
\EndOfBibitem
\bibitem[Yokelson \latin{et~al.}(2022)Yokelson, Tkachenko, Robey, Li, and Dub]{cp2k_gpu}
Yokelson,~D.; Tkachenko,~N.~V.; Robey,~R.; Li,~Y.~W.; Dub,~P.~A. Performance analysis of cp2k code for ab initio molecular dynamics on cpus and gpus. \emph{Journal of Chemical Information and Modeling} \textbf{2022}, \emph{62}, 2378--2386\relax
\mciteBstWouldAddEndPuncttrue
\mciteSetBstMidEndSepPunct{\mcitedefaultmidpunct}
{\mcitedefaultendpunct}{\mcitedefaultseppunct}\relax
\EndOfBibitem
\bibitem[Spiga and Girotto(2012)Spiga, and Girotto]{quantum_espresso_gpu_1}
Spiga,~F.; Girotto,~I. phiGEMM: a CPU-GPU library for porting Quantum ESPRESSO on hybrid systems. 2012 20th Euromicro international conference on parallel, distributed and network-based processing. 2012; pp 368--375\relax
\mciteBstWouldAddEndPuncttrue
\mciteSetBstMidEndSepPunct{\mcitedefaultmidpunct}
{\mcitedefaultendpunct}{\mcitedefaultseppunct}\relax
\EndOfBibitem
\bibitem[Romero \latin{et~al.}(2018)Romero, Phillips, Ruetsch, Fatica, Spiga, and Giannozzi]{quantum_espresso_gpu_2}
Romero,~J.; Phillips,~E.; Ruetsch,~G.; Fatica,~M.; Spiga,~F.; Giannozzi,~P. A performance study of Quantum ESPRESSO’s PWscf code on multi-core and GPU systems. High Performance Computing Systems. Performance Modeling, Benchmarking, and Simulation: 8th International Workshop, PMBS 2017, Denver, CO, USA, November 13, 2017, Proceedings 8. 2018; pp 67--87\relax
\mciteBstWouldAddEndPuncttrue
\mciteSetBstMidEndSepPunct{\mcitedefaultmidpunct}
{\mcitedefaultendpunct}{\mcitedefaultseppunct}\relax
\EndOfBibitem
\bibitem[Gonze \latin{et~al.}(2016)Gonze, Jollet, Araujo, Adams, Amadon, Applencourt, Audouze, Beuken, Bieder, Bokhanchuk, \latin{et~al.} others]{abinit_2016_gpu}
Gonze,~X.; Jollet,~F.; Araujo,~F.~A.; Adams,~D.; Amadon,~B.; Applencourt,~T.; Audouze,~C.; Beuken,~J.-M.; Bieder,~J.; Bokhanchuk,~A. \latin{et~al.}  Recent developments in the ABINIT software package. \emph{Computer Physics Communications} \textbf{2016}, \emph{205}, 106--131\relax
\mciteBstWouldAddEndPuncttrue
\mciteSetBstMidEndSepPunct{\mcitedefaultmidpunct}
{\mcitedefaultendpunct}{\mcitedefaultseppunct}\relax
\EndOfBibitem
\bibitem[Yan \latin{et~al.}(2013)Yan, Li, and O’Grady]{gpaw_gpu}
Yan,~J.; Li,~L.; O’Grady,~C. Graphics Processing Unit acceleration of the Random Phase Approximation in the projector augmented wave method. \emph{Computer Physics Communications} \textbf{2013}, \emph{184}, 2728--2733\relax
\mciteBstWouldAddEndPuncttrue
\mciteSetBstMidEndSepPunct{\mcitedefaultmidpunct}
{\mcitedefaultendpunct}{\mcitedefaultseppunct}\relax
\EndOfBibitem
\bibitem[Jin \latin{et~al.}(2023)Jin, Yu, Govoni, Xu, and Galli]{west_program}
Jin,~Y.; Yu,~V. W.-z.; Govoni,~M.; Xu,~A.~C.; Galli,~G. Excited state properties of point defects in semiconductors and insulators investigated with time-dependent density functional theory. \emph{Journal of Chemical Theory and Computation} \textbf{2023}, \emph{19}, 8689--8705\relax
\mciteBstWouldAddEndPuncttrue
\mciteSetBstMidEndSepPunct{\mcitedefaultmidpunct}
{\mcitedefaultendpunct}{\mcitedefaultseppunct}\relax
\EndOfBibitem
\bibitem[Jones and March(1985)Jones, and March]{yellow_book_eik_summation}
Jones,~W.; March,~N.~H. \emph{Theoretical Solid State Physics, Volume 1: Perfect Lattices in Equilibrium}; Courier Corporation, 1985; Vol.~35; p~12\relax
\mciteBstWouldAddEndPuncttrue
\mciteSetBstMidEndSepPunct{\mcitedefaultmidpunct}
{\mcitedefaultendpunct}{\mcitedefaultseppunct}\relax
\EndOfBibitem
\bibitem[McMurchie and Davidson(1978)McMurchie, and Davidson]{McMurchie_Davidson_original}
McMurchie,~L.~E.; Davidson,~E.~R. One-and two-electron integrals over Cartesian Gaussian functions. \emph{Journal of Computational Physics} \textbf{1978}, \emph{26}, 218--231\relax
\mciteBstWouldAddEndPuncttrue
\mciteSetBstMidEndSepPunct{\mcitedefaultmidpunct}
{\mcitedefaultendpunct}{\mcitedefaultseppunct}\relax
\EndOfBibitem
\bibitem[Boys(1950)]{boys_original}
Boys,~S.~F. Electronic wave functions-I. A general method of calculation for the stationary states of any molecular system. \emph{Proceedings of the Royal Society of London. Series A. Mathematical and Physical Sciences} \textbf{1950}, \emph{200}, 542--554\relax
\mciteBstWouldAddEndPuncttrue
\mciteSetBstMidEndSepPunct{\mcitedefaultmidpunct}
{\mcitedefaultendpunct}{\mcitedefaultseppunct}\relax
\EndOfBibitem
\bibitem[Pinsky(2023)]{poisson_summation_formula_1}
Pinsky,~M.~A. \emph{Introduction to Fourier analysis and wavelets}; American Mathematical Society, 2023; Vol. 102; Chapter 4\relax
\mciteBstWouldAddEndPuncttrue
\mciteSetBstMidEndSepPunct{\mcitedefaultmidpunct}
{\mcitedefaultendpunct}{\mcitedefaultseppunct}\relax
\EndOfBibitem
\bibitem[Ghorbani-Asl \latin{et~al.}(2012)Ghorbani-Asl, Juarez-Mosqueda, Kuc, and Heine]{gamma_point}
Ghorbani-Asl,~M.; Juarez-Mosqueda,~R.; Kuc,~A.; Heine,~T. Efficient quantum simulations of metals within the $\Gamma$-point approximation: Application to carbon and inorganic 1D and 2D materials. \emph{Journal of Chemical Theory and Computation} \textbf{2012}, \emph{8}, 2888--2895\relax
\mciteBstWouldAddEndPuncttrue
\mciteSetBstMidEndSepPunct{\mcitedefaultmidpunct}
{\mcitedefaultendpunct}{\mcitedefaultseppunct}\relax
\EndOfBibitem
\bibitem[Becke(1988)]{becke_weight_original}
Becke,~A.~D. A multicenter numerical integration scheme for polyatomic molecules. \emph{The Journal of Chemical Physics} \textbf{1988}, \emph{88}, 2547--2553\relax
\mciteBstWouldAddEndPuncttrue
\mciteSetBstMidEndSepPunct{\mcitedefaultmidpunct}
{\mcitedefaultendpunct}{\mcitedefaultseppunct}\relax
\EndOfBibitem
\bibitem[Towler \latin{et~al.}(1996)Towler, Zupan, and Caus{\`a}]{becke_weight_periodic}
Towler,~M.~D.; Zupan,~A.; Caus{\`a},~M. Density functional theory in periodic systems using local Gaussian basis sets. \emph{Computer Physics Communications} \textbf{1996}, \emph{98}, 181--205\relax
\mciteBstWouldAddEndPuncttrue
\mciteSetBstMidEndSepPunct{\mcitedefaultmidpunct}
{\mcitedefaultendpunct}{\mcitedefaultseppunct}\relax
\EndOfBibitem
\bibitem[Murray \latin{et~al.}(1993)Murray, Handy, and Laming]{dft_integral}
Murray,~C.~W.; Handy,~N.~C.; Laming,~G.~J. Quadrature schemes for integrals of density functional theory. \emph{Molecular Physics} \textbf{1993}, \emph{78}, 997--1014\relax
\mciteBstWouldAddEndPuncttrue
\mciteSetBstMidEndSepPunct{\mcitedefaultmidpunct}
{\mcitedefaultendpunct}{\mcitedefaultseppunct}\relax
\EndOfBibitem
\bibitem[Gill \latin{et~al.}(1993)Gill, Johnson, and Pople]{dft_standard_grid_1}
Gill,~P.~M.; Johnson,~B.~G.; Pople,~J.~A. A standard grid for density functional calculations. \emph{Chemical Physics Letters} \textbf{1993}, \emph{209}, 506--512\relax
\mciteBstWouldAddEndPuncttrue
\mciteSetBstMidEndSepPunct{\mcitedefaultmidpunct}
{\mcitedefaultendpunct}{\mcitedefaultseppunct}\relax
\EndOfBibitem
\bibitem[Robinson \latin{et~al.}(2024)Robinson, Rettig, Dinh, Chen, and Lee]{periodic_gaussian_review}
Robinson,~P.~J.; Rettig,~A.; Dinh,~H.~Q.; Chen,~M.-F.; Lee,~J. Condensed-Phase Quantum Chemistry. \emph{arXiv preprint arXiv:2403.13207} \textbf{2024}, \relax
\mciteBstWouldAddEndPunctfalse
\mciteSetBstMidEndSepPunct{\mcitedefaultmidpunct}
{}{\mcitedefaultseppunct}\relax
\EndOfBibitem
\bibitem[White and Head-Gordon(1996)White, and Head-Gordon]{j_engine_hgp}
White,~C.~A.; Head-Gordon,~M. A J matrix engine for density functional theory calculations. \emph{The Journal of Chemical Physics} \textbf{1996}, \emph{104}, 2620--2629\relax
\mciteBstWouldAddEndPuncttrue
\mciteSetBstMidEndSepPunct{\mcitedefaultmidpunct}
{\mcitedefaultendpunct}{\mcitedefaultseppunct}\relax
\EndOfBibitem
\bibitem[Shao and Head-Gordon(2000)Shao, and Head-Gordon]{j_prepostprocess_original}
Shao,~Y.; Head-Gordon,~M. An improved J matrix engine for density functional theory calculations. \emph{Chemical Physics Letters} \textbf{2000}, \emph{323}, 425--433\relax
\mciteBstWouldAddEndPuncttrue
\mciteSetBstMidEndSepPunct{\mcitedefaultmidpunct}
{\mcitedefaultendpunct}{\mcitedefaultseppunct}\relax
\EndOfBibitem
\bibitem[Ahmadi and Alml{\"o}f(1995)Ahmadi, and Alml{\"o}f]{j_prepostprocess_family_basis_set}
Ahmadi,~G.~R.; Alml{\"o}f,~J. The Coulomb operator in a Gaussian product basis. \emph{Chemical Physics Letters} \textbf{1995}, \emph{246}, 364--370\relax
\mciteBstWouldAddEndPuncttrue
\mciteSetBstMidEndSepPunct{\mcitedefaultmidpunct}
{\mcitedefaultendpunct}{\mcitedefaultseppunct}\relax
\EndOfBibitem
\bibitem[Luehr \latin{et~al.}(2011)Luehr, Ufimtsev, and Martinez]{terachem_dynamic_precision}
Luehr,~N.; Ufimtsev,~I.~S.; Martinez,~T.~J. Dynamic precision for electron repulsion integral evaluation on graphical processing units (GPUs). \emph{Journal of Chemical Theory and Computation} \textbf{2011}, \emph{7}, 949--954\relax
\mciteBstWouldAddEndPuncttrue
\mciteSetBstMidEndSepPunct{\mcitedefaultmidpunct}
{\mcitedefaultendpunct}{\mcitedefaultseppunct}\relax
\EndOfBibitem
\bibitem[Weigend and Ahlrichs(2005)Weigend, and Ahlrichs]{def2_basis}
Weigend,~F.; Ahlrichs,~R. Balanced basis sets of split valence, triple zeta valence and quadruple zeta valence quality for H to Rn: Design and assessment of accuracy. \emph{Phys. Chem. Chem. Phys.} \textbf{2005}, \emph{7}, 3297--3305\relax
\mciteBstWouldAddEndPuncttrue
\mciteSetBstMidEndSepPunct{\mcitedefaultmidpunct}
{\mcitedefaultendpunct}{\mcitedefaultseppunct}\relax
\EndOfBibitem
\bibitem[Leadbetter \latin{et~al.}(1985)Leadbetter, Ward, Clark, Tucker, Matsuo, and Suga]{ice_xi_structure}
Leadbetter,~A.; Ward,~R.; Clark,~J.; Tucker,~P.; Matsuo,~T.; Suga,~H. The equilibrium low-temperature structure of ice. \emph{The Journal of chemical physics} \textbf{1985}, \emph{82}, 424--428\relax
\mciteBstWouldAddEndPuncttrue
\mciteSetBstMidEndSepPunct{\mcitedefaultmidpunct}
{\mcitedefaultendpunct}{\mcitedefaultseppunct}\relax
\EndOfBibitem
\bibitem[Straumanis and Aka(1952)Straumanis, and Aka]{si_structure}
Straumanis,~M.; Aka,~E. Lattice parameters, coefficients of thermal expansion, and atomic weights of purest silicon and germanium. \emph{Journal of Applied Physics} \textbf{1952}, \emph{23}, 330--334\relax
\mciteBstWouldAddEndPuncttrue
\mciteSetBstMidEndSepPunct{\mcitedefaultmidpunct}
{\mcitedefaultendpunct}{\mcitedefaultseppunct}\relax
\EndOfBibitem
\bibitem[Kisi and Elcombe(1989)Kisi, and Elcombe]{zno_structure}
Kisi,~E.~H.; Elcombe,~M.~M. u parameters for the wurtzite structure of ZnS and ZnO using powder neutron diffraction. \emph{Acta Crystallographica Section C: Crystal Structure Communications} \textbf{1989}, \emph{45}, 1867--1870\relax
\mciteBstWouldAddEndPuncttrue
\mciteSetBstMidEndSepPunct{\mcitedefaultmidpunct}
{\mcitedefaultendpunct}{\mcitedefaultseppunct}\relax
\EndOfBibitem
\bibitem[Bacon \latin{et~al.}(1964)Bacon, Curry, and Wilson]{benzene_structure}
Bacon,~G.; Curry,~N.~T.; Wilson,~S. A crystallographic study of solid benzene by neutron diffraction. \emph{Proceedings of the Royal Society of London. Series A. Mathematical and Physical Sciences} \textbf{1964}, \emph{279}, 98--110\relax
\mciteBstWouldAddEndPuncttrue
\mciteSetBstMidEndSepPunct{\mcitedefaultmidpunct}
{\mcitedefaultendpunct}{\mcitedefaultseppunct}\relax
\EndOfBibitem
\bibitem[Thewlis(1955)]{lif_structure}
Thewlis,~J. Unit-cell dimensions of lithium fluoride made from Li6 and Li7. \emph{Acta Crystallographica} \textbf{1955}, \emph{8}, 36--38\relax
\mciteBstWouldAddEndPuncttrue
\mciteSetBstMidEndSepPunct{\mcitedefaultmidpunct}
{\mcitedefaultendpunct}{\mcitedefaultseppunct}\relax
\EndOfBibitem
\bibitem[Corporation(2024)]{cuda}
Corporation,~N. CUDA Toolkit Documentation. 2024; \url{https://docs.nvidia.com/cuda}, Accessed: March 31, 2024\relax
\mciteBstWouldAddEndPuncttrue
\mciteSetBstMidEndSepPunct{\mcitedefaultmidpunct}
{\mcitedefaultendpunct}{\mcitedefaultseppunct}\relax
\EndOfBibitem
\bibitem[Li \latin{et~al.}(1999)Li, Eddaoudi, O'Keeffe, and Yaghi]{mof5_structure}
Li,~H.; Eddaoudi,~M.; O'Keeffe,~M.; Yaghi,~O.~M. Design and synthesis of an exceptionally stable and highly porous metal-organic framework. \emph{nature} \textbf{1999}, \emph{402}, 276--279\relax
\mciteBstWouldAddEndPuncttrue
\mciteSetBstMidEndSepPunct{\mcitedefaultmidpunct}
{\mcitedefaultendpunct}{\mcitedefaultseppunct}\relax
\EndOfBibitem
\bibitem[NERSC(2021)]{nersc}
NERSC Perlmutter architecture. 2021; \url{https://docs.nersc.gov/systems/perlmutter/architecture/}, Accessed: October 22, 2024\relax
\mciteBstWouldAddEndPuncttrue
\mciteSetBstMidEndSepPunct{\mcitedefaultmidpunct}
{\mcitedefaultendpunct}{\mcitedefaultseppunct}\relax
\EndOfBibitem
\bibitem[Sun \latin{et~al.}(2020)Sun, Zhang, Banerjee, Bao, Barbry, Blunt, Bogdanov, Booth, Chen, Cui, \latin{et~al.} others]{pyscf_recent}
Sun,~Q.; Zhang,~X.; Banerjee,~S.; Bao,~P.; Barbry,~M.; Blunt,~N.~S.; Bogdanov,~N.~A.; Booth,~G.~H.; Chen,~J.; Cui,~Z.-H. \latin{et~al.}  Recent developments in the PySCF program package. \emph{The Journal of Chemical Physics} \textbf{2020}, \emph{153}\relax
\mciteBstWouldAddEndPuncttrue
\mciteSetBstMidEndSepPunct{\mcitedefaultmidpunct}
{\mcitedefaultendpunct}{\mcitedefaultseppunct}\relax
\EndOfBibitem
\bibitem[Darden \latin{et~al.}(1993)Darden, York, and Pedersen]{particle_mesh_ewald}
Darden,~T.; York,~D.; Pedersen,~L. Particle mesh Ewald: An N log (N) method for Ewald sums in large systems. \emph{The Journal of Chemical Physics} \textbf{1993}, \emph{98}, 10089--10092\relax
\mciteBstWouldAddEndPuncttrue
\mciteSetBstMidEndSepPunct{\mcitedefaultmidpunct}
{\mcitedefaultendpunct}{\mcitedefaultseppunct}\relax
\EndOfBibitem
\bibitem[Petersen(1995)]{particle_mesh_ewald_force}
Petersen,~H.~G. Accuracy and efficiency of the particle mesh Ewald method. \emph{The Journal of Chemical Physics} \textbf{1995}, \emph{103}, 3668--3679\relax
\mciteBstWouldAddEndPuncttrue
\mciteSetBstMidEndSepPunct{\mcitedefaultmidpunct}
{\mcitedefaultendpunct}{\mcitedefaultseppunct}\relax
\EndOfBibitem
\bibitem[Alexei(2016)]{particle_mesh_ewald_gpu_1}
Alexei,~I. Implementation of the Particle Mesh Ewald method on a GPU. 2016\relax
\mciteBstWouldAddEndPuncttrue
\mciteSetBstMidEndSepPunct{\mcitedefaultmidpunct}
{\mcitedefaultendpunct}{\mcitedefaultseppunct}\relax
\EndOfBibitem
\bibitem[Harvey and De~Fabritiis(2009)Harvey, and De~Fabritiis]{particle_mesh_ewald_gpu_2}
Harvey,~M.; De~Fabritiis,~G. An implementation of the smooth particle mesh Ewald method on GPU hardware. \emph{Journal of Chemical Theory and Computation} \textbf{2009}, \emph{5}, 2371--2377\relax
\mciteBstWouldAddEndPuncttrue
\mciteSetBstMidEndSepPunct{\mcitedefaultmidpunct}
{\mcitedefaultendpunct}{\mcitedefaultseppunct}\relax
\EndOfBibitem
\bibitem[Corporation(2024)]{cufft}
Corporation,~N. Fast Fourier Transform for NVIDIA GPUs. 2024; \url{https://developer.nvidia.com/cufft}, Accessed: March 31, 2024\relax
\mciteBstWouldAddEndPuncttrue
\mciteSetBstMidEndSepPunct{\mcitedefaultmidpunct}
{\mcitedefaultendpunct}{\mcitedefaultseppunct}\relax
\EndOfBibitem
\bibitem[Xu and Carter(2018)Xu, and Carter]{heterogenous_cat_review_2}
Xu,~S.; Carter,~E.~A. Theoretical insights into heterogeneous (photo) electrochemical CO2 reduction. \emph{Chemical reviews} \textbf{2018}, \emph{119}, 6631--6669\relax
\mciteBstWouldAddEndPuncttrue
\mciteSetBstMidEndSepPunct{\mcitedefaultmidpunct}
{\mcitedefaultendpunct}{\mcitedefaultseppunct}\relax
\EndOfBibitem
\bibitem[Chen \latin{et~al.}(2020)Chen, Xu, and Mavrikakis]{heterogenous_cat_review}
Chen,~B.~W.; Xu,~L.; Mavrikakis,~M. Computational methods in heterogeneous catalysis. \emph{Chemical Reviews} \textbf{2020}, \emph{121}, 1007--1048\relax
\mciteBstWouldAddEndPuncttrue
\mciteSetBstMidEndSepPunct{\mcitedefaultmidpunct}
{\mcitedefaultendpunct}{\mcitedefaultseppunct}\relax
\EndOfBibitem
\bibitem[Kang \latin{et~al.}(2024)Kang, Zhao, Hait, Gauthier, Kempler, Thurman, Boettcher, and Head-Gordon]{corrosion}
Kang,~R.; Zhao,~Y.; Hait,~D.; Gauthier,~J.~A.; Kempler,~P.~A.; Thurman,~K.~A.; Boettcher,~S.~W.; Head-Gordon,~M. Understanding ion-transfer reactions in silver electrodissolution and electrodeposition from first-principles calculations and experiments. \emph{Chemical Science} \textbf{2024}, \emph{15}, 4996--5008\relax
\mciteBstWouldAddEndPuncttrue
\mciteSetBstMidEndSepPunct{\mcitedefaultmidpunct}
{\mcitedefaultendpunct}{\mcitedefaultseppunct}\relax
\EndOfBibitem
\end{mcitethebibliography}
\end{document}


\maketitle

\newpage
\section{Electron Kinetic Energy Integral}

\begin{align}
T_{\mu\nu} &= -\dfrac{1}{2} \iiint_{\infty} d\vec{r} \ \mu_{\vec{k}_\mu}^*(\vec{r}; \vec{i},a,\vec{A}) \left( \frac{\partial^2}{\partial x^2} + \frac{\partial^2}{\partial y^2} + \frac{\partial^2}{\partial z^2} \right) \nu_{\vec{k}_\nu}(\vec{r}; \vec{j},b,\vec{B}) \notag \\
&=\delta_{\vec{k}_\mu, \vec{k}_\nu} \sum_{\vec{R} \in \mathbb{Z}^3} e^{i \vec{k}_\mu \cdot \vec{R}} C_\mu C_\nu 
\left( \frac{\pi}{p} \right)^{3/2}t_{\mu\nu}^{local}(\vec{A}, \vec{B}+\vec{R},p)\\
    t_{\mu\nu}^{local}(\vec{A}, \vec{B}+\vec{R},p) = & \Biggl(-\frac{j_x (j_x-1)}{2} E^{i_x, j_x-2}_{0,x}(A_x, B_x+R_x, p) + (2j_x + 1) b E^{i_x, j_x}_{0,x}(A_x, B_x+R_x, p)  \notag \\ & - 2b^2 E^{i_x, j_x+2}_{0,x}(A_x, B_x+R_x, p) \Biggr) E^{i_y, j_y}_{0,y}(A_y, B_y+R_y, p) E^{i_z, j_z}_{0,z}(A_z, B_z+R_z, p) \notag \\
    & +\Biggl( -\frac{j_y (j_y-1)}{2} E^{i_y, j_y-2}_{0,y}(A_y, B_y+R_y, p) + (2j_y + 1) b E^{i_y, j_y}_{0,y}(A_y, B_y+R_y, p) \notag \\ & - 2b^2 E^{i_y, j_y+2}_{0,y}(A_y, B_y+R_y, p) \Biggr)  E^{i_x, j_x}_{0,x}(A_x, B_x+R_x, p) 
 E^{i_z, j_z}_{0,z}(A_z, B_z+R_z, p) \notag \\
    & + \Biggl(-\frac{j_z (j_z-1)}{2} E^{i_z, j_z-2}_{0,z}(A_z, B_z+R_z, p)+ (2j_z + 1) b E^{i_z, j_z}_{0,z}(A_z, B_z+R_z, p)  \notag \\ & - 2b^2 E^{i_z, j_z+2}_{0,z}(A_z, B_z+R_z, p) \Biggr)E^{i_x, j_x}_{0,x}(A_x, B_x+R_x, p) E^{i_y, j_y}_{0,y}(A_y, B_y+R_y, p)
\label{eq:kinetic_formula}
\end{align}

\pagebreak
\section{Scaling of $\mathbf{J}$ and $\mathbf{K}$}
\begin{figure}[htb!]
\begin{minipage}{0.49\textwidth}
    \includegraphics[width=\linewidth]{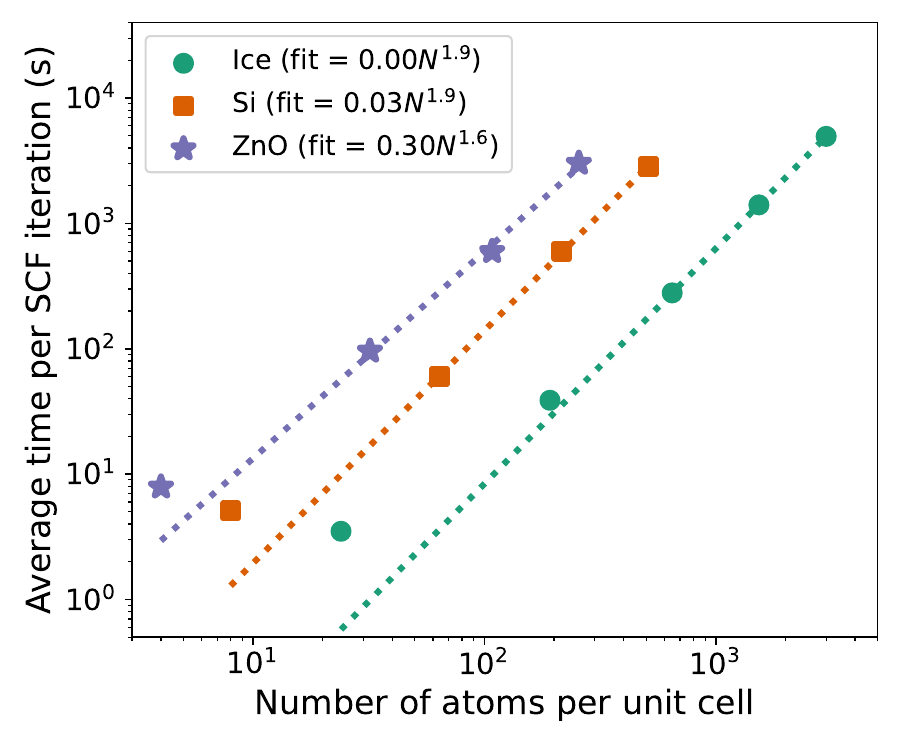}
    \subcaption{$\mathbf{J}$}
\end{minipage}
\begin{minipage}{0.49\textwidth}
    \includegraphics[width=\linewidth]{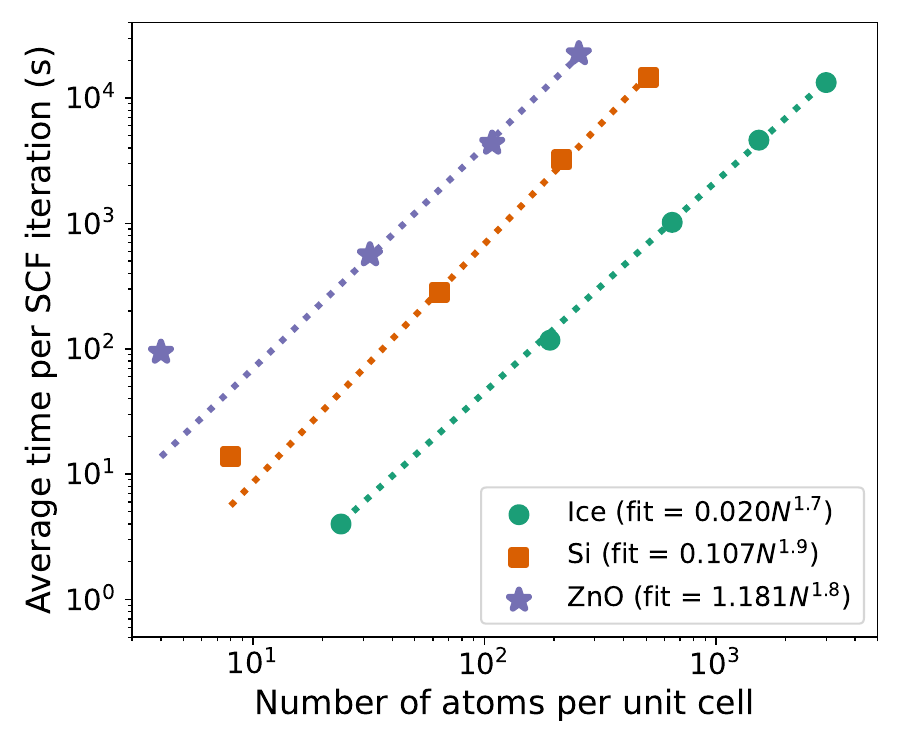}
    \subcaption{$\mathbf{K}$}
\end{minipage}
\caption{Run time per HF SCF iteration (averaged over all cycles) for construction of $\mathbf{J}$ (left) and $\mathbf{K}$ (right) with increasingly large unit cells of ice XI, silicon and wurtzite zinc oxide, on one NVIDIA A100 GPU. The associated power law fits (via fitting to the three largest unit cell sizes) is also shown as a dotted line and the fit parameters are given in the legend. The calculations are all-electron (i.e. pseudopotential free) and utilize the def2-SVP basis (with the $f$ type function on Zn removed). Note that both axes are on log scales.}
\label{fig:material_performance_jk}
\end{figure}

\section{Comparison to NERSC}

\begin{table}[htb!]
\begin{tabular}{l|rrrr|rrrr}
\hline 
     & \multicolumn{4}{c|}{Local Cluster} & \multicolumn{4}{c}{NERSC}      \\ \hline 
GPUs & $\mathbf{J}$      & $\mathbf{K}$       & XC    & Total  & $\mathbf{J}$     & $\mathbf{K}$      & XC   & Total  \\ \hline 
1    & 682.0  & 3457.4  & 90.2  & 4232.0 & 681.9 & 3399.3 & 89.2 & 4171.5 \\
2    & 344.6  & 1738.4  & 46.5  & 2131.7 & 344.3 & 1708.1 & 46.1 & 2099.5 \\
3    & 233.3  & 1161.6  & 32.6  & 1429.7 & 232.9 & 1142.0 & 32.4 & 1408.3 \\
4    & 177.1  & 874.6   & 26.1  & 1079.9 & 177.1 & 858.5  & 25.7 & 1062.4\\\hline 
\end{tabular}
\caption{Average time over all SCF iterations (in s) for construction of $\mathbf{J}$, $\mathbf{K}$, local exchange-correlation (XC) and total time (including other operations like matrix diagonalization) for the $3\times3\times3$ supercell of Si (216 atoms) with PBE0/def2-SVP. The performance on our local cluster is compared to a single NERSC machine. Both systems use NVIDIA A100 GPUs, while the associated CPUs are very similar (AMD EPYC 7742 for the local cluster and AMD EPYC 7763 for NERSC), with 128 threads each.}
\label{tab:nersc_comparison}
\end{table}